\documentclass[aps,twocolumn,nobalancelastpage,amsmath,amssymb,
    nofootinbib,longbibliography]{revtex4}
\usepackage[utf8]{inputenc}
\usepackage{float}
\usepackage{amsmath }
\usepackage{tikz}
\usepackage{graphicx}
\usepackage{dcolumn}
\usepackage{multirow}

\usepackage{bm}
\newcommand{\bZ}{\mathbb{Z}}
\usepackage[colorlinks=true,citecolor=blue,urlcolor=blue]{hyperref}
\begin{document}
\begin{abstract}
We study effectively one-dimensional systems that emerge at the edge of a two-dimensional topologically ordered state, or at the boundary between two topologically ordered states. We argue that anyons of the bulk are associated with emergent symmetries of the edge, which play a crucial role in the structure of its phase diagram. Using this symmetry principle, transitions between distinct gapped phases at the boundaries of Abelian states can be understood in terms of symmetry breaking transitions or transitions between symmetry protected topological phases. Yet more exotic phenomena occur when the bulk hosts non-Abelian anyons. To demonstrate these principles, we explore the phase diagrams of the edges of a single and a double layer of the toric code, as well as those of domain walls in a single and double-layer Kitaev spin liquid (KSL). In the case of the KSL, we find that the presence of a non-Abelian anyon in the bulk enforces Kramers-Wannier self-duality as a symmetry of the effective boundary theory. 
These examples illustrate a number of surprising phenomena, such as spontaneous duality-breaking, two-sector phase transitions, 
and unfreezing of marginal operators at a transition between different gapless phases. 
\end{abstract}

\title{Bulk Anyons as Edge Symmetries:\\ Boundary Phase Diagrams of Topologically Ordered States}

\author{Tsuf Lichtman$^1$} \author{Ryan Thorngren$^{1,2}$} \author{Netanel H. Lindner$^3$} \author{Ady Stern$^1$} \author{Erez Berg$^1$}
\affiliation{$^1$Department of Condensed Matter Physics, The Weizmann Institute of Science, Rehovot, 76100, Israel}
\affiliation{$^2$Center of Mathematical Sciences and Applications, Harvard University, Cambridge, MA 02138}
\affiliation{$^3$Physics Department, Technion, Haifa, 320003, Israel}

\date{\today}
\maketitle

\section{Introduction}
The study of topological phases deals with phenomena which are beyond the Landau symmetry breaking paradigm. However, symmetry still plays a ubiquitous role. The two most studied examples are symmetry protected topological (SPT) phases and topologically ordered phases. SPT phases are gapped phases with an unbroken symmetry, which are distinct from the trivial symmetric phase---as long as the symmetry is preserved, one cannot adiabatically interpolate to a trivial phase without closing the spectral gap. SPT phases may be characterized in several equivalent ways: by the existence of degenerate or gapless edge modes \cite{affleck1988}, by the interplay of symmetry and entanglement in the system \cite{pollmann_entanglement_2010,chen_symmetry_2013}, and by non-local order parameters \cite{kennedytasaki,pollmann_symmetry_2012}.

Topologically ordered phases are non-trivial gapped phases which are stable without imposing any microscopic symmetry. They are distinguished from SPT and symmetry breaking phases by a ground state degeneracy (GSD) which depends on the topology of space. For example, a system on a sphere has a unique ground state in the absence of extra symmetry breaking, while a system on a torus has several \cite{PhysRevB.41.9377}. This torus GSD is associated with superselection sectors of anyons---special quasiparticles which are in general neither bosons nor fermions but may instead have exotic spin, braiding, and fusion rules which characterize the topological order \cite{kitaev_anyons_2005,nayak_non-Abelian_2007}.

As bizarre a phenomenon as it is, topological order may actually also be understood via symmetry principles: these phases spontaneously break emergent higher form symmetries, whose order parameters are the quasiparticle string operators \cite{Nussinov_2009,gukov2013topological,Gaiotto_2015,Feng_2007}. Since the order parameters are loop-like rather than point-like, the number of independent ones, and hence the GSD, depends on the topology of space. For instance, the sphere has no loops, while the torus has effectively one loop (the other constitutes a canonically conjugate basis of ground states \cite{kapustin_coupling_2014}) and so the torus GSD counts the number of symmetry generators, i.e. the anyon superselection sectors.

Boundaries between topological phases are fascinating physical systems. For topologically ordered systems, gapped boundaries have been primarily classified in terms of how the anyons behave at the boundary---some are condensed while others are confined. The consistency conditions for the anyon condensate are by now well-understood \cite{bais_condensate_2009,Kapustin_2011,Kong_2014,Davydov_2013,levin_protected_2013,Fuchs_2013}. More general boundary phases, such as critical points between different condensates and stable gapless boundaries have been studied in specific models \cite{feiguin_interacting_2007,gils_topology_2009,M_nsson_2013,finchain,buican_anyonic_2017} and a more systematic picture of the constraints imposed on conformally-invariant boundaries by modularity is slowly emerging \cite{aasen_topological_2016,Vanhove_2018,chen_topological_2019,ji_non-invertible_2019,kong2019mathematical}.

In this work, we focus on the role of the emergent symmetries, associated with the anyon string operators, for the phase diagrams of general boundaries. We find that the emergent symmetries are a powerful tool for understanding the phase diagram of the boundary. For instance, the gapped anyon condensates can be understood as spontaneous symmetry breaking and SPT phases for the emergent symmetries. Transitions between different boundaries realize familiar Landau universality classes, as well as more exotic ones, such as transitions between SPT phases and deconfined quantum critical points~\cite{Roberts_2019}. The emergent-symmetry principle places the study of boundaries of topological order on the same footing as the study of boundaries of SPT phases, which are understood by anomaly in-flow, and correspond precisely to the case that the topological order is a finite gauge theory~\cite{dijkgraaf1990,Kitaev_2003}. For a complementary approach to ours, motivated by the analogy to gauge theory, see~\cite{thorngren2019fusion}.

We demonstrate these principles in several model systems of increasing complexity. Section \ref{secphyspicture} presents the general physical picture and reviews some of the results. Section \ref{secKSLtricrit} discusses the phase diagram of a domain wall in the Kitaev spin liquid. Section \ref{sectoriccodebilayer} discusses the phase diagram of a bilayer toric code edge. In Section \ref{secKSLgenon} we discuss phase diagram of a domain wall in a Kitaev spin liquid bilayer. In Section \ref{secdiscuss} we summarize and discuss our results. The paper is followed by several appendices discussing technical details.

 \section{Physical Picture and Overview of Results}\label{secphyspicture}

In this section we illustrate the general physical picture using the example of the toric code, and review some of the results of later sections. We will explain how emergent symmetries, non-local order parameters, and dualities in the effective 1d boundary theory are inherited from the topologically ordered bulk.

 We consider a system on a wide cylinder, with two circular edges on the top and bottom. The bottom is held in a reference gapped phases and the interactions at the top edge are tuned. We consider the top edge of the cylinder as an effectively one-dimensional (1d) system, in a limit where the circumference of the cylinder is much larger than its height, while the height is still much larger than the bulk correlation length:
  \begin{equation}\label{eqngeometry}
  L_{\rm circumference} \gg L_{\rm height} \gg \xi_{\rm bulk}.\end{equation}
  The system is shown schematically in Fig. \ref{figcylinder}.

\subsection{Boundaries of the Toric Code} \label{subsecPhysTC}
    To illustrate the general principles, let us consider the example of the toric code topological order. This topological order supports two distinct gapped boundaries with the vacuum, corresponding to the $e$ and $m$ anyon condensates~\cite{bravyi_quantum_1998}. There is a third non-trivial anyon, $em$, but being a fermion it cannot condense.
\subsubsection{Symmetries}\label{subsec2symmetries}

 \begin{figure}
    \centering
    \hspace*{0.5cm}\includegraphics[width=\columnwidth]{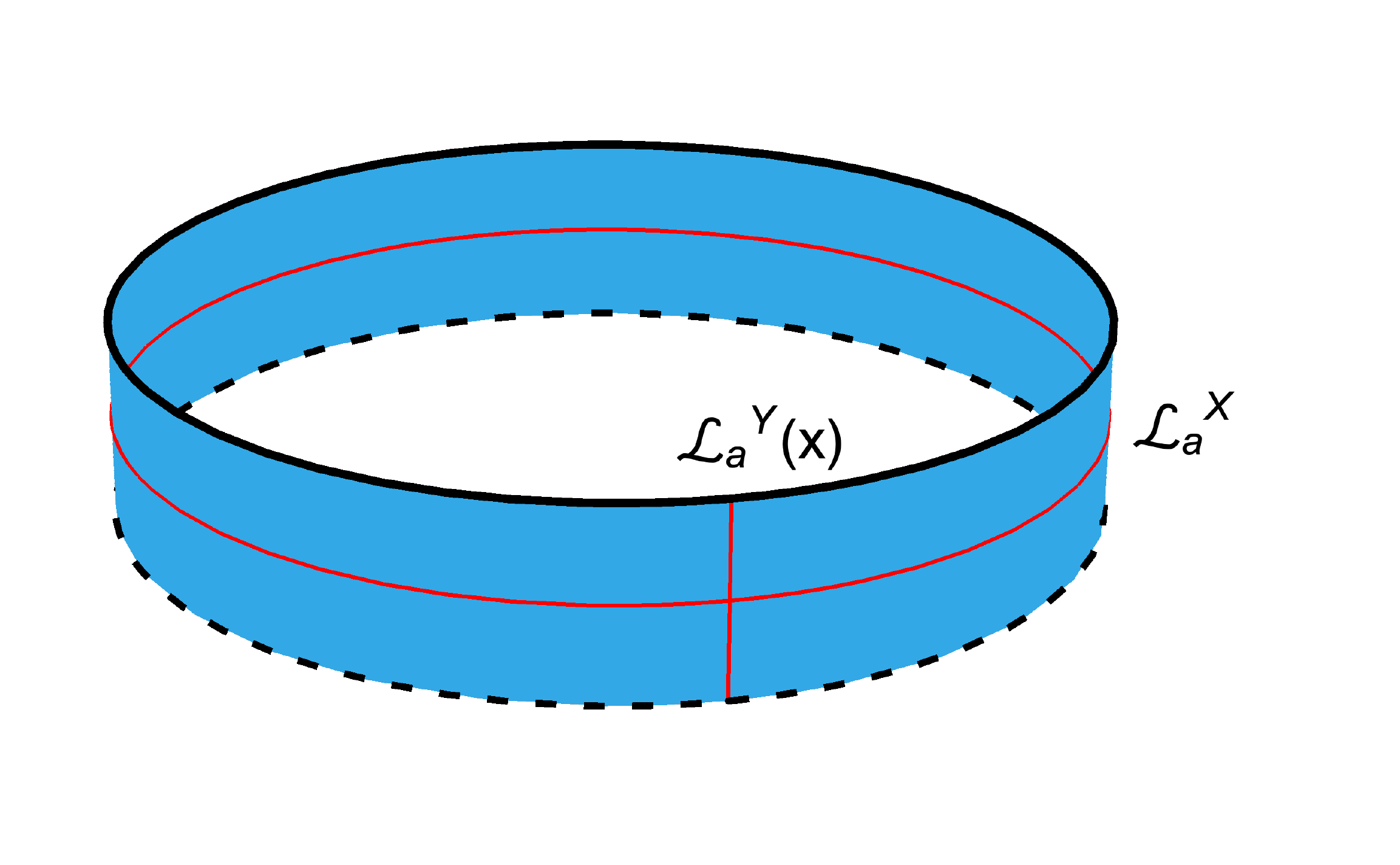}
    \caption{A topological order is placed on a cylinder geometry, where the bottom edge (dashed) is fixed to be in some reference gapped boundary condition, and the top edge (solid) is tuned. We explore the phase diagram of the top edge using emergent symmetries which arise from anyon lines (red) encircling the cylinder, $\mathcal{L}_a^X$. Any anyon which is not condensed at the bottom edge defines a global symmetry of the top edge. Meanwhile, anyon tunneling operators between the two edges act as local operators $\mathcal{L}_a^Y(x)$ in this quasi-1d system. Such operators are charged under the emergent symmetries, according to the braiding rules. A tunneling operator for an anyon which is condensed at both edges corresponds to a local operator with a nonzero VEV, and implies spontaneous symmetry breaking of the emergent symmetries. This way, we can understand transitions between gapped phases at the top edge as symmetry-breaking transitions in this quasi-1d system.}
    \label{figcylinder}
\end{figure}

The symmetry operators in the quasi-1d system are the string operators $\mathcal{L}^X_{e}$, $\mathcal{L}^X_{m}$ encircling the cylinder in Fig. \ref{figcylinder}. These operators are implemented by the creation of a pair of $e$ (resp. $m$) particles, moving one of them adiabatically along the circumference loop $X$, and annihilating them again~\cite{Kitaev_2003}. Note that in the geometry of Fig. \ref{figcylinder}, these operators are considered non-local from the point of view of the quasi-1d system, since they wrap the longest cycle of the cylinder. Because these operators are topological, they can be applied far away from both boundaries. When applied far away from the boundary, $\mathcal{L}^X_{e}$, $\mathcal{L}^X_{m}$ commute with the Hamiltonian up to corrections that decay exponentially with the system size. Thus, $\mathcal{L}^X_{e}$ and $\mathcal{L}^X_{m}$ define emergent symmetries of the effectively 1d system.
 
Let us fix the bottom edge of the cylinder to be gapped via $m$ condensation and consider the phase diagram of the top edge.
The operator $\mathcal{L}^X_m$ acts trivially in this setup. Indeed, being a topological operator we can freely move it to the bottom edge, where it is absorbed into the $m$ condensate.

The operator $\mathcal{L}_{e}^{X}$ is also topological, but cannot be absorbed into the $m$ condensate. This means that we expect it to act non-trivially in the low energy Hilbert space. Further, the $\bZ_2$ fusion rule of $e$ implies $(\mathcal{L}^X_{e})^2 = 1$, so $\mathcal{L}_{e}^{X}$ generates a $\bZ_2$ global symmetry. Moreover, since the $e$ anyon is a self-bosons, open string operators of the $e$ particle commute with each other. Thus, $\mathcal{L}_{e}^{X}$ can be thought of as a product of local unitary symmetry operators at the edge.

The existence of this symmetry in the quasi-1d system rests only upon having a boundary of the toric code. Indeed, symmetries defined in this way cannot be explicitly broken by any local perturbation, unless the bulk goes through a phase transition. The phases of a 1d system with such global symmetry are in one-to-one correspondence with the possible boundary conditions of the toric code~\cite{thorngren2019fusion}.
Thus, the study of the toric code boundary phase diagram is the same as the study of phase diagrams of 1d systems with a $\bZ_2$ symmetry.\footnote{
If one repeats the same construction for the double semion model, one finds a $\bZ_2$ symmetry with only one gapped phase where the symmetry is spontaneously broken. This is because in this case, the $\bZ_2$ symmetry cannot be thought of as a product of local operators. In other words, the symmetry is anomalous---the ends of the symmetry string carry (fractional) charge under the symmetry. This charge is inherited from the topological spin of the semions, which braid non-trivially with themselves, unlike the $e$ and $m$ quasiparticles which are self bosons. It is more precise to say that the bulk specifies not just the fusion algebra of the symmetry lines, but also their crossing relations (i.e. $F$-symbols), which encode their anomaly in the group-like case, and a fusion category in the general case \cite{aasen_topological_2016,chang_topological_2019,thorngren2019fusion}.}


Tracking the fate of the $\bZ_2$ symmetry can help us in classifying the phases of the edge. For example, when the top edge is in the $m$ condensate (same as the bottom edge), the topological ground state degeneracy (GSD) of the system is two, and we consider the $\bZ_2$ symmetry to be spontaneously broken. Indeed, the two ground states in this case are distinguished by the presence of a long $e$ string wrapping the circumference of the cylinder in one state but not in the other. The operator $\mathcal{L}^X_{e}$ exchanges these two ground states by adding an extra such $e$ string, hence it is a spontaneously broken symmetry.

On the other hand, when the top edge is in the $e$ condensate, there is no GSD, and we consider the $\bZ_2$ symmetry to be preserved. This is so because in this case, the global symmetry operator $\mathcal{L}^X_e$ can be moved to the top edge (the $e$ condensate) where it is absorbed and acts trivially.

As we tune parameters on the top edge, we may encounter phase transitions between those two condensates. A second order phase transition between the $e$ and the $m$ condensates is characterized by a spontaneous breaking of the above $\bZ_2$ symmetry, and is known to be in the $c=1/2$ Ising universality class~\cite{Yang_2014,chen_topological_2019,ji_non-invertible_2019,ji_categorical_2020,Barkeshli_2014}. It is also possible to induce a first order transition between the two gapped phases. In the phase diagram of the edge, the lines of first and second order transitions meet at a $c=7/10$ tricritical Ising point.

\subsubsection{Order parameters}\label{subsecorderpara}

To further understand the breaking of the $\mathcal{L}^X_{e}$ symmetry, we determine the relevant order parameter. Consider an $m$ string operator $\mathcal{L}_m^Y(x)$ which brings an $m$ particle from the $m$ condensate at the bottom edge to the top edge \cite{levin_protected_2013,hung_ground-state_2015}, at fixed circumferential coordinate $x$. In our quasi-1d setup, this is considered as a local operator. We observe
\begin{equation}\mathcal{L}^X_{e}\mathcal{L}_m^Y(x) = -\mathcal{L}_m^Y(x)\mathcal{L}^X_{e},\end{equation}
so $\mathcal{L}_m^Y(x)$ may serve as an order parameter for the $\bZ_2$ symmetry breaking.

Indeed, in the case where the top edge is also in the $m$ condensate, the short $m$ string operator $\mathcal{L}_m^Y(x)$ is long range ordered:
\begin{equation}\langle \mathcal{L}_m^Y(x) \mathcal{L}_m^Y(0)\rangle \rightarrow \text{const.} \neq 0\end{equation}
for large $|x|$. The associated symmetry $\mathcal{L}^X_{e}$ is spontaneously broken and the two ground states can be distinguished by the sign of the vacuum expectation value (VEV) $\langle \mathcal{L}_m^Y(x)\rangle$.

On the other hand, if the top edge supports the $e$ condensate, $m$ particles are confined there because they braid non-trivially with $e$. Thus, we expect $\langle \mathcal{L}_m^Y(0) \mathcal{L}_m^Y(x) \rangle$ to decay exponentially quickly with increasing $x$. The operator $\mathcal{L}_m^Y(x)$ does not develop a VEV, and the symmetry $\mathcal{L}^X_{e}$ is preserved. Hence the GSD in this case is one.

In general, if there is an anyon $a$ which is condensed at both the top and the bottom edges, the tunneling string operator $\mathcal{L}_a^Y(x)$ is long range ordered. Each global symmetry $\mathcal{L}_b^X$ which is associated with an anyon $b$ that braids non-trivially with $a$ is spontaneously broken.
 
\subsubsection{Dualities}

If we exchanged $e$ with $m$ everywhere in the above discussion, all our claims would still be accurate. Indeed, the topological order of the toric code has a well-known anyon permuting symmetry (sometimes called a ``duality") which exchanges $e$ and $m$ \cite{Nussinov_2005,bombin_topological_2010}. In general, such dualities are not symmetries of the microscopic Hamiltonian, but are nonetheless a robust feature of topologically ordered states \cite{barkeshli_symmetry_2014,Etingof_2010}. 

Each anyon permuting symmetry has a twist defect associated with it.
When an anyon crosses such a twist defect line, it may change its type. For example, the toric code supports twist defects which exchange $e$ and $m$ \cite{bombin_topological_2010}.

In our quasi-1d setup, anyon permuting symmetries of the bulk act as dualities on the edge. This is done by wrapping a bulk defect line around the circumference of the cylinder, and then fusing it onto one of the edges. 
In the toric code example, applying $e\leftrightarrow m$ to one of the edges interchanges the symmetry-broken and symmetry-preserving phases. Evidently, applying the duality to the top edge exchanges the $m$ and $e$ condensates, where $\mathcal{L}^X_{e}$ symmetry is broken and unbroken, respectively. On the other hand, applying $e\leftrightarrow m$ to the bottom edge transforms the reference boundary condition from the $m$ to the $e$ condensate. Now, the relevant symmetry line is $\mathcal{L}^X_{m}$ instead of $\mathcal{L}^X_{e}$, and the $e$ boundary of the top edge is identified with the symmetry broken phase, instead of the symmetric one.

In the first case described above, the dynamical edge has changed, and the symmetry operator remained the same, while in the second case the identification of global symmetries has changed, while the dynamical edge remained the same. In both cases, the symmetry breaking labels of the phases are exchanged, so this bulk defect acts as a Kramers-Wannier duality transformation of the quasi-1d system \cite{ho_edge-entanglement_2015}.

\subsection{Boundaries of Non-Abelian Topological Phases}

We are also interested in analyzing boundaries of non-Abelian topologically ordered phases. For these boundaries, we restrict ourselves to non-chiral phases, and again employ the same cylinder geometry of Fig. \ref{figcylinder}, using a reference gapped boundary condition. We will also study the related problem of domain walls in chiral phases below.

As before, for every bulk anyon $a$ we have a topological line operator $\mathcal{L}_a^X$. The reference boundary condition may be described as a condensate of some subset of these anyons \cite{bais_condensate_2009,Kong_2014}. If $a$ is condensed at the reference boundary, the operators $\mathcal{L}_a^X$ act trivially because they may be absorbed at the bottom edge. However, for anyons $a$ \emph{not} in the condensate, $\mathcal{L}_a^X$ defines a non-trivial topological operator. Because these operators commute with the Hamiltonian, we regard them as global symmetries.

The algebra of the nontrivial operators $\mathcal{L}_a^X$ is closed, but it is not a group algebra. Instead, the operators satisfy the fusion algebra of the associated bulk anyons:
\begin{equation}\label{eqnfusionalgsymms}
    \mathcal{L}_a^X \mathcal{L}_b^X = \sum_c N_{ab}^c \mathcal{L}_a^X.
\end{equation}
Where $N_{ab}^c$ are the fusion coefficients of the bulk theory.
These so-called ``fusion category symmetries" have been studied from a number of different perspectives, see for instance \cite{feiguin_interacting_2007,aasen_topological_2016,buican_anyonic_2017,Vanhove_2018,chang_topological_2019,thorngren2019fusion}. They are more difficult to work with than ordinary symmetries, but many familiar concepts such as spontaneous symmetry breaking, SPT phases, and anomalies still apply.

As before, these symmetries are unbreakable by local operators, and we find a general correspondence between the phase diagram of the boundary and the phase diagram of 1d systems enjoying this associated fusion category symmetry.\footnote{We note that the symmetry depends crucially on the phase of the reference boundary. For example, let us consider $D_8$ gauge theory. This is a non-chiral, non-Abelian phase whose gapped boundaries can be described as 1d systems enjoying a $D_8$ global symmetry. On the other hand, each boundary phase has an equivalent description as a 1d system with $\bZ_2 \times \bZ_2 \times \bZ_2$ global symmetry and the cubic anomaly---or as a 1d system with a certain symmetry algebra called the Tambara-Yamagami fusion category \cite{thorngren2019fusion}.}

\subsection{Example: Domain Walls of the Kitaev Spin Liquid}\label{subsecKSLpic}

Perhaps the simplest example of a boundary of a non-Abelian phase is the case of a domain wall in the Kitaev spin liquid (KSL) \cite{kitaev_anyons_2005}. The KSL is a gapped, chiral, non-Abelian topological order. Domain walls are 1d boundaries between this phase and itself. A domain wall is equivalent, by folding the system around the domain wall, to a boundary of a bilayer of a KSL and its chiral partner $\overline{\text{KSL}}$ with the vacuum.

\begin{figure}
    \centering
    \hspace*{-1cm}\includegraphics[width=\columnwidth]{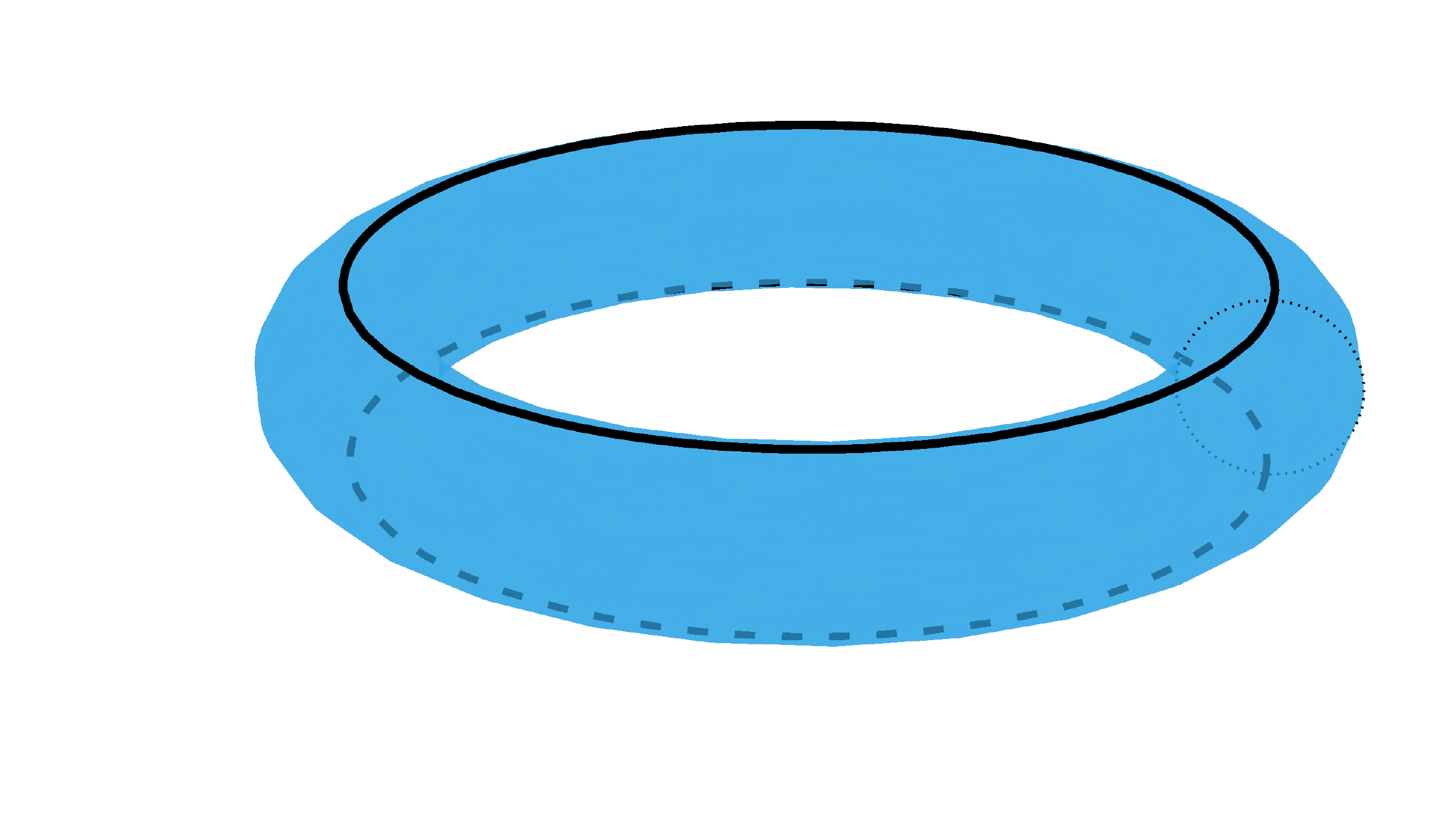}
    \caption{In the study of domain walls in topological order, a natural quasi-1d geometry is given by a thin torus with the domain wall (solid black) running along the top cycle. We can relate this geometry to the cylinder geometry by squashing the cross-section of the torus. If one begins with a chiral theory, the squashed geometry is non-chiral, being a product of the original theory with its anti-chiral partner. The domain wall is placed at the top of the cylinder, while the bottom boundary (dashed) is in the ``fold" gapped boundary condition. Then, we may invoke the symmetry methods described in Fig. \ref{figcylinder}.}
    \label{figtorus}
\end{figure}

We place the KSL on a torus, and study the phase diagram of a domain wall by tuning the Hamiltonian along the top cycle, see Fig. \ref{figtorus}. The torus geometry is related to the cylinder geometry of the bilayer (Fig. \ref{figcylinder}) by squashing the cross-section of the torus. 
On the bottom edge of the cylinder we obtain the ``fold" boundary condition that ``glues" the KSL and $\overline{\text{KSL}}$. With respect to this boundary condition, using the procedure described in \ref{subsecPhysTC} we obtain a symmetry operator for each anyon in the original chiral theory. The symmetry operators satisfy the same fusion rules and crossing relations (F-moves) of the anyons in the KSL bulk \cite{kitaev_anyons_2005}.


In the case of the KSL, there are two nontrivial anyons: a fermion $\psi$, and a non-Abelian anyon $\sigma$. In the cylindrical geometry, we denote the operators that wind a $\psi$ particle along the circumference of the cylinder in the KSL and $\overline{{\text{KSL}}}$ as $\mathcal{L}_\psi^X$ and $\mathcal{L}_{\bar{\psi}}^X$, respectively. Since the bottom edge of the cylinder is in the ``fold'' boundary conditions, these two operators act in the same way, as a $\bZ_2$ symmetry: $(\mathcal{L}_\psi^X)^2=1$. On the other hand, as it will turn out, $\mathcal{L}_\sigma^X$ (as well as $\mathcal{L}_{\bar{\sigma}}^X$) act as the Kramers-Wannier (KW) duality associated with $\mathcal{L}_\psi^X$. This will be argued qualitatively in Sec. \ref{secKSLtricrit} and derived from the fusion rules in Appendix \ref{appKWduality}. We stress that here, in contrast to the toric code edge case, the Kramers-Wannier duality acts as a \emph{physical symmetry} of the system. This is since, unlike the defect of the toric code, the $\sigma$ anyon of the KSL is a deconfined excitation, and hence $\mathcal{L}^X_\sigma$ commutes with the Hamiltonian.

The domain wall phases of the KSL correspond to the phases of a 1d system which is invariant under a $\bZ_2$ symmetry and its associated KW duality. 
There are two such stable phases: one is gapped, and the other is gapless. The gapped phase is analogous to a coexistence phase at a first-order transition between a $\bZ_2$ preserving and a $\bZ_2$ breaking phase, which are swapped under KW duality. This phase corresponds to the trivial ``gluing" domain wall of the KSL boundary.

The stable gapless phase corresponds to the ``cut open" domain wall of the KSL, described by the (non-chiral) $c = 1/2$ Ising  conformal field theory (CFT). This theory enjoys its usual $\bZ_2$ spin flip symmetry and its Kramers-Wannier duality. The generic continuous phase transition between the gapped and the gapless phases respecting the full symmetry algebra is the $c = 7/10$ tricritical Ising model \cite{chang_topological_2019}. The generic phase diagram of a KSL domain wall is shown in Fig.~\ref{figtricrit}.

This picture can be related to the microscopic Kitaev Honeycomb model. In that model, the local spin degrees of freedom are decomposed into Majorana fermions, and the chiral edge carries a $c=1/2$ Majorana mode. The cut open domain wall corresponds to two counter-propagating Majorana modes. In the fermion variables, the KW duality is implemented by the chiral fermion parity operator.

The presence of the duality symmetry $\mathcal{L}^X_\sigma$ stabilizes the ``cut open" domain wall. One might expect any small inter-edge tunneling to generate a two-body mass term between the two edge modes, gapping the domain wall immediately. However, the two-Majorana mass term is not a local operator in the spin degrees of freedom of the honeycomb model. 
Indeed, this mass term is odd under the KW duality $\mathcal{L}^X_\sigma$. This additional symmetry is seen to encode the appropriate notion of locality on the edge degrees of freedom.

\subsection{General Principles and Further Examples}

To summarize the method we outlined so far, we introduce three general principles to analyze boundary phase diagrams of topologically ordered states. Those principles apply both to the Abelian and non-Abelian cases. We refer to the cylindrical geometry of Fig. \ref{figcylinder}, where the bottom boundary is gapped in some reference anyon condensate, and the top edge is dynamical.
\begin{enumerate}
\item Anyon line operators of the bulk act as emergent symmetries of the effectively 1d low energy theory, constraining the allowed edge interactions. Anyons which are condensed at the reference boundary give rise to line operators which act trivially (gauge symmetries), while line operators of anyons which are confined at the reference boundary act as global symmetries. In the case of a non-Abelian topological order in the bulk, some of the symmetries of the edge may act as duality operations; we refer to these as ``duality symmetries''.
\item For each anyon which is condensed at the reference boundary, we can define a line operator which creates it at the reference boundary and drags it to the boundary we study. From the point of view of the quasi-1d system, this is a local operator, that carries a certain charge under the emergent symmetries. Such operators serve as order parameters for the dynamical edge, and can detect spontaneous breaking of the emergent symmetries. 
\item Anyon permuting symmetries of the bulk topological order act as dualities on the phase diagram of the edge (these are \emph{not} symmetries of the Hamiltonian, unlike the duality symmetries defined above). The ``duality frame" of the system is determined by the boundary condition on the reference edge. The identification of the symmetries and order parameters of the quasi 1d system depends on the choice of reference boundary condition.
\end{enumerate}

In the rest of this subsection, we outline how these principled are applied to two additional examples, highlighting some additional aspects.

\subsubsection{Boundaries of a Toric Code Bilayer}
In section \ref{sectoriccodebilayer} we analyze an edge of a toric code bilayer. This edge has an interesting phase diagram, and is equivalent to a one-dimensional system with a  $\bZ_2 \times \bZ_2$ symmetry. In addition to the various spontaneously symmetry broken phases, such a system supports a SPT phase. After fixing a boundary condition, each of these gapped phases corresponds to a particular anyon condensate at the boundary of the toric code bilayer. Moreover, the rich phase diagram of the toric code bilayer boundary can be read off directly from the known phase diagram of a $\bZ_2 \times \bZ_2$ symmetric 1d system~\cite{verresen_gapless_2019}. The phase diagram enjoys various dualities, corresponding to the various twist defects of the toric code bilayer.

\subsubsection{Domain Walls in a Kitaev Spin Liquid Bilayer}

In Section \ref{secKSLgenon} we turn to our main example: domain walls in a Kitaev spin liquid bilayer. The relevant symmetry algebra is dubbed the $\textnormal{Ising} \times \textnormal{Ising}$ category symmetry. This symmetry algebra consists of a $\bZ_2 \times \bZ_2$ symmetry, along with the two associated KW dualities acting as additional symmetries. The system supports a stable gapless phase, corresponding to the ``cut open'' domain wall, in which both layers are cut, and two pairs of counter-propagating $c=1/2$ Majorana edge modes are exposed. On the gapped side, the system supports three distinct phases. The first corresponds to a trivial domain wall, in which each layer is ``healed'' separately. The second is the genon twist defect line~\cite{barkeshli_genons_2012}, in which the healing is performed in a swapped fashion. Interestingly, there is a third phase that corresponds to an intermediate toric code region connecting the two layers. This phase may be understood by recalling that the toric code can be obtained from a bilayer of KSL and $\overline{\text{KSL}}$ by condensing a bound state $\psi\overline{\psi}$, of fermions from each layer~\cite{bais_condensate_2009,burnell_anyon_2018}. 

Applying the methods developed in this work, 
we are able to construct several diagrams around different critical points of this system. This includes an elaborate analysis of the possible transitions between the different phases, and the effective field theories describing them. 
Some of the highlights of this analysis are as follows:
\begin{itemize}
    \item In addition to the $c=1$ cut-open gapless phase, there are several additional stable gapless phases. In particular, there is a $c=1$ gapless phase that is distinct from the cut-open one. This phase, described by an orbifold CFT, is distinguished from the cut-open phase in two main ways: first, it has a doubly degenerate ground state, while the cut-open phase has only one; second, it has a symmetry allowed marginal perturbation, whereas the cut-open phase has none. 
    \item The phase transition between the orbifold and the cut-open phases occurs via a $c=3/2$ critical point. At this point, the two KW duality symmetries which are preserved in the cut-open phases become spontaneously broken, while their product is preserved. The spontaneous breaking of those duality symmetries is responsible for the phenomenology of the orbifold phase.
    \item  The $c=1$ orbifold also describes the phase transition between the genon and trivial domain walls of the KSL bilayer. The transitions between the ``toric code gluing" domain wall and the two other gapped domain walls are described by a $c=1/2$ Ising CFT with multiple degenerate ground states. For instance, for our choice of boundary conditions, the transition between the toric code gluing (GSD 6) and the trivial domain wall (GSD 9) is described by an Ising CFT with 5 ground states.
\end{itemize}

To make contact with a concrete microscopic model we study a chain of genon twist defects in a KSL bilayer. A related model has been studied as a string-net with tension, primarily numerically, in the context of topology-changing phase transitions by Gils et al. \cite{gils_topology_2009,gils_ashkinteller_2009,gils_anyonic_2013}. The emergent symmetries allow us to derive the phase diagrams for this model and argue for their robustness to arbitrary perturbations. We are also able to identify the operators responsible for the various phase transitions. This model provides explicit realizations for the gapped phases and some of the possible phase transitions at a domain wall in a KSL bilayer.

\section{Slicing the Kitaev Spin Liquid and The Tricritical Ising CFT}\label{secKSLtricrit}

In this section, we expand on the phase diagram of a domain wall in the Kitaev spin liquid (KSL). This system has a trivial domain wall as well as a stable gapless one, hosting the $c = 1/2$ Ising CFT. We will describe the phase diagram from an emergent symmetries perspective as well as from a lattice perspective.

\subsection{Domain Walls of the Kitaev Spin Liquid and Ising Symmetry}\label{subsec4symm}

First, following Section \ref{subsecKSLpic}, we identify the emergent symmetries that act in the quasi-1d system associated with the domain wall. The topological order of the KSL is a TQFT described by the Ising braided fusion category with $\nu = 1$ in \cite{kitaev_anyons_2005}, which we call Ising for short. It has three anyons, $1$ (the trivial anyon), $\psi$ (a fermion), and $\sigma$ (the non-Abelian vortex) with the fusion rules
\begin{equation}\label{eqnisingfusion}
\begin{gathered}
\psi^2 = 1 \\
\psi \sigma = \sigma \\
\sigma^2 = 1 + \psi.
\end{gathered}
\end{equation}

We start with a torus geometry as in Fig. \ref{figtorus}. A squashing procedure yields a cylinder whose bulk carries the doubled topological order ${\textnormal{Ising}} \times \overline{\textnormal{Ising}}$, while its bottom edge is in the fold reference boundary condition. This implies that the fusion algebra \eqref{eqnisingfusion} acts as a global symmetry of the top edge.

The fusion rule $\psi^2 = 1$ implies that $\mathcal{L}_\psi^X$ generates a $\bZ_2$ global symmetry of the top edge. The fusion rules for $\sigma$, however, imply that $\mathcal{L}_\sigma^X$ defines a \emph{non-invertible} global symmetry. Indeed, $(\mathcal{L}_\sigma^X)^2=1+\mathcal{L}_\psi^X$ is proportional to the operator that projects onto $\mathcal{L}_\psi^X$-even states. The operator $\mathcal{L}_\sigma^X$ is a duality symmetry, associated with a Kramers-Wannier duality (see Appendix \ref{appgaugingKW}). We will motivate the identification of $\mathcal{L}^X_\sigma$ with the KW duality further below.

\begin{figure} 
    \centering
    \includegraphics[width=\columnwidth]{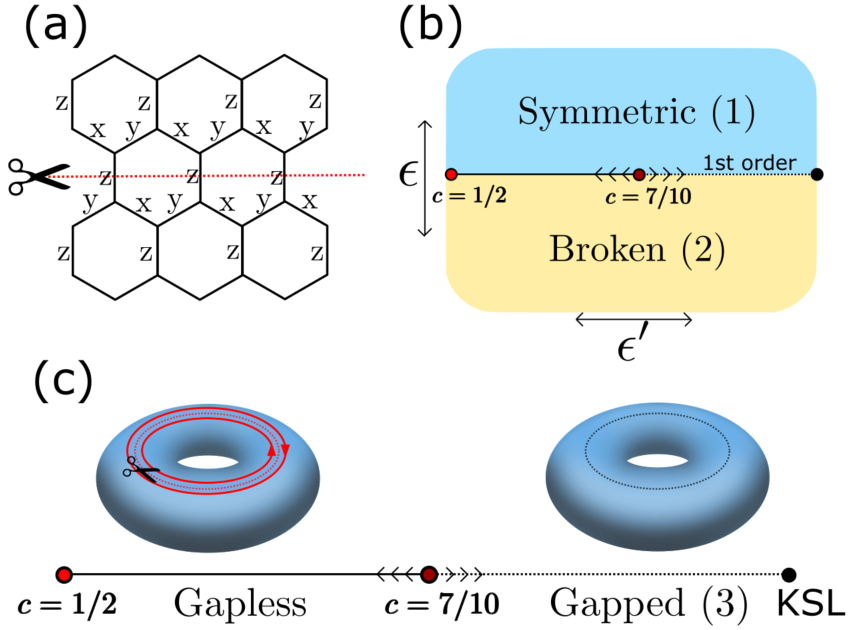}  
    \caption{(a) The Kitaev honeycomb model, with letters indicating the direction of the ferromagnetic couplings on each edge. We gradually turn off some of the bonds, which results in a ``decoupling transition".  (b) The tricritical Ising model phase diagram. The c=7/10 point has 2 relevant operators: $\epsilon$ is duality odd, and tunes between the 2 gapped phases, while $\epsilon'$ is duality even and tunes between the c=1/2 Ising CFT and a threefold degenerate gapped coexistence line. (c) The phase diagram of the 1 dimensional KSL-KSL boundary is the self dual part of the tricritical Ising model phase diagram. This is so since KW duality is enforced as a symmetry of the quasi 1d system.}
\label{figtricrit}
\end{figure}

Thus, we can identify the domain walls of the Kitaev spin liquid with 1d systems which are invariant under a $\bZ_2$ global symmetry, and are also KW self-dual. We refer to this structure as \emph{Ising-category symmetry}, where Ising refers to a fusion category with the fusion rules \eqref{eqnisingfusion}.\footnote{More precisely, there are two fusion categories associated with the Ising fusion rules, which differ by their Frobenius-Schur indicator $(-1)^{(\nu^2-1)/8}$ \cite{kitaev_anyons_2005}. We study symmetric phases associated with the positive indicator. The other category is $SU(2)_2$, see \cite{chang_topological_2019} for more details.}
There is a unique stable gapped phase, with three degenerate ground states. This three ground states in this phase corresponds to the direct sum of a $\bZ_2$ symmetry preserving state and two $\bZ_2$ symmetry breaking states.
Under the duality symmetry, the $\bZ_2$ symmetric state transforms into a linear combination of the $\bZ_2$ breaking states. 
We denote this phase as
\begin{equation}{\rm Brok} + {\rm Symm}
\label{eqbroksymm}
\end{equation}
This gapped phase is equivalent to a system tuned to a first order phase transition between the $\bZ_2$ ordered and disordered phases, see Fig. \hyperref[figtricrit]{\ref{figtricrit}b}. Since KW duality is enforced as physical symmetry, the first order transition line becomes a stable phase. I.e., the ordered and disordered ground states are degenerate without any fine tuning. This degeneracy cannot be lifted by any local perturbation, because favouring one over the other breaks the duality symmetry. The $\text{Brok}+\text{Symm}$ phase can be viewed as a phase where the duality symmetry is spontaneously broken, in the sense that it has ground states that are not invariant under the duality symmetry.

The $\text{Brok}+\text{Symm}$ gapped phase corresponds to the trivial domain wall of the KSL. Indeed, the KSL on the torus with a trivial domain wall has three ground states, corresponding to the superselection sector of an anyon $1$, $\psi$, or $\sigma$ encircling the long (horizontal) direction. Let us denote these three states as $|1\rangle$, $|\psi\rangle$, $|\sigma\rangle$. The symmetries $\mathcal{L}_a^X$ act on these states by fusion. We see that $|1\rangle$ and $|\psi\rangle$ form a closed orbit under $\mathcal{L}_\psi^X$, corresponding to two $\bZ_2$ symmetry breaking states, while $|\sigma\rangle$ is $\bZ_2$ symmetry preserving. This matches our description above. Further, $\mathcal{L}_\sigma^X|\sigma\rangle = |1\rangle + |\psi\rangle$, in accordance with the expected action of the KW duality (see Appendix \ref{appgaugingKW}).

There exists another stable Ising symmetric phase which is gapless and is described by the critical Ising CFT. This phase has a unique ground state and is famously KW self-dual. To see why this is a stable phase, recall that the Ising critical point has two relevant perturbations: a longitudinal magnetic field, and a transverse field, which is the tuning parameter for the critical point. The longitudinal field is forbidden by the $\bZ_2$ symmetry, while the transverse field is forbidden by the KW duality symmetry~\cite{kramersandwannier}.

One can understand the duality transformation of the transverse field operator as follows. With one sign of the deviation of the transverse field from its critical value, this term perturbs the critical point into the disordered phase, while with the other sign perturbs it into the ordered phase. Since these two phases are related by KW duality, this operator must change sign when we apply $\mathcal{L}_\sigma^X$.

We interpret this stable gapless phase as the ``cut-open'' domain wall in the KSL, which hosts two counter-propagating $c = 1/2$ chiral modes, matching the operator content of the critical Ising CFT.

While the critical gapless phase has no symmetric relevant operators, there are symmetric \emph{irrelevant} operators. One such operator eventually tunes the theory to the tricritical Ising point, a CFT of central charge $c=7/10$, beyond which the symmetry breaking transition becomes first order. This first order line is the Ising-category symmetric stable gapped phase $\text{Brok}+\text{Symm}$ described above. The Ising-category symmetry of the $c = 7/10$ CFT was studied in \cite{chang_topological_2019}. There is one $\bZ_2$ and KW duality-invariant relevant perturbation, known as $\epsilon'$ \cite{francesco_conformal_1997}, which drives us into either of the two phases depending on its sign, see Fig. \hyperref[figtricrit]{\ref{figtricrit}b} and \hyperref[figtricrit]{\ref{figtricrit}c}. Thus, we conclude that the transition between the gapless and gapped phases at a domain wall in a KSL is generically of the tricritical Ising universality class.

We will further interpret both of these domain wall phases and the continuous transition between them in the context of the Kitaev honeycomb model below.

\subsection{The Honeycomb Model}

To make contact with a concrete microscopic model, we consider the Kitaev honeycomb model in its non-Abelian phase \cite{kitaev_anyons_2005}. This is a spin model on the honeycomb lattice, with the spins sitting at the vertices. The Hamiltonian consists of spin-spin interactions and a transverse field term, and is given by:
\begin{equation}\label{eqnhoneycomb}
    H=-\sum_{\alpha}\sum_{\alpha\  {\rm links}\ jk}J_\alpha \sigma^{\alpha}_{j} \sigma^{\alpha}_{k} -\sum_{i}h_\alpha \sigma^{\alpha}.
\end{equation}
Where ${\alpha=x,y,z}$, see Fig. \hyperref[figtricrit]{\ref{figtricrit}a}. The non-Abelian phase may be accessed with the isotropic choice $J_{\alpha}=J$, $h_{\alpha}=h$, for sufficiently small but non-zero $h$ \cite{kitaev_anyons_2005}.

The model is solved by a transformation to Majorana fermion variables along with a $\bZ_2$ gauge field. In these variables, for $h = 0$, the system is equivalent to a free Majorana fermion moving in the background of a $\bZ_2$ gauge field. The ground state is in the sector with no $\bZ_2$ fluxes, and in this sector, the Majorana fermion is massless. 
Turning on a small $h$ breaks time reversal symmetry, and the spectrum becomes that of a gapped $p+ip$ topological superconductor. In terms of the physical degrees of freedom, this phase is actually topologically ordered---it supports two nontrivial anyons, a vortex $\sigma$ and a fermion $\psi$ realizing the fusion rules \eqref{eqnisingfusion} above.

If one terminates the Hamiltonian \eqref{eqnhoneycomb} at an edge, one finds gapless modes, which are described in the Majorana variables as the $c = 1/2$ chiral mode of a $p+ip$ superconductor \cite{read_paired_2000}. Bringing two such $c = 1/2$ edges of the KSL near each other, one chiral and one anti-chiral, and considering the system as a single non-chiral Majorana mode, one might expect that a local interaction can turn on a mass term, coupling the two edges together and creating a gapped domain wall. However, because the Majorana operators are non-local in the physical spin degrees of freedom, it turns out that the mass term is forbidden, and instead the gapless domain wall is stable.\footnote{In a recent work \cite{aasen2020electrical}, similar results were described (see their Section IIIA). Moreover, a scheme to detect the KSL using a superconducting ``bridge" to turn electrons into emergent fermions was discussed.} This matches the gapless Ising-category symmetric phase we observed above. Below, we describe how the emergent symmetry forbids the mass term.

\subsection{Two edges and the cut torus}

We place the honeycomb model on a torus and consider a circular edge between the model and itself by turning off the bond interactions along a cycle. We will then perturbatively turn these interactions back on. The effective domain wall Hamiltonian consists of two counter-propagating Majoranas:
\begin{equation}
    H=iv\int dx (\gamma_{R}\partial_{x}\gamma_{R}-\gamma_{R}\partial_{x}\gamma_{R}).
\end{equation}
While the Majorana operators $\gamma_{L,R}$ themselves are non-local in terms of the original spin variables, polynomials in the bilinears $\gamma_L \partial_x \gamma_L$ and $\gamma_R \partial_x \gamma_R$ are local. We can think of the $\gamma_{L,R}$ operators as living at the ends of string operators with $\bZ_2$ fusion rules (these are the Wilson lines of the $\bZ_2$ gauge field mentioned above). As long as all the strings can be connected without any string going off to infinity, the operator is local \cite{kitaev_anyons_2005,Chen_2018}. An operator like the mass term $\gamma_L \gamma_R$ is non-local since the string for $\gamma_L$ and the string for $\gamma_R$ are located in different halves of the system and cannot be connected locally.

According to these rules, the most relevant operator which may be generated by local spin interactions across the cut is the four-fermion term
\begin{gather}\label{eqn4fermipert}
    \delta H=g\int dx \left(\gamma_{R}\partial_{x}\gamma_{R}\gamma_{L}\partial_{x}\gamma_{L}\right),
\end{gather}
which matches the spin-spin bond coupling when we translate back to the lattice variables \cite{kitaev_anyons_2005}. This perturbation is irrelevant with scaling dimension 4, ensuring stability of the $c=1/2$ gapless domain wall to any small enough local perturbation.

The locality rules for the Majorana fields can be succinctly stated: local operators are those which have both an even number of $\gamma_L$'s and $\gamma_R$'s, i.e. operators which are invariant under both chiral fermion parities \cite{aasen2020electrical}. Let us connect these rules with the Ising-category symmetry perspective in Section \ref{subsec4symm} above.

The Majorana field theory and the critical Ising CFT are related by bosonization. The emergent $\bZ_2$ symmetry line $\mathcal{L}_{\psi}$ (we suppress the $X$ label in the loop operator), which acts as the spin-flip symmetry of the Ising model, is associated with fermion parity, while the KW duality associated with $\mathcal{L}_{\sigma}$ acts in the Majorana variables as a chiral fermion parity \cite{thorngren2018anomalies,jones20191d,Karch_2019}
\begin{equation}\label{eqnchiralparity}
\begin{gathered}
    \gamma_L \to - \gamma_L \\
    \gamma_R \to \gamma_R.
\end{gathered}
\end{equation}
Indeed, the mass term $i\gamma_L\gamma_R$ corresponds to the transverse field operator of the Ising model, which perturbs the system into the ordered and disordered phases, depending on the sign of its coefficient. These correspond to the trivial and topological phases of the Majorana fermions and are also exchanged by the $\sigma$ line operators from either side of the cut~\cite{thorngren2018anomalies}. Combining \eqref{eqnchiralparity} with the usual fermion parity, we can generate the other chiral fermion parity, so the Ising-category symmetric operators exactly match the operators in the Majorana field theory we identified as local operators in the KSL above. Intuitively, the string operators associated with the $\gamma$'s braid non-trivially with $\mathcal{L}_{\psi}$ and $\mathcal{L}_{\sigma}$, so their charges are tied to their non-locality.

For some critical value of the perturbation \eqref{eqn4fermipert} we encounter a $c = 7/10$ multicritical point associated with the tricritical Ising theory in the spin degrees of freedom \cite{zamolodchikov_tricritical_1991,Rahmani_20152,Rahmani_2015}. The operator \eqref{eqn4fermipert} is approximately the scaling operator $\epsilon'$ at the tricritical point. As we tune \eqref{eqn4fermipert} beyond this critical point, the cut bonds in the honeycomb model are effectively restored and we find ourselves in the phase of the trivial domain wall. This is consistent with our anticipation based on symmetry considerations in Section \ref{subsec4symm}.

\subsection{Electric-Magnetic Duality and Ising Anyons}

Let us make contact with our discussion in Section \ref{secphyspicture} of the toric code boundary phase diagram. Recall that studying these boundaries is equivalent to studying $\bZ_2$-symmetric 1d systems. KW duality for these theories is associated with the electric-magnetic duality which exchanges the $e$ and $m$ anyons of the toric code topological order. In this context, it is an accidental symmetry of the Ising critical point between the $e$ and $m$ condensates on the top edge.

If we enforce this extra duality as a global symmetry in both the bulk and the boundary of the toric code, we will find the same phase diagram as for the ${\textnormal{Ising}} \times \overline{\textnormal{Ising}}$ boundaries. Indeed, there is a gauging procedure of the electromagnetic duality in the bulk of the toric code which directly relates it to the ${\textnormal{Ising}} \times \overline{\textnormal{Ising}}$ system. In general one can take such anyon permuting symmetry and attempt to promote its associated twist defect to a deconfined anyon, resulting in a new topological order \cite{barkeshli_symmetry_2014,teo_theory_2015,Etingof_2010}. If one performs this procedure for the electric-magnetic duality of the toric code, one obtains the ${\textnormal{Ising}} \times \overline{\textnormal{Ising}}$ topological order~\cite{bais_condensate_2009,barkeshli_symmetry_2014,teo_theory_2015}.\footnote{Note that the fusion category one obtains from the gauging procedure is not fixed uniquely by specifying the permutation of the anyons. This ambiguity is resolved in any microscopic model of the bulk. To relate the phase diagram for $\bZ_2$-symmetric theories with KW self-duality to the boundary phase diagram of the ${\textnormal{Ising}} \times \overline{\textnormal{Ising}}$ topological order, one has the ensure that the Frobenius-Schur indicator of the Ising fusion category obtained is positive~\cite{chang_topological_2019}.} Conversely, if one condenses the $\bar\psi\psi$ anyon in the Ising string-net one obtains the toric code again \cite{burnell_anyon_2018,teo_theory_2015}.


\section{Toric Code Bilayer Boundaries and $\bZ_2 \times \bZ_2$ Symmetric Spin Chains}\label{sectoriccodebilayer}

In this section we study boundaries of the toric code bilayer, which we denote (TC)$^2$. These boundaries are equivalent to domain walls in a single-layer toric code by unfolding. The bilayer theory has six distinct gapped boundary conditions with the vacuum \cite{lan_gapped_2015}, listed in Table \ref{tab:tcbilayer} as anyon condensates.

The boundaries of the bilayer match the following six domain walls of a single toric code: the first is the trivial domain wall between TC and itself, corresponding to a plain fold boundary for the bilayer. This boundary condition is the anyon condensate $\{e_1 e_2, m_1 m_2\}$ in Table \ref{tab:tcbilayer}. The second is the electromagnetic duality defect of the toric code, such that an $e$ from one layer may pass through the domain wall and come back as an $m$ from the other layer (the condensate $\{e_1 m_2, m_1 e_2\}$). The other four boundary conditions correspond to composite domain walls, where the single toric code is cut open along the domain wall and then one creates a condensate on each side of the domain wall. One may choose to condense $e$ or $m$ on either side, yielding four different gapped domain walls.~\footnote{Those four domain walls are not invertible under stacking, and are similar to the surface operators studied in \cite{Kapustin_surf2011}.}

We will discuss the associated boundary conditions in terms of the emergent $\bZ_2 \times \bZ_2$ symmetry, two copies of the $\bZ_2$ we described in Section \ref{secphyspicture} for a single toric code. In particular, the symmetry point of view will help us understand the critical points between these gapped boundaries and make contact with the rich phase diagram of $\bZ_2 \times \bZ_2$ symmetric spin chains \cite{verresen_gapless_2019}.

\subsection{Matching Gapped Boundaries (TC)$^2$ with $\bZ_{2}\times \bZ_{2}$ Symmetric 1d Phases}

We now study the (TC)$^2$ topological order on a cylinder as in Section \ref{secphyspicture}. Two circular boundaries are exposed, and for now we will fix the bottom boundary to be gapped by $m$ condensation in each layer, a boundary condensate which we indicate as $\{m_{1},m_{2}\}$ (in general anyons of type $a$ from layer $i$ will be denoted $a_i$). As we discussed in Section \ref{secphyspicture}, the operators $\mathcal{L}_{e_i}$ 
define two $\bZ_2$ global symmetries for the quasi-1d system. These symmetries commute and have no anomaly, i.e., the ends of one symmetry string are not charged under the other.

We wish to identify the six gapped boundary conditions with different gapped $\bZ_2\times\bZ_2$ symmetric phases in 1d. In fact there are also six of the latter! To wit, we have one trivial symmetric phase which we denote as ``Symm", one SPT phase ``SPT", three partial symmetry breaking phases $I_x$, $I_y$, $I_z$, each preserving a different $\bZ_2$ subgroup in $\bZ_2 \times \bZ_2$, and one phase ``Brok" which breaks the $\bZ_2 \times \bZ_2$ symmetry completely. The notation for $\bZ_2 \times \bZ_2$ elements is meant to reflect their realization as $\pi$-rotations in $SO(3)$, such that $R_x$ is a $\pi$ rotation around the $x$ axis, etc. Relative to the $\{m_1,m_2\}$ condensate on the bottom edge, $\mathcal{L}_{e_1}$ generates $R_x$, $\mathcal{L}_{e_2}$ generates $R_y$, and their product generates $R_z$. The phase $I_x$ preserves $R_x$ but breaks $R_y$ and $R_z$, etc. In an $S = 1$ spin chain with $R_x$ and $R_y$ as symmetries, the SPT phase is the well known Haldane gapped state \cite{P_rez_Garc_a_2008}.

To match the above phases with the gapped boundary conditions, we may identify the order parameters which are long range ordered and determine which symmetries are broken. For instance, with the $\{m_1,m_2\}$ condensate on the both edges of the cylinder, both of the $\mathcal{L}_{e_i}$ symmetries are broken, and we identify the $\{m_1,m_2\}$ condensate (on top) as the phase Brok with GSD 4.

The two condensates which preserve the full $\bZ_2 \times \bZ_2$ symmetry are $\{e_1,e_2\}$ and $\{e_1 m_2, e_2 m_1\}$. Those two condensates correspond to the phases Symm and SPT. To distinguish the trivial symmetric phase from the SPT, it is known that one must consider non-local string order parameters~\cite{pollmann_symmetry_2012}. In our case, these are the open symmetry strings (symmetry twist operators) where the $\mathcal{L}_{e_i}$ string begins and ends on the top edge. 


In $\{e_1,e_2\}$ phase, the condensate contains $e_i$, so the long range ordered $\mathcal{L}_{e_i}$ string can end on an $e_i$ particle. Observe that $e_1$ and $e_2$ have trivial braiding, so the ends of the operator are uncharged under both $\bZ_2$ symmetries. Thus, the $\{e_1,e_2\}$ condensate is identified with the trivial symmetric (Symm) phase.

For the $\{e_1 m_2, e_2 m_1\}$ phase, the condensate does not contain $e_i$, but an $\mathcal{L}_{e_1}$ string can end on an $e_1 m_2$ particle, and likewise for $\mathcal{L}_{e_2}$, giving us a different open symmetry string operator with long range order on the top edge. Now, since $e_2$ has nontrivial braiding with $m_2$, the ends of the of the first $\bZ_2$ string are charged under the second $\bZ_2$ and vice versa, and we recognize this condensate as the SPT phase. The remaining identifications are summarized in Table 1.


\begin{table}
\caption{Gapped boundaries of (TC)$^2$ and their labeling in terms of anyon condensates. We match those with gapped phases of $\bZ_2 \times \bZ_2$ spin chains, by fixing the bottom edge to the $\{m_1,m_2\}$ condensate.}
\begin{tabular}{ll} 
\begin{tabular}{ |c|c|c|c| } 
\hline
Condensed Anyons & GSD &Unbroken Symm. & $\bZ_2 \times \bZ_2$ Label \\
\hline
$e_{1},m_{2}$ & 2 &$\mathcal{L}_{e_1}$ & $I_x$ \\
$m_{1},e_{2}$ & 2 &$\mathcal{L}_{e_2}$ & $I_y$ \\
$e_{1}e_{2},m_{1}m_{2}$ & 2 &$\mathcal{L}_{e_1 e_2}$ & $I_z$\\
$m_{1},m_{2}$ & 4 & none & Brok\\
$e_{1},e_{2}$ & 1 & all & Symm\\
$e_{1}m_{2},m_{1}e_{2}$ & 1 & all & SPT \\
 \hline
\end{tabular} &

\end{tabular}
\label{tab:tcbilayer}
\end{table}

\subsection{Bulk Defects and Edge Dualities}\label{subsecZ2Z2duality}

In the previous discussion we fixed the boundary condition of the bottom edge to be in the $\{m_{1},m_{2}\}$ condensate. If we chose a different boundary condition, the identification between the gapped boundary conditions on the top edge and the $\bZ_2 \times \bZ_2$ phases of the quasi-1d system would change. For example, choosing a $\{e_{1},e_{2}\}$ boundary condition at the bottom edge, one would identify $\{e_{1},e_{2}\}$ as the symmetry broken phase, etc. If one is only given the Hamiltonian at the top edge, there is an ambiguity in identifying the condensates with symmetry breaking phases. We will now quantify this ambiguity, and show the relation to dualities of $\bZ_2 \times \bZ_2$ spin chains.

The (TC)$^2$ topological order enjoys a wide array of anyon permuting symmetries, 72 in total, which were studied in \cite{yoshida_topological_2015,kesselring_boundaries_2018}. These form a group, generated by three defects, with the following action on the anyons:
\begin{align} 
 {\rm Swap} &:(e_{1},m_{1})  \leftrightarrow (e_{2},m_{2})\\ S_1 &:e_{1} \leftrightarrow m_{1}\\
E &:(e_1,e_2) \leftrightarrow (e_1 m_2,e_2 m_1)
\label{eq:TCDefects}
\end{align}

The defect group includes the duality ${\rm Swap}$ which exchanges the two layers ($e_1 \leftrightarrow e_2$, $m_1 \leftrightarrow m_2$ etc.), as well as the electric-magnetic dualities of either layer, $S_1$ and $S_2 = {\rm Swap} S_1 {\rm Swap}$. There are also order 3 elements such as $T = E {\rm Swap} S_1 S_2$, which corresponds to a triality recently studied in Refs. \cite{thorngren2019fusion}.

By applying these dualities to the condensate on the top edge and using Table \ref{tab:tcbilayer} (fixing the $\{m_1,m_2\}$ condensate on the top edge), we can read off a corresponding action on the $\bZ_2 \times \bZ_2$ symmetric phases. We find $S_1$ and $S_2$ act as the KW duality transformations associated with the two $\bZ_2$ symmetries. Explicitly, $S_1$ and $S_2$ act on the $\bZ_2 \times \bZ_2$ gapped phases by:
\begin{equation}
\begin{split}
    S_{1}:(I_{x},I_{y},I_{z})\leftrightarrow ({\rm Brok},{\rm Symm},{\rm SPT})\\
    S_{2}:(I_{x},I_{y},I_{z})\leftrightarrow ({\rm Symm},{\rm Brok},{\rm SPT})
\end{split}
\label{eq:Gauging X}
\end{equation}
C.f. Appendix C of \cite{verresen_gapless_2019}. Meanwhile $E$ is the SPT entangler
\begin{equation}
    E:({\rm Symm},{\rm SPT}) \leftrightarrow ({\rm SPT},{\rm Symm}).
\end{equation}
There are also simple dualities related to automorphisms of $\bZ_2 \times \bZ_2$, such as $S_1 E S_2$ which is the order 3 automorphism and acts on the phases by
 \begin{equation}
     S_1 E S_2 :I_{x} \to I_{y} \to I_{z} \to I_{x}.
 \end{equation}

To summarize, by fixing a boundary condition on the bottom edge of the cylinder, we fix the duality frame for the emergent symmetries. The bulk twist defects may be dragged either to the bottom boundary condition, changing the reference boundary condition, or onto the dynamical edge on top, implementing a duality.

\begin{figure}
    \centering
    \includegraphics[width=\columnwidth]{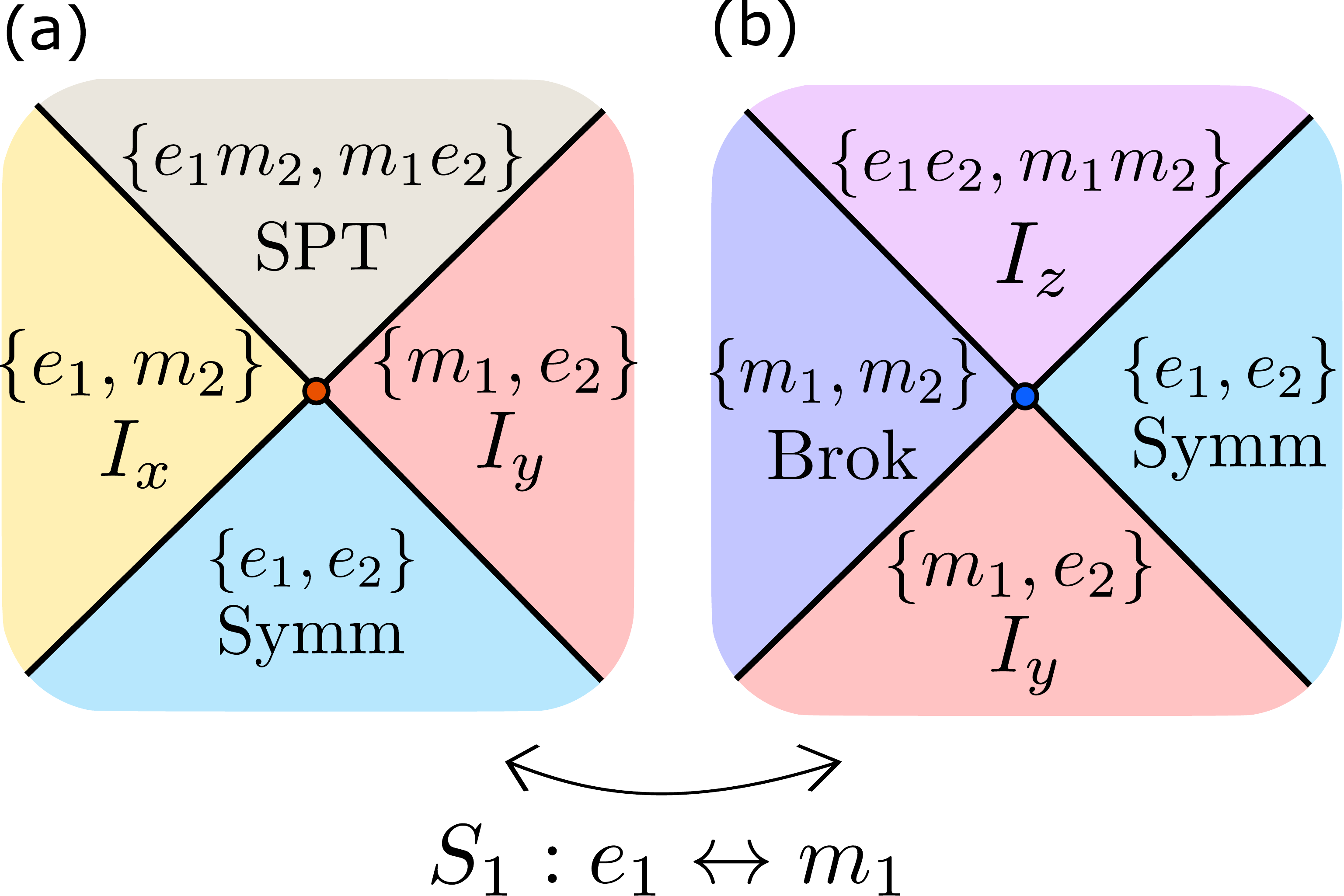}
    \caption{(a) A two-parameter phase diagram of a boundary of a toric code bilayer near a $c = 1$ multicritical point. The multicritical point corresponds to a free fermion theory. 
    The phase diagram of the boundary maps onto that of a $\bZ_2 \times \bZ_2$ symmetric one-dimensional system, where the correspondence between the gapped phases is summarized in Table~\ref{tab:tcbilayer}. 
    The four $c = 1/2$ lines in black are all distinguished by the emergent $\bZ_2 \times \bZ_2$ charges of their order and disorder operators~\cite{verresen_gapless_2019}. Additional phase diagrams can be obtained by applying dualities associated with anyon permuting symmetries in the bulk. (b) A dual phase diagram obtained by applying $S_1 :e_{1} \leftrightarrow m_{1}$. The $c=1$ theory in the middle is a product of two Ising critical theories. }
    \label{fig:TC2}
\end{figure}

\subsection{Phase Diagrams}

After fixing a boundary condition and identifying the gapped boundaries with the $\bZ_2 \times \bZ_2$ phases, one can draw phase diagrams describing transitions between the different gapped phases. One can approach the problem by finding CFTs with a $\bZ_2 \times \bZ_2$ symmetry and a small number of symmetric relevant operators. For instance, in Fig. \ref{fig:TC2} we draw the nearby phase diagram of two $c = 1$ multicritical points, each with two symmetric relevant operators. These points were recently discussed in \cite{verresen_gapless_2019} in the context of spin chains, where accessing the critical point required enforcing the symmetry as well as tuning two parameters. In our setting the emergent symmetry is enforced by the bulk topological order, so one only has to tune two parameters to access these points. 

The group of 72 duality transformations acts non-trivially on these phase diagrams. For instance, the two phase diagrams in Fig. \hyperref[fig:TC2]{\ref{fig:TC2}a} and Fig. \hyperref[fig:TC2]{\ref{fig:TC2}b} are related by the duality $S_1$ in \eqref{eq:Gauging X}. The $c=1$ point in Fig. \hyperref[fig:TC2]{\ref{fig:TC2}a} is a compact boson CFT at the free fermion point $K = 1$ (see Section \ref{secKSLgenon} below) while the $c=1$ point in Fig. \hyperref[fig:TC2]{\ref{fig:TC2}b} corresponds to a product of two decoupled Ising CFTs. 

\section{The Kitaev Spin Liquid Bilayer and Genon Chain}\label{secKSLgenon}

We now turn our attention to our main example: the domain wall phase diagram of a bilayer of KSLs with the same chirality, labeled (KSL)$^2$. By folding, this domain wall is equivalent to the boundary of a bilayer of non-chiral Ising$\times \overline{\rm Ising}$ topological orders.
The corresponding quasi-1d system enjoys the Ising$\times$Ising category symmetry, the symmetry operators being the bulk anyon lines in both layers. As we argue below, there are three gapped domain walls: a trivial domain wall, a genon domain wall that swaps the layers, and a third domain wall that we refer to as a ``toric code gluing" of the two layers (see below). In addition, we find several kinds of stable gapless domain walls. 
In the following, we derive these phases and the critical points that separate them from the properties of the Ising$\times$Ising category symmetry. 

To substantiate these results, we study a microscopic model of interacting bilayer twist defects known as the genon chain. We use this model to investigate the phase transition between the trivial and layer swap domain walls, and the nearby phase diagram. The analysis of this model is quite complex, but the symmetry principles we have outlined can be used to understand the phase diagram in great detail and make contact with previous numerical and analytic results on the genon chain model~\cite{gils_ashkinteller_2009}.

As it turns out, the genon chain cannot describe the gapless cut-open domain wall of the bilayer, which hosts two pairs of decoupled counter-propagating $c = 1/2$ modes (compare Fig. \ref{figtricrit}), such that the total central charge is $c = 1$. However, the system supports a different $c = 1$ stable gapless domain wall, where the layers are strongly interacting. The two types of $c=1$ gapless phases can be connected by a direct continuous transition, described by a $c = 3/2$ theory. 
We will characterize the three-dimensional phase diagram that connects the phases of the genon chain with the cut-open domain wall.

We also study two $c = 7/5$ multicritical points enjoying the full Ising$\times$Ising category symmetry, making contact with the discussion of Section \ref{secKSLtricrit}. One of those points is the direct product of two $c=7/10$ points discussed in that section. The other point is a kind of twisted product of the two $c=7/10$ points, and is related to the first by a duality transformation. The duality relating the two theories is associated with the genon defect of the bulk (KSL)$^2$. We discuss the action of this duality on the phase diagram of the domain wall.

\subsection{(KSL)$^2$ Gapped Domain Walls and Ising$\times$Ising Category Symmetry}\label{subsec5bilayerphases}

\begin{figure}
    \centering
    \includegraphics[width=\columnwidth]{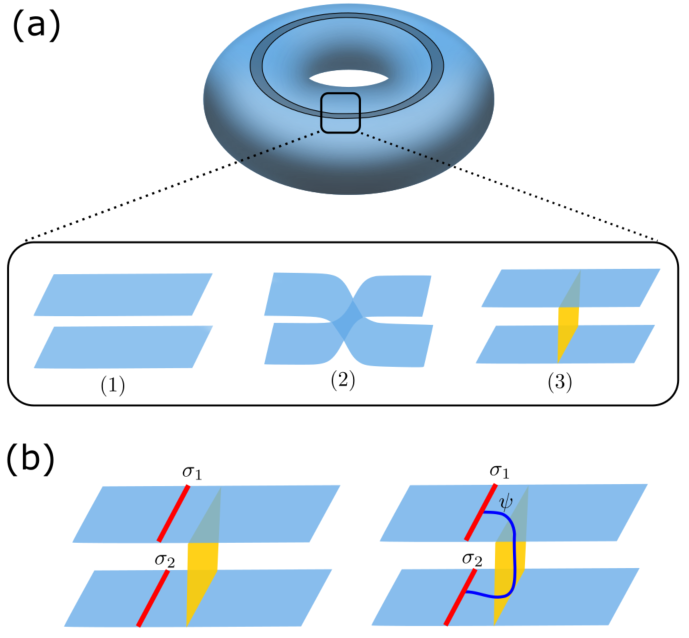}
    \caption{(a) The three domain walls in the Kitaev spin liquid bilayer, inserted along the top cycle of a torus (compare Fig. \ref{figtorus}). The view is from the direction parallel to the domain wall. (1) is the trivial domain wall with GSD 9, (2) is the genon domain wall with GSD 3, and (3) has the two Ising layers glued with a toric code (shown in yellow) with GSD 6. The gluing of each single KSL to a toric code may be understood when the gluing line is regarded as a gapped boundary between a KSL$\times \overline{\text{KSL}}$ and a toric code \cite{burnell_anyon_2018}. (b) In the toric code gluing domain wall, the state $|\sigma,\sigma\rangle$ is split into $|\sigma_1\sigma_2\rangle$ (left) and $|\sigma_1\sigma_2*\rangle$ (right). The red lines denote $\sigma$ lines wrapping around the tori. The state are distinguished by a $\psi$ line going through the domain wall, ending on the $\sigma$ lines.}
    \label{figKSLbilayerdomainwalls}
\end{figure}

Let us begin by describing the three gapped domain walls between (KSL)$^2$ and itself, illustrated graphically in Fig. \ref{figKSLbilayerdomainwalls}. Each of these domain walls may be also thought of as a gapped boundary condition for (Ising)$^2\times \overline{\text{(Ising)}}^2$ topological order by folding. We relate these domain walls to symmetric phases of the (Ising)$^2$ fusion category, generated by $\mathcal{L}_{\psi_j}$, $\mathcal{L}_{\sigma_j}$, $j = 1,2$, with each pair satisfying the fusion rules \eqref{eqnisingfusion}.

First, we have the trivial domain wall (Fig. \hyperref[figKSLbilayerdomainwalls]{\ref{figKSLbilayerdomainwalls}a1}) at which each of the two layers is healed separately. The ground states of the system correspond to those of the (Ising)$^2$ topological order on a torus. The basis state can be labelled by the possible anyons encircling the long ($x$ direction) cycle. There are nine such states, labeled as $|a_1a_2\rangle$ with $a_{i}=1,\psi,\sigma$. The (Ising)$^2$ category symmetry acts in this basis according to the fusion rules\footnote{
Eq.~\eqref{eq:Lb1b2} is written in a particular gauge choice for the basis states $|a_1,a_2\rangle$, such that there are no additional phase factors. 
To see that such a gauge choice is possible, we start from a basis $|\Psi_a\rangle$ of eigenstates of $\mathcal{L}_{b}$, such that
$\mathcal{L}_b|\Psi_a\rangle=\frac{S_{ab}}{S_{Ia}} |\Psi_a\rangle$, where S is the topological S-matrix
of the theory. We construct another basis according to: $\vert a \rangle = \sum_b S_{ab} \vert \Psi_b\rangle$. The Verlinde formula~\cite{VERLINDE1988360} then assures that in the $|a\rangle$ basis,  Eq.~\eqref{eq:Lb1b2} is satisfied with no additional phases.}:
    \begin{equation}\mathcal{L}_{b_{1}}\mathcal{L}_{b_2}|a_1, a_2\rangle=\sum_{c_1,c_2} N^{c}_{a_{1}b_{1}}N^{c_2}_{a_{2}b_{2}}|c_1, c_2\rangle.\label{eq:Lb1b2}\end{equation}
Where $N_{ab}^c$ are the fusion coefficients~\cite{kitaev_anyons_2005} of the Ising theory. Since every element of the (Ising)$^2$ category symmetry acts non-trivially in the ground state subspace, the category symmetry is completely broken in the trivial domain wall phase.
    
Next is the layer swap or ``genon" domain wall (Fig. \hyperref[figKSLbilayerdomainwalls]{\ref{figKSLbilayerdomainwalls}a2}). This domain wall reconnects the two layers, so that an anyon which crosses through the domain wall is transported to the other layer. The system is topologically equivalent to a single Ising torus (as can be seen by noticing that there are two inequivalent cycles), and thus has GSD 3. From the symmetry point of view this phase corresponds to a symmetry breaking pattern where the two Ising fusion categories act on a set of three states $\widetilde{|a\rangle}$, with $a = 1, \psi,\sigma$, according to:
    \begin{equation}\mathcal{L}_{b_j}\widetilde{|a\rangle}=\sum_c N^{c}_{ab}\widetilde{|c\rangle},\end{equation}
independently of the layer index $j$.
    
There is a third gapped domain wall, which we refer to as the ``toric code gluing" domain wall. To understand this domain wall, we recall that ${\textnormal{Ising}} \times \overline{\textnormal{Ising}}$ becomes the toric code after condensing the boson $\psi \bar\psi$ (the bound state of the fermion from each layer) \cite{burnell_anyon_2018,teo_theory_2015}. Therefore, there is a three-way junction where a toric code can end on a Kitaev spin liquid. We can use this to construct a domain wall in a bilayer of this chiral phase by connecting the two layers with a toric code as in Fig. \hyperref[figKSLbilayerdomainwalls]{\ref{figKSLbilayerdomainwalls}a3}. In the anyon condensation framework, this gapped domain walls corresponds to a boundary of Ising$^2\times \overline{\text{Ising}}^2$ with the vacuum obtained by condensing the anyons $\psi_{i}\psi_{j}$ with $i,j\in\{1,2,\bar{1},\bar{2}\}$ as well as $\sigma_1\sigma_2\sigma_{\bar{1}}\sigma_{\bar{2}}$. All other anyons are confined.
    
To characterize the ground states of the system in the presence of this domain wall, we start with the nine states of two disconnected layers, that belong to nine distinct topological sectors. We consider the fate of these nine sectors after connecting the two layers by a toric code domain wall. First, $\psi$'s from both layers may enter the toric code and annihilate each other, and therefore the states $|1,1\rangle$ and $|\psi,\psi\rangle$ are identified, and can be labeled as a single state $|1\rangle$. Similarly, any pair of states $|a_1,a_2\rangle$, $|b_1,b_2\rangle$ such that $a_1a_2 = \psi_1 \psi_2 b_1 b_2$ are identified for the same reason. The state $|\sigma,\sigma\rangle$ is special since it is invariant under fusion with $\psi_1\psi_2$. We claim that $|\sigma,\sigma \rangle$ splits into two different ground states in the presence of the toric code gluing domain wall, see Fig. \hyperref[figKSLbilayerdomainwalls]{\ref{figKSLbilayerdomainwalls}b}. This gives a total of 6 ground states: $|1\rangle$, $|\psi_1\rangle$, $|\sigma_1\rangle$, $|\sigma_2\rangle$, $|\sigma_1\sigma_2\rangle$, $|\sigma_1\sigma_2*\rangle$, summarized in Table \ref{tabletoriccodegluing}.

We now discuss the splitting of $\vert \sigma, \sigma\rangle$. In the presence of the toric code gluing domain wall, there are two versions of this state, distinguished by the presence of a $\psi$ line that connects the $\sigma_1$ and $\sigma_2$ lines through the domain wall (see Fig.~\ref{figKSLbilayerdomainwalls}b). These are the states $\vert \sigma_1 \sigma_2 \rangle$, $\vert \sigma_1 \sigma_2 *\rangle$ in Table~\ref{tabletoriccodegluing}. To show that these are distinct states, we compute the action of $\mathcal{L}_{\psi_j}$ on the two states using the fusion rules and F-moves of the Ising theory. This gives: 

\begin{equation}
\begin{split}
     \mathcal{L}_{\psi_j}|\sigma_1\sigma_2\rangle & = |\sigma_1\sigma_2\rangle, \\
  \mathcal{L}_{\psi_j}|\sigma_1\sigma_2*\rangle & = -|\sigma_1\sigma_2*\rangle. \\
\end{split}
\label{eq:Fmove}
\end{equation}
Hence, the state $|\sigma,\sigma\rangle$ of the trivial domain wall corresponds to two distinct states of the toric code gluing domain wall.

\begin{table}
\caption{Ground states of the toric code gluing domain wall and their corresponding states in the trivial domain wall, before connecting the layers.}
\begin{tabular}{ll} 
\begin{tabular}{ |c|c| } 
\hline
Toric code gluing & Trivial domain wall \\
\hline
$|1\rangle$ & $|1,1\rangle$,$|\psi,\psi\rangle$ \\
$|\psi_1\rangle$ & $|\psi,1\rangle$,$|1,\psi\rangle$ \\
$|\sigma_1\rangle$& $|\sigma,1\rangle$,$|\sigma,\psi\rangle$ \\
$|\sigma_2\rangle$& $|1,\sigma\rangle$,$|\psi,\sigma\rangle$\\
$|\sigma_1\sigma_2\rangle$,$|\sigma_1\sigma_2*\rangle$ & $|\sigma,\sigma\rangle$\\
 \hline
\end{tabular} &
\end{tabular}
\label{tabletoriccodegluing}
\end{table}

The three domain walls and their ground states can be viewed in relation to the $\bZ_2 \times \bZ_2$ subgroup of the Ising$\times$Ising category symmetry, generated by $\mathcal{L}_{\psi_1},\mathcal{L}_{\psi_2}$. We may label the states according to the action of the $\bZ_2 \times \bZ_2$ subgroup; each state is a ground state of one of the six gapped phases of a $\bZ_2 \times \bZ_2$ symmetric 1d system, described in Sec.~\ref{sectoriccodebilayer}, $\{\text{Symm},\text{SPT},\text{Brok},I_x,I_y,I_z\}$. Then, we study the action of the operators $\mathcal{L}_{\sigma_1},\mathcal{L}_{\sigma_1}$, associated with the KW dualities of the $\bZ_2$ symmetries $\mathcal{L}_{\psi_1},\mathcal{L}_{\psi_2}$, corresponding to $R_x$, $R_y$ respectively. 

We find two closed orbits under the action of the KW dualities\footnote{Those orbits can be thought of as first order transitions in a $\bZ_2 \times \bZ_2$ symmetric system, which become stable phases after promoting the KW dualities to symmetries, as in the case of the single KSL, see Sec. \ref{secKSLtricrit}} (Eq.~\ref{eq:Gauging X}),
\begin{equation}{\rm Brok} + I_x + I_y + {\rm Symm}\end{equation}
of total GSD 9, and
\begin{equation}I_z + {\rm SPT}\end{equation}
of total GSD 3. Since both phases in the second orbit are self-dual under gauging the full $\bZ_2 \times \bZ_2$ symmetry (corresponding to acting with the symmetry line $\mathcal{L}_{\sigma_1\sigma_2}$), this orbit gives rise to two phases - one of GSD 3 where $\mathcal{L}_{\sigma_1\sigma_2}$ is preserved, and one of GSD 6 where $\mathcal{L}_{\sigma_1\sigma_2}$ is spontaneously broken. Thus we have completely accounted for the three gapped domain walls described above.

Let us elaborate on this point. When $\mathcal{L}_{\sigma_1\sigma_2}$ is spontaneously broken, we can use the operator $\mathcal{L}^Y_{\psi_2{\psi}_{\bar{2}}}$ as an order parameter.  
This operator can be described using Fig. \hyperref[figKSLbilayerdomainwalls]{\ref{figKSLbilayerdomainwalls}b} as a horizontal $\psi$ line in layer 2, transverse to the $\sigma$ line. Note that the left and right edges of each layer are glued together, thus this $\psi$ line is a closed loop.  
$\mathcal{L}^Y_{\psi_2{\psi}_{\bar{2}}}$ is an order parameter for the broken symmetry $\mathcal{L}_{\sigma_1\sigma_2}$ (see the discussion in \ref{subsecorderpara} for a reminder). This follows from the fact that $\mathcal{L}^Y_{\psi_2{\psi}_{\bar{2}}}$ anti-commutes with $\mathcal{L}_{\sigma_1\sigma_2}$. We label the six ground states in the symmetry broken phase according to the sign of $\langle \mathcal{L}^Y_{\psi_2{\psi}_{\bar{2}}} \rangle$. There are two triplets with opposite signs of $\langle \mathcal{L}^Y_{\psi_2{\psi}_{\bar{2}}} \rangle$, which we can identify in our previous description of those states in Table \ref{tabletoriccodegluing} as:

\begin{equation}
\begin{split}
    \langle \mathcal{L}^Y_{\psi_2{\psi}_{\bar{2}}} \rangle>0 &: {|1\rangle,|\psi_1\rangle,|\sigma_1\rangle}\\
    \langle \mathcal{L}^Y_{\psi_2{\psi}_{\bar{2}}} \rangle<0 &: {|\sigma_1\sigma_2\rangle \pm |\sigma_1\sigma_2*\rangle,|\sigma_2\rangle}  
\end{split}
\label{eqtwotriplets}
\end{equation}


Within this subspace, the action of the symmetries $\mathcal{L}_{\psi_1}$ and $\mathcal{L}_{\psi_2}$ is identical (since the toric gluing domain wall allows a $\psi$ line to pass between the layers). 
Both symmetries switch the two first states of each triplet in \eqref{eqtwotriplets}, and fix the third. Hence, we identify the first two states of each triplet as the two ground states of $I_z$. On the third state, open strings of $\mathcal{L}_{\psi_{1,2}}$ are charged under each other, as can be shown using the F-moves of the Ising theory [in a similar manner to the step that leads to Eq.~(\ref{eq:Fmove})]. Therefore, we identify the two states $\vert\sigma_{1}\rangle$ and $\vert\sigma_{2}\rangle$ as ground states of the SPT phase. We can therefore write each triplet in \eqref{eqtwotriplets} as ``$I_z+\text{SPT}$" in terms of the $\bZ_2 \times \bZ_2$ action.

The total ground state degeneracies of the three types of gapped domain walls, and their unbroken symmetries, are summarized in Table~\ref{table369}. As one may expect, the more symmetries are broken, the larger the ground state degeneracy.

\begin{table}
\caption{KSL$^2$ phases in terms of unbroken symmetry.}
\begin{tabular}{ll} 
\begin{tabular}{ |c|c|c| } 
\hline
(KSL)$^2$ Label & GSD &  Unbroken Symm. \\
\hline
Trivial & 9 & None\\
TC gluing & 6 &$\mathcal{L}_{\psi_1\psi_2}$ \\
Genon & 3 & $\mathcal{L}_{\psi_1\psi_2}$,$\mathcal{L}_{\sigma_1\sigma_2}$\\
 \hline
\end{tabular} &
\end{tabular}
\label{table369}
\end{table}

\subsection{Ising Genon Chain}\label{subsec5genonchain}

We will now present a microscopic model that realizes the three domain walls discussed above, and serves as a natural setup to study the phase transitions between them. The model consists of a 1d chain of interacting genons, which are the point defects sitting at the ends of layer flip domain walls \cite{barkeshli_genons_2012,barkeshli_symmetry_2014}, see Fig. \hyperref[figpants]{\ref{figpants}a}.
When the genons are far away from each other, the system has a ground state degeneracy that scales exponentially with the number of genons. Those ground states form the Hilbert space of the genon chain. Coupling between the genons arising from their finite separation introduces a Hamiltonian within this Hilbert space, and such couplings are modeled as tunneling events of anyons around and between the genons. The genon chain we discuss may also be realized as a chain of lattice dislocations in a certain crystalline-symmetry-enriched version of the toric code bilayer \cite{Knapp_2019}.

The Ising$\times$Ising fusion category symmetry must be respected by any local Hamiltonian for the genon chain. We will use this category symmetry in our analysis of the system, and interpret the phases of the chain as different patterns of symmetry breaking. We will construct an effective field theory of the genon chain near its phase transitions, and show that the Ising$\times$Ising category symmetry plays a crucial role in stabilizing the phase diagram and identifying the different order parameters.

\subsubsection{Hilbert Space and Symmetries}

\begin{figure}[!htbp]
    \centering
    \includegraphics[width=\columnwidth]{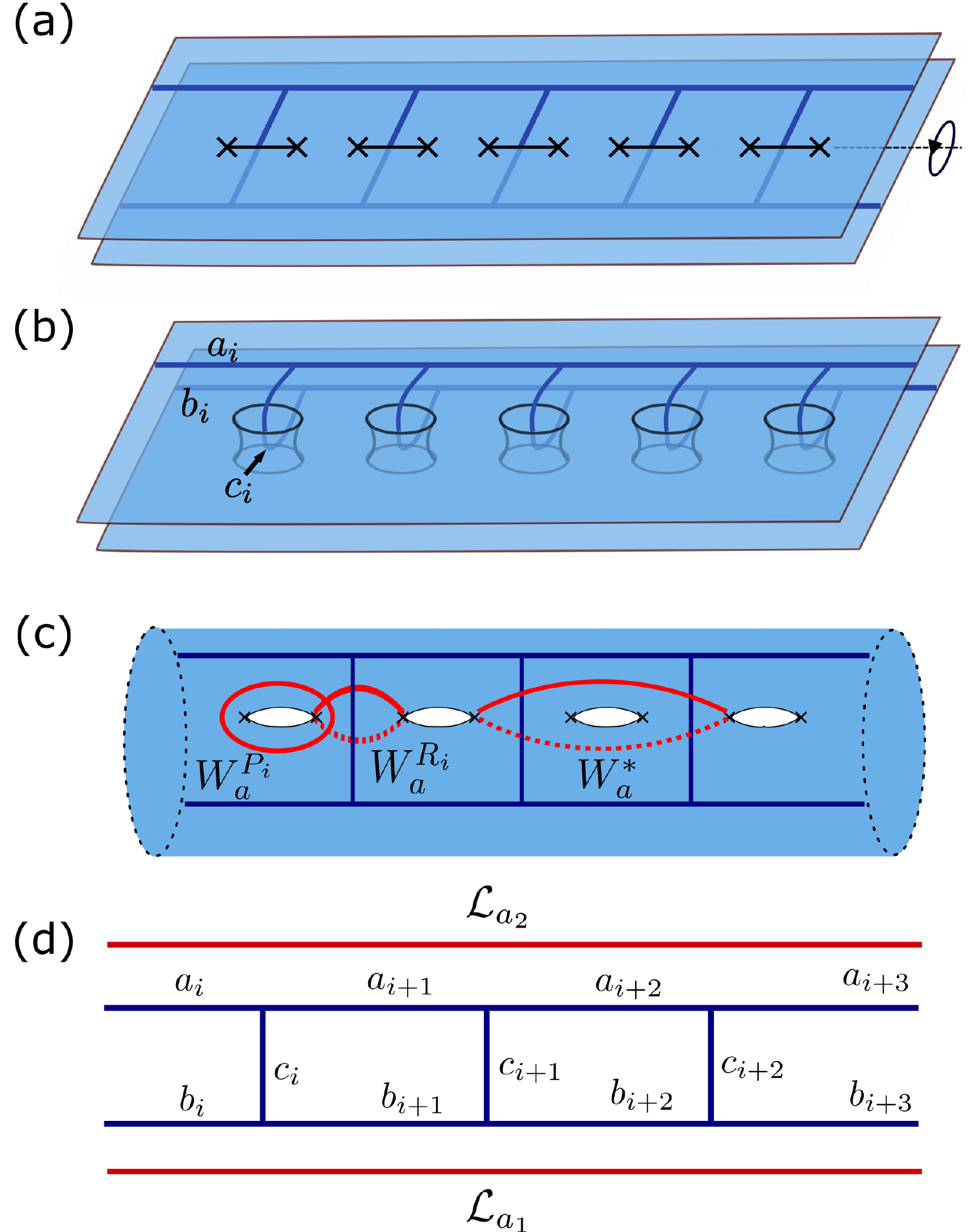}
    \caption{(a) The genon chain is placed on a bilayer of two KSLs with the same chirality. The dark blue lines are used as a fusion graph labeling the states in the Hilbert space of the chain. (b) After a $\pi$ rotation of the bottom layer around the axis indicated by a dashed line in (a), the two layers have opposite chiralities, and the defect lines become tubes which reflect anyons as they pass from one layer to the other. 
    The system is equivalent to a monolayer system of Ising anyons on a covering space of high genus, shown in (c). In this construction, we have used the fact that each of the two layers in (b) has periodic boundary conditions. The fusion graph in (c) is associated with a ``pants'' decomposition of the surface as explained in Appendix \ref{appgappedphases}. In (c), from left to right, we have drawn a plaquette operator, a rung operator (both are terms in the Hamiltonian \eqref{eqH}) and a next nearest neighbour loop operator, not in the Hamiltonian.
    (d) Each edge of the fusion graph is labeled by an anyon charge: $a_i,b_i,c_i = 1,\psi,\sigma$. Note that the labels $a_i$, $b_i$, $c_i$ appear also in panel (b). The Ising fusion rules are respected at each vertex of the graph. The emergent symmetries coming from the bulk anyon lines act by fusion of lines $\mathcal{L}_{a_{1,2}}$ with the top and bottom of this graph.}
    \label{figpants}
\end{figure}

To understand the structure of the Hilbert space of a multi-genon system, it is useful to map the system onto a monolayer topological order on a higher genus surface \cite{barkeshli_genons_2012}.
The mapping can be described as flipping the orientation of one of the two layers. After this procedure, an anyon crossing the defect line from one layer is reflected back in the other layer. Effectively, the defect line is equivalent to a ``hole" in the geometry, or more precisely a tube connecting the two layers, see Fig \ref{figpants}.

Bases for the ground state subspace of a topologically ordered system on a high genus surface are given in terms of fusion graphs \cite{moore_classical_1989,bonderson_anyonic_2017,nayak_non-Abelian_2007}. This is done using the so-called ``pants decompositions" \cite{hatcher1999pants}. The idea is as follows: one picks a decomposition of the surface into pairs of pants. Each pair of pants is then drawn as a trivalent vertex of a graph. The edges of the graph are assigned anyon labels, and the fusion rules are enforced at each vertex. It is useful to have several different pants decompositions at one's disposal in analyzing the genon chain. The different bases are related to each other by applying the F and S moves \cite{hatcher1999pants}, see Appendix \ref{appgappedphases}.

In our case of interest, the labels are taken from the Ising fusion algebra, and the Ising$\times$Ising symmetry lines act by fusion with the graph, such that $\mathcal{L}_{\psi_1}$, $\mathcal{L}_{\sigma_1}$ act by fusion from the top of the graph and $\mathcal{L}_{\psi_2}$, $\mathcal{L}_{\sigma_2}$ act by fusion from the bottom of the graph, see Fig. \hyperref[figpants]{\ref{figpants}d}.\footnote{This presentation is special to double-cover genons, however we expect a similarly rich structure to arise for permutation defects in $n$-layer systems.} 

The Hilbert space of the chain turns out to have $2^g(2^{g+1}+1)$ states per $2g$ genons on the torus, corresponding to the GSD of the Ising TQFT on a genus $g$ surface, see \cite{read_paired_2000,oshikawa_topological_2007}. In particular, the Hilbert space cannot be decomposed into a tensor product on sites.

\subsubsection{Hamiltonian and Solvable Points}

\begin{figure*}
 \includegraphics[width=\textwidth]{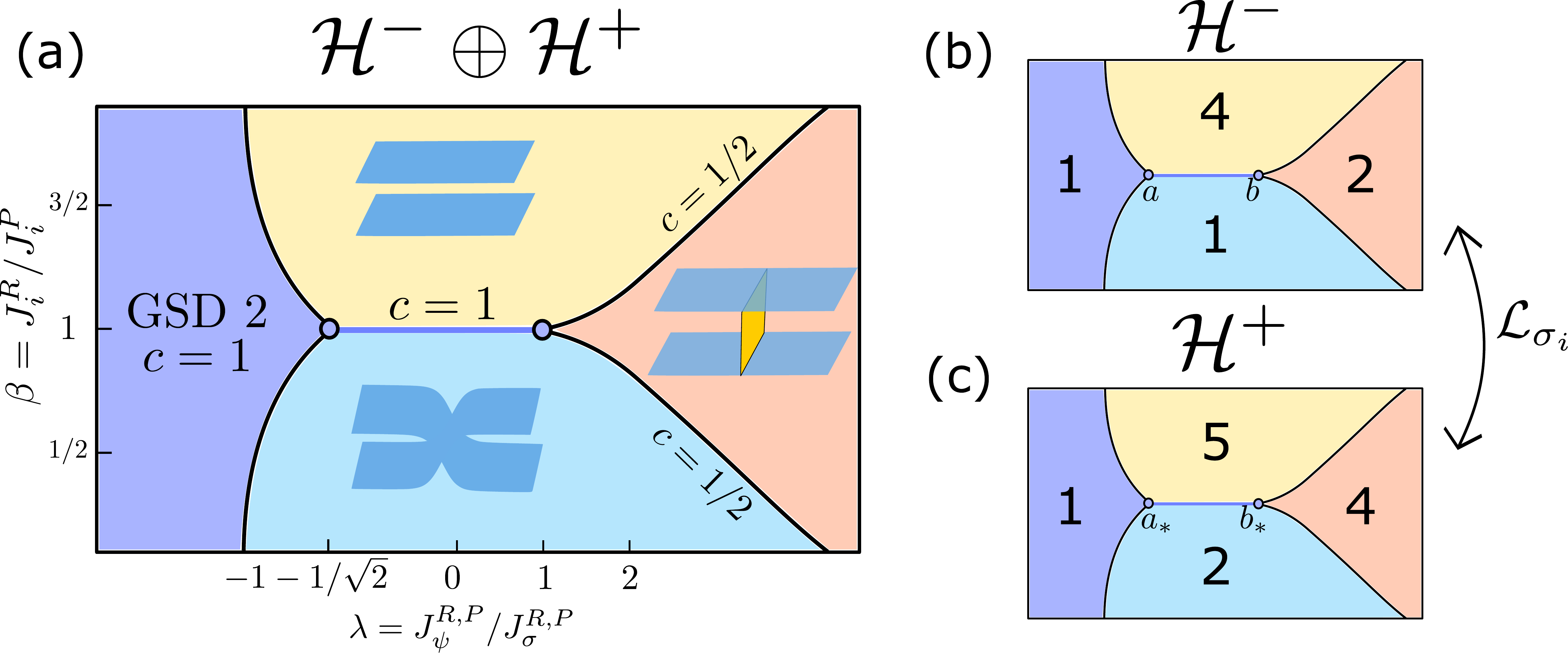}
    \caption{The phase diagram of the genon chain. The Hilbert space can be divided into sectors, $\mathcal{H}^+$ and $\mathcal{H}^-$, discussed in Sec.~\ref{subsecsymmetrybreaking}. In (a), we consider both sectors together as an Ising$\times$Ising symmetric system and find three gapped phases of GSD's 9, 6, and 3 (clockwise from top) and one gapless phase (left, in purple), corresponding to the trivial, TC gluing, genon, and a stable gapless phase with GSD 2, as explained in the text. In (b), we consider the two sectors of the broken symmetries $\mathcal{L}_{\sigma_j}$. Each of the sectors is treated separately as a $\bZ_2 \times \bZ_2$ Tambara-Yamagami symmetric system, and the two sectors are related by $\mathcal{L}_{\sigma_j}$. We have labelled the phases by their GSD. Observe that they add up to the proper values in (a) and compare the splitting to sectors with Table \ref{tableTYPhases}. The points $a,b$ in (b) are described by orbifold theories with $K=1/2$ and $K=2$, while their partners $a_*,b_*$ in $\mathcal{H}^{+}$ are orbifolds with $\tilde K=2$ and $\tilde K=8$ respectively.}
    \label{figgenonphasediagram}
\end{figure*}

Let us now discuss a particular family of Hamiltonians of the genon chain. 
Following Ref.~\cite{gils_ashkinteller_2009}, we consider Hamiltonians that include two types of terms, that we refer to as ``rung'' and ``plaquette'' operators. The rung operators $W_{\sigma}^{R_i}$ and $W_{\psi}^{R_i}$ correspond to creating a pair of $\sigma$ or $\psi$ anyons from the vacuum, winding one of the two anyons around the $i$th rung of the ``ladder'' (see Fig.~\hyperref[figpants]{\ref{figpants}c,d}), and annihilating the pair. Since the rung has topological charge $c_i$ (Fig.~\hyperref[figpants]{\ref{figpants}d}), this yields the statistical phase that corresponds to winding $\psi$ or $\sigma$ around the anyon $c_i=1,\psi,\sigma$. The plaquette operators $W_{\psi}^{P_i}$ and $W_{\sigma}^{P_i}$ similarly create a pair of anyons, wind one of the anyons around the $i$th hole of the surface in Fig.~\hyperref[figpants]{\ref{figpants}c}, and annihilate the pair. This corresponds to nucleating a small $\psi$ or $\sigma$ loop from the vacuum and fusing it into the plaquette $P_i$ of the ladder, whose edges carry charges $c_{i-1}$, $a_i$, $b_i$, $c_i$ (Fig.~\hyperref[figpants]{\ref{figpants}d}). The Hamiltonian has the form

\begin{equation}
\begin{split}
H=-\sum_{R}\left(J_{\sigma}^{R}W_{\sigma}^{R}+J_{\psi}^{R}W_{\psi}^{R}\right)\\
-\sum_{P}\left(J_{\sigma}^{P}W_{\sigma}^{P}+J_{\psi}^{P}W_{\psi}^{P}\right),
\label{eqH}
\end{split}
\end{equation}
with parameters $J_\sigma^R$, $J_\psi^R$, $J_\sigma^P$, $J_\psi^P$. See Fig.~\hyperref[figpants]{\ref{figpants}c} for a graphical representation of the terms in the above Hamiltonian.

Clearly, (\ref{eqH}) is not the most general Hamiltonian that commutes with the Ising$\times$Ising symmetry lines. In principle, we should allow for any finite-range loop operator that does not wrap around the upper or lower edge of the surface in Fig.~\hyperref[figpants]{\ref{figpants}c} (the edges are assumed to be very far from the genon chain). See the loop operator $W^{*}_a$ in Fig.~\hyperref[figpants]{\ref{figpants}c} for one such possible long-range term. We note also that, in addition to the Ising$\times$Ising category symmetry, the Hamiltonian (\ref{eqH}) is invariant under translation by one rung of the ladder, and under flipping the two legs of the ladder (corresponding to interchanging the two layers of the Ising$\times$Ising system). 
However, as we shall argue below, perturbing the Hamiltonian (\ref{eqH}) with terms that break these two symmetries will not change any of the qualitative features of the phase diagram, as long the Ising$\times$Ising category symmetry is maintained. 

Importantly, the model realizes all three gapped (KSL)$^2$ domain walls discussed in Sec.~\ref{subsec5bilayerphases}. For each of the three, there is a representative exactly solvable point in the space of parameters of (\ref{eqH}), where the Hamiltonian is a sum of commuting terms.
\begin{enumerate}
 \item The \textbf{trivial domain wall} (GSD 9) is realized for $J_{\psi}^{P}=J_{\psi}^{R}=J_{\sigma}^{P}=0, J_{\sigma}^{R}>0$. The $J^R_\sigma$ term forces all the rung labels to be $c_i = 1$ in the ground states, effectively separating the top and bottom legs of the ladder. By the fusion rules, we are forced to have $a_i = a$ and $b_i = b$ for all $i$, so we find 9 ground states for the three choices of $a$ and $b$ from $1$,$\psi$,$\sigma$. In terms of the covering surface, we can picture the cycles dual to the rungs 
 as pinching off, turning our surface into two disconnected tori. 
 Each torus contributes a factor of 3 to the GSD. Note that this is distinct from the ``cut open'' domain wall, which is gapless.

    \item The \textbf{genon domain wall} (GSD 3) is realized for $J_{\psi}^{P}=J_{\psi}^{R}=J_{\sigma}^{R}=0, J_{\sigma}^{P}>0$. 
    In this case, the ground states are states with the plaquettes invariant under fusion with a $\sigma$ loop. In terms of the covering surface, this pinches off the cycles encircled by the plaquettes, and we are left with a single torus topology, with GSD 3 and a diagonal action of the Ising$\times$Ising fusion algebra. Explicit ground states in terms of a fusion graph are given in Appendix \ref{appgappedphases}.

 \item The \textbf{toric code gluing domain wall} (GSD 6) is realized for $J_{\sigma}^{R}=J_{\sigma}^{P}=0, J_{\psi}^{R}>0,J_{\psi}^{P}>0$. 
    Note that $[J^R_\psi, J^P_\psi]=0$ for any rung and plaquette. This Hamiltonian favors superpositions of 1 and $\psi$'s along both rungs and plaquettes. 
    In Appendix \ref{appgappedphases} we verify that there are indeed six ground state and give their explicit form. 
  
\end{enumerate}

Some features of the phase diagram of the model (\ref{eqH}) can be inferred from qualitative considerations. The ratio of rungs versus plaquettes $\sigma$ terms controls the transition between genon and trivial phase. Those two phases correspond to two different dimerization patterns of the genons. Therefore, if we set $J^{R,P}_{\psi}=0$, we expect a phase transition between the trivial and genon domain wall phases at $J^R_\sigma = J^P_\sigma$. In addition, when $J_{\psi}^{R}$ and $J_{\psi}^{P}$ are both large and positive, we should enter the toric code gluing phase. A two-dimensional cut through the phase diagram, to be discussed in detail below, is shown in Fig.~\hyperref[figgenonphasediagram]{\ref{figgenonphasediagram}a}.

\subsubsection{$\mathcal{L}_{\sigma_{j}}$ Symmetry Breaking and Splitting into Sectors}\label{subsecsymmetrybreaking}

Interestingly, the symmetries $\mathcal{L}_{\sigma_{j=1,2}}$ are spontaneously broken {\it in all three gapped phases} (see Table~\ref{table369}). This can be seen explicitly by noting that none of the ground states of any of the gapped phases is invariant under $\mathcal{L}_{\sigma_j}$. 

The $\mathcal{L}_{\sigma_{j}}$ symmetry breaking can be understood within the genon chain model. Note that the operator $\mathcal{L}^Y_{\psi_1 \psi_2 \psi_{\bar{1}} \psi_{\bar{2}}} (l)$ that wraps two $\psi$ loops around two legs $a_l$ and $b_l$ of a single plaquette of the ladder (Fig.~\hyperref[figpants]{\ref{figpants}d}) anti-commutes with $\mathcal{L}_{\sigma_j}$ (cf. Section \ref{secphyspicture}). Hence, $\mathcal{L}^Y_{\psi_1 \psi_2 \psi_{\bar{1}} \psi_{\bar{2}}} (l)$ may serve as an order parameter for the breaking of $\mathcal{L}_{\sigma_j}$. Every state in the basis of the Hilbert space labelled by $a_l,b_l,c_l$ in Fig.~\hyperref[figpants]{\ref{figpants}d} is an eigenstate of $\mathcal{L}^Y_{\psi_1 \psi_2 \psi_{\bar{1}} \psi_{\bar{2}}} (l)$, and the fusion rules dictate that the eigenvalues are independent of $l$. We henceforth drop the label $l$. Moreover, any local term that acts within this Hilbert space cannot flip the eigenvalue of $\mathcal{L}^Y_{\psi_1 \psi_2 \psi_{\bar{1}} \psi_{\bar{2}}}$, as this would require threading a $\sigma$ line across the entire system. Therefore, each ground state of the genon chain is characterized by $\langle \mathcal{L}^Y_{\psi_1 \psi_2 \psi_{\bar{1}} \psi_{\bar{2}}}\rangle = \pm 1$, and the symmetries $\mathcal{L}_{\sigma_j}$ are spontaneously  broken throughout the phase diagram of the model\footnote{The fact that $\langle \mathcal{L}^Y_{\psi_1 \psi_2 \psi_{\bar{1}} \psi_{\bar{2}}}\rangle = \pm 1$ is specific to the genon chain model. One can add local terms at the (KSL)$^2$ domain wall that would make
$|\langle \mathcal{L}^Y_{\psi_1 \psi_2 \psi_{\bar{1}} \psi_{\bar{2}}} \rangle|<1$. However, the symmetry breaking is robust, implying that $\langle \mathcal{L}^Y_{\psi_1 \psi_2 \psi_{\bar{1}} \psi_{\bar{2}}}\rangle \ne 0$ throughout the gapped phases, as well as at any direct phase transition between them.}. We hence split the Hilbert space of the genon chain into two sectors, $\mathcal{H}=\mathcal{H}^{-}\oplus\mathcal{H}^{+}$ with $\langle \mathcal{L}^Y_{\psi_1 \psi_2 \psi_{\bar{1}} \psi_{\bar{2}}}\rangle = \mp 1$ respectively.  

\subsubsection{Phase diagram within $\mathcal{H^-}$: Ashkin-Teller model}
{As was noticed in \cite{gils_ashkinteller_2009}, the $\mathcal{H}^-$ sector factorizes into a tensor product on sites, and the Hamiltonian \eqref{eqH} may be mapped onto the Ashkin-Teller model. This mapping is very helpful in constructing the phase diagram of the genon chain. Concretely, one sets $J_{\psi}^{P,R}=\lambda J_{\sigma}^{P,R}$, $J_{i}^{P}=\beta J_{i}^{R}$ and identifies $\beta,\lambda$, as the two parameters in the Ashkin-Teller model as formulated in \cite{kohmoto_hamiltonian_1981}. This allows us to draw a phase diagram for this sector, shown in Fig. \hyperref[figgenonphasediagram]{\ref{figgenonphasediagram}b}. There are three gapped phases, with GSD's 1,2 and 4. We will show how these combine with the ground states of $\mathcal{H}^+$ sector (related to the ground states of $\mathcal{H}^-$ by acting with either $\mathcal{L}_{\sigma_1}$ or $\mathcal{L}_{\sigma_2}$) to form the three gapped phases of the (KSL)$^2$ domain wall above, see Fig.~\ref{figgenonphasediagram}.

The model also realizes a stable gapless orbifold phase with $c=1$, which shrinks upon increasing $\lambda$ until it collapses into a $c=1$ transition line separating the GSD 1 and 4 phases. For even larger $\lambda$, this $c=1$ line splits into two $c=1/2$ lines, one separating the GSD 1 and 2 phases, while the other separating the GSD 2 and 4 phases (see Fig. \hyperref[figgenonphasediagram]{\ref{figgenonphasediagram}b}).
} 

\subsubsection{Remaining symmetry within $\mathcal{H}^{\pm}$}
\label{subsubremainingsymm}

Let us discuss the remaining unbroken symmetry algebra within each sector, which will be used to understand the structure of the phase diagram and the nature of the phase transitions. In fact, analyzing one sector will assist us in analyzing the other since the two are related by application of the broken symmetry $\mathcal{L}_{\sigma_j}$.
 
The remaining symmetries within each sector generate a fusion subalgebra
\begin{equation}\label{eqnTYfusionalg}
\begin{gathered}
    \mathcal{L}_{\psi_1}^2 = \mathcal{L}_{\psi_2}^2=1 \\
    \mathcal{L}_{\psi_1} \mathcal{L}_{\sigma_1 \sigma_2} = \mathcal{L}_{\psi_2} \mathcal{L}_{\sigma_1 \sigma_2} = \mathcal{L}_{\sigma_1 \sigma_2} \\
    (\mathcal{L}_{\sigma_1 \sigma_2})^2 = (1 + \mathcal{L}_{\psi_1})(1 + \mathcal{L}_{\psi_2}),
\end{gathered}
\end{equation}
known as a $\bZ_2 \times \bZ_2$ Tambara-Yamagami (TY) fusion algebra \cite{Tambara1998TensorCW} (the relevant fusion category is known as ${\rm Rep}(H_8)$, which was recently studied in a related context in \cite{thorngren2019fusion}).

To construct gapped phases for this symmetry algebra, we follow the same logic as we used in the (Ising)$^2$ case in \ref{subsec5bilayerphases}. The product of the duality symmetries $\mathcal{L}_{\sigma_1\sigma_2}$ acts on the $\bZ_2 \times \bZ_2 $ phases as:
\begin{equation}
    (I_x,I_y,\text{Brok},\text{Symm})\leftrightarrow(I_y,I_x,\text{Symm},\text{Brok})
    \label{eqOrbitL12}
\end{equation}
While $I_z$ and SPT are fixed. (To see, for example, the transformation laws of the $I_z$ and SPT phase under $\mathcal{L}_{\sigma_1 \sigma_2}$, recall that these phases correspond to the three ground states of the genon domain wall, see Sec.~\ref{subsec5bilayerphases}.) There are four closed orbits under the operation of $\mathcal{L}_{\sigma_1 \sigma_2}$: $\{I_x,I_y\}$, \{Brok,Symm\}, $\{I_z\}$ and \{SPT\}.


\begin{table}
\caption{Splitting of the ground states of the genon chain (Eq.~\eqref{eqH}) into the $\mathcal{H}^{\pm}$ sectors. 
}
\begin{tabular}{ll} 
\begin{tabular}{ |c|c|c|c| } 
\hline
$\mathcal{H}^{-}$ Sector&$\mathcal{H}^{+}$ Sector& GS Splitting &(KSL)$^2$\\
\hline
$I_x+I_y$ & Brok+Symm & 4+5 & Trivial\\
$\langle \mathcal{L}^Y_{\psi_2{\psi}_{\bar{2}}}\rangle$SPT & $\langle \mathcal{L}^Y_{\psi_2{\psi}_{\bar{2}}}\rangle I_z$ & 2+4 & TC gluing\\
SPT & $I_z$ & 1+2 & Genon\\
\hline 
\end{tabular} &
\end{tabular}
\label{tableTYPhases}
\end{table}

By comparing with the discussion in Sec.~\ref{subsec5bilayerphases}, we can understand how the ground states of the three gapped phases of the (Ising)$^2$ domain wall are split between the two sectors $\mathcal{H}^{\pm}$.
This is summarized in Table \ref{tableTYPhases} (to be compared with Table \ref{table369}). We label the phases that build up the TC gluing domain wall by $\langle \mathcal{L}^Y_{\psi_2{\psi}_{\bar{2}}}
\rangle I_z$ and $\langle \mathcal{L}^Y_{\psi_2{\psi}_{\bar{2}}}
\rangle{\rm SPT}$. This labeling indicates, as discussed in \ref{subsec5bilayerphases}, that in the TC gluing phase $\mathcal{L}_{\sigma_1\sigma_2}$ is spontaneously broken and each ground state can be labeled by the sign of the VEV $\langle \mathcal{L}^Y_{\psi_2{\psi}_{\bar{2}}}\rangle
$.

In Fig.~\ref{figgenonphasediagram} and the corresponding Hamiltonians \eqref{eqH}, the phases described in Table \ref{tableTYPhases} are realized. 
In Appendix~\ref{app:genonchain} we present a different family of local Hamiltonians for the genon chain for which the first and second columns of Table \ref{tableTYPhases} are interchanged. E.g., the $I_x + I_y$ phase is realized in the $\mathcal{H}^+$ sector, and so forth.

\subsubsection{Effective Field Theory: $\mathcal{H}^{-}$ Sector}\label{subsubseccritline}
Our goal is to develop an effective field theory for the Hamiltonian \eqref{eqH} near the $c = 1$ critical line (the line between the points $a$ and $b$ in Fig.~\hyperref[figgenonphasediagram]{\ref{figgenonphasediagram}b}) and use it to establish the stability of the nearby phase diagram. Let us focus on the Hilbert space sector $\mathcal{H}^-$. In Subsection \ref{subsubsecothersector} we will derive the field theory of the $c=1$ critical line in the $\mathcal{H}^+$ sector, using our knowledge of $\mathcal{H}^-$.

A key observation is that the $c=1$ critical line is described by an $S^1/\bZ_2$ orbifold theory~\cite{ginsparg_applied_1988,francesco_conformal_1997}. This is well-known for the corresponding critical line in the phase diagram of the Ashkin-Teller model~\cite{ginsparg_applied_1988}, and was confirmed numerically for the genon chain Hamiltonian~\eqref{eqH} in the $\mathcal{H}^{-}$ sector in Ref.~\cite{gils_ashkinteller_2009}. We will describe the action of the TY symmetry operators \eqref{eqnTYfusionalg} within this critical theory, which would allow us to identify the symmetry-allowed relevant perturbations and the nearby gapped phases.

Let us briefly review the properties of the $c=1$ orbifold theory. Its construction begins with a $U(1)$ compact boson (Luttinger liquid), which is conveniently described as a pair of $2\pi$-periodic fields $\phi$ and $\theta$, satisfying 
\begin{equation}\left[\partial_{x}\phi\left(x\right),\theta\left(y\right)\right]=2\pi i\delta\left(x-y\right).\end{equation}
The Hamiltonian is
\begin{equation}
H=\frac{1}{2\pi}\int\left(K\left(\partial_{x}\phi\right)^{2}+\frac{1}{4K}\left(\partial_{x}\theta\right)^{2}\right) dx,
\end{equation}
where $K$ is the Luttinger parameter. This theory has a primary vertex operator $V_{n,m} = e^{i n \phi}e^{i m \theta}$ for every $m,n \in \bZ$ of dimension
\begin{equation}\label{eqndimformula}
    \Delta_{n,m} = \frac{n^2}{4K} + Km^2.
\end{equation}
It also has a pair of $U(1)$ currents $\partial \phi$, $\partial \theta$ from which we construct the remaining operators in the spectrum \cite{francesco_conformal_1997}. 

The theory at $K = 1$ is equivalent to a free Dirac fermion. The moduli of theories enjoy the so-called $T$ duality:
\begin{equation}
\begin{split}
    K & \leftrightarrow \frac{1}{4K}\\
    \phi & \leftrightarrow \theta,
\end{split}
\label{eqntduality}
\end{equation}
for which $K = 1/2$ is the self-dual point, equivalent to the $SU(2)$-symmetric spin-$1/2$ Heisenberg chain. Finally, $K = 2$ is the Berezinskii-Kosterlitz-Thouless point. For more details, see \cite{ginsparg_applied_1988}.

This theory has two $U(1)$ symmetries acting as shifts of $\phi$ and $\theta$, as well as a charge conjugation symmetry \begin{equation}
C:\begin{cases} \phi \to -\phi, \\ \theta\to -\theta. \end{cases}
\end{equation} The $S^1/\bZ_2$ orbifold is obtained from this theory by gauging $C$. This procedure projects out all states that are charged under $C$, and adds the states from the $C$ twisted sectors (see Appendix~\ref{appgauging}). Intuitively, this identifies $\theta \sim - \theta$ and $\phi \sim - \phi$, effectively restricting the range of $\theta$ and $\phi$ to $[0,\pi]$. All $C$-odd operators such as $\sin\theta \sim i(V_{0,1} - V_{0,-1})$ and $\partial \phi$ are projected out, while $C$-even combinations such as $\cos\phi \sim V_{1,0} + V_{-1,0}$ remain and have dimensions determined by \eqref{eqndimformula}. The marginal operator $(\partial \phi)^2$, which tunes $K$, also remains in the spectrum.

Gauging $C$ also introduces four additional primaries in the twisted sector: $\sigma_i$ of dimension $1/8$, and $\tau_i$ of dimension $9/8$, with $i = 1,2$ corresponding to the two $C$-fixed points $\theta = 0, \pi$. The dimensions of the twist operators are insensitive to the marginal parameter $K$ \cite{francesco_conformal_1997}.

Let us elaborate on this point. After the gauging procedure, the Hilbert space includes states defined by configurations of $\theta$ or $\phi$ with symmetry-twisted boundary conditions. Configurations with constant $\theta = 0, \pi$ define states that satisfy either periodic or anti-periodic boundary conditions for both $\theta$ and $\phi$. 
The twist operator $\sigma_1$ has eigenvalue $\pm 1$ when acting on the twisted/untwisted state with $\theta=0$, respectively. Similarly, $\sigma_2$ returns $\pm 1$ when acting on the two states with  $\theta=\pi$. 


There exists a point in parameter space where the Hamiltonian decouples into two critical Ising models, matching the $K=1$ point along the orbifold line. At this point, the twists fields are identified with the familiar spin operators of the Ising critical point $s_1$, $s_2$, respectively. 

\begin{figure}
    \centering
    \includegraphics[width=0.9\columnwidth]{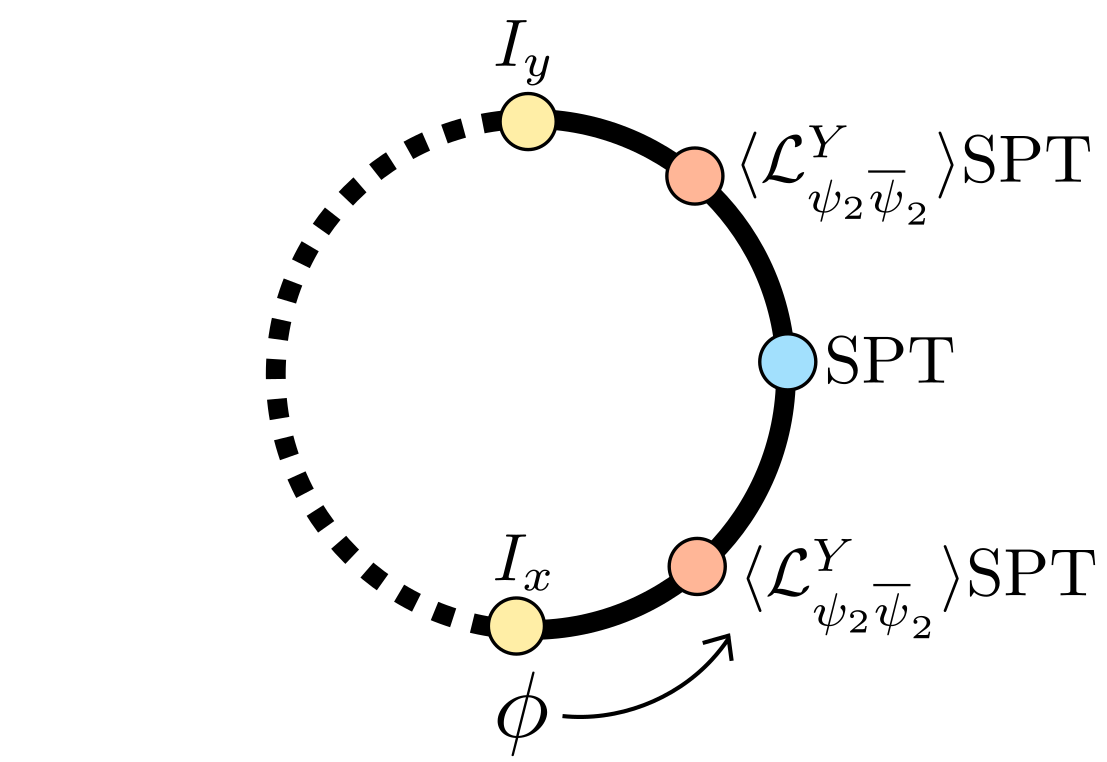}
    \caption{The target space of the orbifold is visualized as half of a circle, with special points at the top and bottom, corresponding to $\phi = 0,\pi$. Each of the special points contributes two ground states, since they are compatible with both possible boundary conditions. Consider the perturbation $\pm\cos 2\phi$. With the positive sign, $\phi$ settles to the minima $\phi = \pm \pi/2$, both identified with the blue point representing a unique ground state, which is an SPT phase. This is the phase on the bottom of Fig. \hyperref[figgenonphasediagram]{\ref{figgenonphasediagram}b}. With the negative sign, $\phi$ settles into the minima $\phi = 0,\pi$ (yellow points). The values $\phi = 0,\pi$ each contributes two ground states, to a four-ground-state TY phase identified as $I_x+I_y$, as explained in the main text. This is the top phase in Fig. \hyperref[figgenonphasediagram]{\ref{figgenonphasediagram}b}. The two signs of $\langle\cos\phi\rangle$ distinguish $I_x$ and $I_y$. The red points at $\phi=\pi/4$ and $\phi=3\pi/4$ represent ground states in the TC gluing phase, which may be accessed by perturbing with the operator $\cos4\phi$.}
    \label{figvacua}
\end{figure}

In the genon chain Hamiltonian~\eqref{eqH}, this point corresponds to $J^P_\sigma =  J^R_\sigma$, $J^P_\psi=J^R_\psi=0$, where $H$ can be written explicitly as a sum of two decoupled transverse Ising models (Appendix \ref{appATmapping}). By examining the action of the line operators $\mathcal{L}_{\psi_{1,2}}$ on the genon chain operators at this point we find the action of the emergent symmetries on the scaling fields:
\begin{equation}
\begin{split}
        \mathcal{L}_{\psi_{1}}&:(s_1,s_2,\epsilon_1,\epsilon_2)\mapsto (s_2,s_1,\epsilon_2,\epsilon_1),\\ \mathcal{L}_{\psi_{2}}&:(s_1,s_2,\epsilon_1,\epsilon_2)\mapsto (-s_2,-s_1,\epsilon_2,\epsilon_1).
\end{split}
\label{eqsymmetryIsing2}
\end{equation}
These transformation rules are derived in Appendix \ref{appATmapping}. The symmetry action may be extended to the rest of the orbifold by matching the (Ising)$^2$ fields with the orbifold ones via their scaling dimensions at the $K=1$ point. The spin fields are identified with the twist operators with scaling dimension $1/8$. The product of the spins $s_1s_2$ is identified with $\cos\phi$ with scaling dimension $1/4$. The operators $\epsilon_1 + \epsilon_2$ and $\epsilon_1 - \epsilon_2$ are identified with $\cos2\phi$ and $\cos\theta$, respectively, with scaling dimension $1$. This can be checked by using these operators to perturb the system to nearby gapped phases. For example, tuning the coefficient of a perturbation of the form $g\cos2\phi$ from $g<0$ to $g>0$ tunes the system between a phase with 4 ground states, and a phase with a unique ground state, as $\epsilon_1 + \epsilon_2$ does. See Fig. \ref{figvacua} for details. The extension of \eqref{eqsymmetryIsing2} to the rest of the orbifold is\footnote{The symmetries and fields can be matched to the Ising$^2$ operators using Section 4 of \cite{thorngren2019fusion}, although note that our conventions are $T$-duals of each other, so $\phi$ and $\theta$ are switched. As a subgroup of $D_8$ in that reference, the $\bZ_2 \times \bZ_2$ symmetry we consider in the $\mathcal{H}^+$ (resp. $\mathcal{H}^-$) sector is $\langle r^2, s\rangle$ (resp. $\langle r^2, sr\rangle$).}:
 
\begin{equation}
\begin{split}
        \mathcal{L}_{\psi_{1}}&:(\theta,\phi,\sigma_1,\sigma_2)\mapsto (\theta+\pi,\phi,\sigma_2,\sigma_1),\\ \mathcal{L}_{\psi_{2}}&:(\theta,\phi,\sigma_1,\sigma_2)\mapsto (\theta+\pi,\phi,-\sigma_2,-\sigma_1).
\end{split}
\label{eqnZ2symmetry}
\end{equation}

We also need to understand the $\bZ_2 \times \bZ_2$ action in the twisted sectors. In other words, we need to know the charges of open string operators of  $\mathcal{L}_{\psi_1}$ and $\mathcal{L}_{\psi_2}$ in the microscopic genon chain model. 
There are two possibilities given the above action of $\mathcal{L}_{\psi_{1,2}}$,  distinguished either by the $\mathcal{L}_{\psi_2}$ charge of the open $\mathcal{L}_{\psi_1}$ string operator with the smallest scaling dimension for $K>1/2$, or by analyzing the nearby gapped phase obtained by perturbing by $\cos 2\phi$. This gapped phase is either a trivial or SPT phase, depending on whether that charge of the end of the $\mathcal{L}_{\psi_1}$ is trivial or nontrivial under $\mathcal{L}_{\psi_2}$ (the latter has a non-zero VEV in this phase). We verify in Appendix \ref{appATmapping} using the microscopic model that, in the gapped phase, the ends of the $\mathcal{L}_{
\psi_1}$ string are charged under $\mathcal{L}_{\psi_2}$, and hence this phase is indeed an SPT. (We refer to this property of the $\bZ_2 \times \bZ_2$ symmetry as discrete torsion.) In fact, without this property, the theory would not be self-dual under gauging $\bZ_2 \times \bZ_2$, which we know it must be to have $\mathcal{L}_{\sigma_1 \sigma_2}$ symmetry. See \cite{thorngren2019fusion} for a related discussion at the Ising$^2$ point $K = 1$.


Next, we would like to determine the action of the duality-symmetry $\mathcal{L}_{\sigma_1\sigma_2}$ on the local operators of the theory. It is useful to keep in mind the case of the usual KW duality in the Ising CFT, where the energy density operator $\epsilon$ is charged under KW duality. This can be seen from the fact that $\epsilon$ perturbs the CFT into either the ordered or disordered phases, neither of which is KW self-dual. Similarly, we can perturb the orbifold by local operators, and check whether the theory flows to a phase which is self-dual under $\mathcal{L}_{\sigma_1\sigma_2}$. Any operator that triggers a flow to a phase which is not self-dual is symmetry disallowed. In other words, if the flow ends at one of the phases in Table~\ref{tableTYPhases}, the operator is $\mathcal{L}_{\sigma_1\sigma_2}$ even, and if we end up in another $\bZ_2 \times \bZ_2$-symmetric gapped phase, then it is odd.

Consider, for example, the operator $\pm \cos2\phi \sim \epsilon_1+\epsilon_2$. Perturbing the theory with this operator with a negative sign yields a ferromagnetic phase with four ground states. In terms of the fields, the potential has two minima, $\phi = 0$ and $\phi = \pi$. Since both of these points are fixed by the gauge symmetry $C$, they each contribute two ground states where the magnetic symmetry $\sigma_j \mapsto -\sigma_j$ is spontaneously broken. Considering the action of the $\bZ_2$ symmetries in Eq. \eqref{eqsymmetryIsing2} on these four ground states, we find that they transform as the ground states of $I_x+I_y$ with the two $\phi = 0$ states transforming as $I_x$ and the two $\phi = \pi$ states transforming as $I_y$. For example, the $\phi = 0$ states are both invariant under $\mathcal{L}_{\psi_1}$ ($\phi$ is the eigenvalue of the $\theta$ shift operator), while $\mathcal{L}_{\psi_2}$ is $\mathcal{L}_{\psi_1}$ times the magnetic symmetry and is therefore broken, along with $\mathcal{L}_{\psi_1} \mathcal{L}_{\psi_2}$. Comparing with Table \ref{tab:tcbilayer}, this is $I_x$. With the positive sign $\cos2\phi$ perturbation on the other hand, the theory flows into a phase with one symmetric ground state, which can be verified to be an SPT state (as we mentioned above). Since both $I_x+I_y$ and SPT are self-dual under $\mathcal{L}_{\sigma_1\sigma_2}$ (see Eq. \ref{eqOrbitL12}), we conclude that the operator $\cos2\phi$ is symmetry allowed.




Next, let us consider the operator $\cos\phi \sim s_1s_2$. Considering the action of this operator on the four ground states of $I_x+I_y$, we find that $\langle\cos\phi\rangle$ has an opposite sign for the ground states of $I_x$ and $I_y$. Since those ground states interchange under $\mathcal{L}_{\sigma_1\sigma_2}$, we conclude that $\cos\phi$ is symmetry disallowed.

In Appendix \ref{appdualmapping} we derive a general formula for the action of $\mathcal{L}_{\sigma_1 \sigma_2}$ on all the vertex operators
\begin{equation}\label{eqndualityactionminussector}
\mathcal{L}_{\sigma_1 \sigma_2} |\Psi_{m,n}\rangle =  i^{m+2n} \frac{1}{2}(1 + \mathcal{L}_{\psi_1})(1 + \mathcal{L}_{\psi_2}) |\Psi_{m,n}\rangle,
\end{equation}
where $m,n$ are the momentum and winding numbers of the state $|\Psi_{m,n}\rangle$ related by operator-state correspondence to the operator $V_{m,n}$. The state $|\Psi_{m,n}\rangle$ is sent to zero if $m$ is odd, otherwise it gets a factor $2 (-1)^{m/2+n}$. Meanwhile all twist operators are sent to zero. Observe that
\begin{equation}
    (\mathcal{L}_{\sigma_1 \sigma_2})^2 = (1 + \mathcal{L}_{\psi_1})(1 + \mathcal{L}_{\psi_2})
\end{equation}
as required by the TY fusion algebra \eqref{eqnTYfusionalg}. From Eq. \eqref{eqndualityactionminussector} and the above considerations, one finds the charges of all the vertex operators in the theory under the elements of the TY symmetry. The charges are summarized in Table \ref{tablecharges} in Appendix \ref{appdualmapping}.

In Fig. \ref{figvacua}, we show the target space of the orbifold theory, and identify the phases discussed above. The vacua at $\phi=0$ and $\phi=\pi$ belong to the phases $I_x$ and $I_y$ respectively, while $\phi=\pm\pi/2$ belongs to the SPT phase. The two ground states of $\langle \mathcal{L}^Y_{\psi_2{\psi}_{\bar{2}}}\rangle$SPT (belonging to the TC gluing domain wall) correspond to $\phi=\pi/4$ and $\phi=3\pi/4$. Indeed, $\langle \cos\phi\rangle$ has opposite sign at those two points, and serves as an order parameter for the spontaneous breaking of $\mathcal{L}_{\sigma_1 \sigma_2}$. Those ground states are accessed by the operator $\cos4\phi$, which can be checked to be TY symmetric by the methods above. We have thus accounted for all the gapped states in the $\mathcal{H}^-$ sector, appearing in Table~\ref{tableTYPhases}.

We will now derive the phase diagram in Fig. \hyperref[figgenonphasediagram]{\ref{figgenonphasediagram}b} and its stability. We start from the critical orbifold line, consider all symmetry allowed operators (see charge assignments of Appendix \ref{appdualmapping}), and identify the relevant perturbations using Eq. \eqref{eqndimformula}.

For the critical line between the points $a$ and $b$ in Fig. \hyperref[figgenonphasediagram]{\ref{figgenonphasediagram}b}, corresponding to $1/2 < K < 2$, there is a single symmetry allowed relevant operator $\cos 2\phi$. Along this line $\pm\cos 2\phi$ perturbs into the phases SPT and $I_x + I_y$ respectively, See Fig. \ref{figvacua}. Those are the expected TY symmetric phases in Fig. \hyperref[figgenonphasediagram]{\ref{figgenonphasediagram}b}.

For $K>2$, the operator $\cos 4\phi$ is relevant. With a positive coefficient [realized in the microscopic genon chain Hamiltonian, Eq.~\eqref{eqH}] this leads to the spontaneous duality breaking phase $\langle \mathcal{L}^Y_{\psi_2{\psi}_{\bar{2}}}\rangle$SPT, as explained above. With negative coefficient, the critical line becomes a first order line (this is not realized in this Hamiltonian).

At $K = 2$, the $c = 1$ critical line ends and splits off into two $c = 1/2$ lines (the point $b$ in Fig. \hyperref[figgenonphasediagram]{\ref{figgenonphasediagram}b}). The lower line in Fig. \hyperref[figgenonphasediagram]{\ref{figgenonphasediagram}b} is a transition between the phases SPT and $\langle \mathcal{L}^Y_{\psi_2{\psi}_{\bar{2}}} \rangle{\rm SPT}$. This critical line is in the Ising universality class, where the order parameter $\cos \phi$ is charged under $\mathcal{L}_{\sigma_1\sigma_2}$.

The upper critical line coming off the $K = 2$ point is a $c = 1/2$ transition from the 2 ground state phase $\langle \mathcal{L}^Y_{\psi_2{\psi}_{\bar{2}}} \rangle{\rm SPT}$ to the 4 ground state phase $I_x + I_y$. The critical theory has two sectors, each an Ising CFT but with one order parameter charged under $\mathcal{L}_{\psi_1}$ and the other charged under $\mathcal{L}_{\psi_2}$, so that individually each looks like a transition from the SPT to $I_x$ or $I_y$, respectively. The two sectors are exchanged by the already broken symmetry $\mathcal{L}_{\sigma_1\sigma_2}$.

For $1/8<K<1/2$, there is a $c=1$ gapless region with no relevant operators. The boundaries between this region and the gapped phases are of Berezinskii-Kosterlitz-Thouless universality class.

As discussed in Sec.~\ref{subsubremainingsymm}, these are the only phases realized in $\mathcal{H}^{-}$
within the model \eqref{eqH}. These are the phases appearing in the first column of Table~\ref{tableTYPhases}.  Within the field theory description, the possibility is open to find also the phases corresponding to the second column of Table~\ref{tableTYPhases} in $\mathcal{H}^{-}$.  
At $K \le 1/8$, the TY symmetric operator $\cos 4\theta$ becomes relevant. In that region, one can verify that a positive perturbation flows to the phase $I_z$ while a negative perturbation flows to ${\rm Symm} + {\rm Brok}$. 

As we show below, these phases have simple realizations in the $\mathcal{H}^+$ sector, and can also be accessed in $\mathcal{H}^{-}$ by augmenting the genon chain Hamiltonian with longer-range terms (see  Appendix \ref{appgappedphases}). If one keeps decreasing $K$, one accesses a phase diagram that is a mirror to Fig. \ref{figgenonphasediagram}, with the phases in the two sectors swapped, and the effective field theories related by $T$-duality.

\subsubsection{Effective Field Theory : $\mathcal{H}^{+}$ Sector}\label{subsubsecothersector}

We now turn our attention to the $\mathcal{H}^+$ sector of the Hilbert space. Recall that the two sectors are related by applying $\mathcal{L}_{\sigma_j}$.  This is equivalent to gauging either of the $\bZ_2$ symmetries (see Appendix~\ref{appgauging}). The gapped phases are matched with those of $\mathcal{H}^{-}$ according to Table \ref{tableTYPhases}.

Along the $c = 1$ critical line it is convenient to gauge $\mathcal{L}_{\psi_1}$, represented in the effective field theory as a shift symmetry \eqref{eqnZ2symmetry}. Gauging this symmetry results in an $S^1/\bZ_2$ orbifold with fields $\tilde \theta$, $\tilde \phi$ at $\tilde K = 4K$. Intuitively, gauging the symmetry in Eq. \eqref{eqnZ2symmetry} effectively restricts the range of $\theta$ to $[0,\pi/2]$, and rescaling this range back changes the radius of the orbifold and its parameter $K$ accordingly, see Appendix \ref{appshiftsymmetry}. However, because of the discrete torsion, the symmetries $\mathcal{L}_{\psi_j}$ act differently on the $\tilde \theta$, $\tilde \phi$ fields than on the $\theta$, $\phi$ fields in \eqref{eqnZ2symmetry}. Instead, we have:
\begin{equation}
\begin{split}
        \mathcal{L}_{\psi_{1}}&:(\tilde\theta,\tilde\phi,\sigma_1,\sigma_2)\mapsto (\tilde\theta,\tilde\phi+\pi,-\sigma_1,\sigma_2),\\ \mathcal{L}_{\psi_{2}}&:(\tilde\theta,\tilde\phi,\sigma_1,\sigma_2)\mapsto (\tilde\theta,\tilde\phi+\pi,\sigma_1,-\sigma_2).
\end{split}
\label{eqnZ2symmetry2}
\end{equation}
Likewise we find (see Appendix \ref{appdualmapping})
\begin{equation}\label{eqndualityactionplussector}
\mathcal{L}_{\sigma_1 \sigma_2} |\Psi_{m,n}\rangle =  i^{n+2m} \frac{1}{2}(1 + \mathcal{L}_{\psi_1})(1 + \mathcal{L}_{\psi_2}) |\Psi_{m,n}\rangle,
\end{equation}
compare Eq. \eqref{eqndualityactionplussector} to \eqref{eqndualityactionminussector}---the two equations are $T$-duals, see also Section \ref{subsecisingcubed} below.

The $c = 1$ line in the $\mathcal{H}^+$ sector extends from $\tilde K = 2$ to $\tilde K = 8$ (the points $a_*$ and $b_*$ in Fig. \hyperref[figgenonphasediagram]{\ref{figgenonphasediagram}c}). This matches the critical line in the $\mathcal{H}^-$ sector, by recalling $\tilde K = 4K$. Along this line, we find a single symmetry allowed relevant operator $\cos 4 \tilde \phi$, which tunes from the phase $I_z$ to the phase ${\rm Symm} + {\rm Brok}$ (see Appendix \ref{appdualmapping}). Those are the GSD 2 and GSD 5 phases in Fig. \hyperref[figgenonphasediagram]{\ref{figgenonphasediagram}c}.

For $\tilde K > 8$, the operator $\cos 8 \tilde \phi$ becomes relevant and drives spontaneous breaking of the symmetry $\mathcal{L}_{\sigma_1\sigma_2}$, leading to the phase $\langle \mathcal{L}^Y_{\psi_2{\psi}_{\bar{2}}} \rangle I_z$. This phase has GSD 4, and is the representative in $\mathcal{H}^{+}$  of the toric code gluing domain wall, see Fig. \hyperref[figgenonphasediagram]{\ref{figgenonphasediagram}c}.

The lower $c = 1/2$ line emerging from the $\tilde K = 8$ point ($b_{*}$ in Fig. \hyperref[figgenonphasediagram]{\ref{figgenonphasediagram}c}) is in the Ising universality class, with order parameter charged under the TY symmetry $\mathcal{L}_{\sigma_1\sigma_2}$. This line is the transition between the phases $I_z$ and $\langle\mathcal{L}^Y_{\psi_2{\psi}_{\bar{2}}}\rangle I_z$.

The upper $c = 1/2$ line is an Ising CFT with three degenerate ground states, representing two transition sectors corresponding to the two possible signs of the VEV $\langle\mathcal{L}^Y_{\psi_2{\psi}_{\bar{2}}}\rangle$. Those two sectors are toggled by the duality symmetry $\mathcal{L}_{\sigma_1\sigma_2}$. One sector undergoes an $\langle\mathcal{L}^Y_{\psi_2{\psi}_{\bar{2}}}\rangle I_z$ (with a fixed sign of $\langle\mathcal{L}^Y_{\psi_2{\psi}_{\bar{2}}}\rangle$) to ${\rm Symm}$ transition, while the other undergoes an $\langle\mathcal{L}^Y_{\psi_2{\psi}_{\bar{2}}}\rangle I_z$ to Brok transition.\footnote{We find a disagreement with the numerics of \cite{gils_ashkinteller_2009} with regards to the spectrum of of this $c=1/2$ theory. We find that both the vacuum and the spin field are threefold degenerate, while in \cite{gils_ashkinteller_2009} the reported degeneracies are 1 for the vacuum and 3 for the spin field.} Notice that interestingly, one sector undergoes symmetry breaking while in the other sector a symmetry is restored. 

For $\tilde K < 2$, the operator $\cos 4 \tilde \phi$ becomes irrelevant and we have a stable gapless phase with GSD 1 which is eventually destroyed at $\tilde K = 1/2$. For smaller $\tilde K$ the operator $\cos 2\tilde \theta$ becomes relevant and perturbs to the phases $I_x + I_y$ or SPT, depending on its sign (this regime is not realized in the model \eqref{eqH}, and is the analogue to the regime accessed by $\cos4\theta$ in the $\mathcal{H}^{-}$ sector, see previous subsection).

\subsection{$c = 3/2$ Point
Restoring the Symmetries $\mathcal{L}_{\sigma_j}$}\label{subsecisingcubed}

The (KSL)$^2$ topological order supports a stable gapless domain wall with $c=1$, corresponding to ``cutting open" both layers. This domain wall is not realized in the genon chain model. In this section we will propose a field theoretical description for the phase transition connecting this domain wall with the $c=1$ orbifold gapless phase of the genon chain.

In the genon chain, the symmetries $\mathcal{L}_{\sigma_j}$ were spontaneously broken throughout the whole phase diagram, as explained in \ref{subsecsymmetrybreaking}. In fact, this was built into the very nature of the Hilbert space. One could imagine enlarging this Hilbert space to access some gapped degrees of freedom that eventually drive the symmetry breaking transition of $\mathcal{L}_{\sigma_j}$. The phase transition point connects the two-sector stable gapless phase of Fig. \ref{figgenonphasediagram} to another stable gapless phase with a unique ground state, corresponding to the ``cut open" domain wall of the KSL bilayer.

To describe this critical point, we consider the $c = 3/2$ theory (Ising)$^3$, a tensor product of three decoupled critical Ising chains. This theory has energy and spin operators $\epsilon_i$, $s_i$, $i = 1,2,3$. We take the Ising$\times$Ising category symmetry to act so that $\mathcal{L}_{\psi_j}$ act as the usual $\bZ_2$ spin flip symmetries of the $j$th chain, $j = 1,2$. This forces $\mathcal{L}_{\sigma_j}$ to act by KW duality on the $j$th chain. However, we will also choose both $\mathcal{L}_{\sigma_j}$ to act on the third chain as a $\bZ_2$ symmetry $s_3 \mapsto -s_3$.

The resulting theory has one symmetric relevant operator $\epsilon_3$ and no symmetric marginal operators. Perturbation by $\epsilon_3$ with positive coefficient gaps the third chain, resulting in a stable gapless (Ising)$^2$ phase with a unique ground state and all symmetries preserved. This phase corresponds to the ``cut open" domain wall of the KSL bilayer.

With a negative sign of the perturbation $\epsilon_3$, $s_3$ obtains a non-zero VEV and the symmetries $\mathcal{L}_{\sigma_j}$ spontaneously break. The result is a two-sector (Ising)$^2$ theory, characterized by the sign of $\langle s_3\rangle$, where the sectors are nothing but $\mathcal{H}^{\pm}$ discussed above. In each individual sector, the unbroken TY symmetry $\mathcal{L}_{\sigma_1 \sigma_2}$ acts as the diagonal KW duality. Meanwhile, the operators $\mathcal{L}_{\sigma_j=1,2}$ exchange the sectors and gauge the corresponding $\mathcal{L}_{\psi_j}$ symmetries. Note that there are no symmetric relevant perturbations, but there is an exactly marginal operator:
\begin{equation}\label{eqnmargpert3/2}
\epsilon_1 \epsilon_2 \langle s_3\rangle
\end{equation}
which is symmetry allowed. One should think of this operator as coming from the dangerously irrelevant symmetric operator $\epsilon_1 \epsilon_2 s_3$ of the (Ising)$^3$ point, which becomes exactly marginal once $s_3$ obtains a VEV. 
The $\epsilon_3<0$ phase corresponds to the stable gapless region of the $c=1$ orbifold theory discussed in Sections~\ref{subsubseccritline}, \ref{subsubsecothersector}, where the sign of $\langle s_3 \rangle$ determines whether the system is in the  $\mathcal{H}^{+}$ or $\mathcal{H}^{-}$ sector. 
Crucially, while in the cut open gapless phase the system was pinned to the (Ising)$^2$ point, in this gapless phase we can tune the marginal parameter~\eqref{eqnmargpert3/2}, which changes the orbifold radius. The $\epsilon_3<0$ phase corresponds to the stable gapless region of the $c=1$ orbifold theory discussed in Sections~\ref{subsubseccritline}, \ref{subsubsecothersector}. 

In order to relate the $\bZ_2 \times \bZ_2$ symmetry of the two sectors of the orbifold theory [Eqs.~\eqref{eqnZ2symmetry},\eqref{eqnZ2symmetry2}] to the symmetries of the effective theory of three Ising chains discussed here, 
it is convenient to perform $T$-duality [see Eq.~\eqref{eqntduality}] to one of the sectors in the symmetry broken ($\langle s_3 \rangle\ne 0$) phase, say $\mathcal{H}^{-}$. This has two pleasant effects. First, tuning the allowed marginal operator will tune the parameter of the orbifold in the $\mathcal{H}^{-}$ sector, $K$, and that of the $\mathcal{H}^{+}$ sector, $\tilde K = 4K$, in the same direction. Second, before the T-duality, the $\bZ_2 \times \bZ_2$ symmetry acted differently in the two sectors. 
After performing $T$-duality in the $\mathcal{H}^{-}$ sector, 
the $\bZ_2 \times \bZ_2$ in both sectors act as spin flips for the two spin fields [Eq.~\eqref{eqnZ2symmetry2}], and can be matched with the symmetries that flip $s_1$ and $s_2$.

\begin{figure}
    \centering
    \includegraphics[width=\columnwidth]{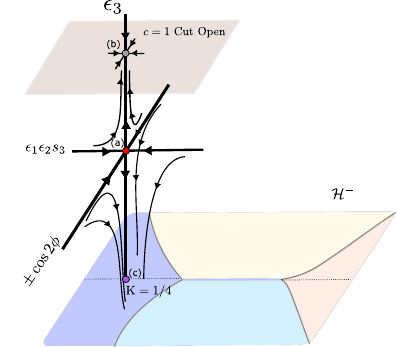}
    \caption{The 3-parameter RG flow surrounding the $c = 3/2$ critical point (a). This critical point has a single relevant direction labelled by $\epsilon_3$. With one sign, this perturbation triggers a flow to the cut open $c = 1$ gapless phase, with no marginal directions, point (b). With the other sign, this perturbation triggers a flow from the fixed point to the $K = 1/4$ point (c) in the $c = 1$ orbifold phase of the genon chain (cf. Fig. \ref{figgenonphasediagram}; the corresponding point in $\mathcal{H}^{+}$ is $\tilde K=1$). At this point, $s_3$ obtains a nonzero VEV and the operator $\epsilon_1\epsilon_2s_3$ becomes marginal.
    }
    \label{fig3/2}
\end{figure}

In Fig. \ref{fig3/2}, we present a three dimensional phase diagram describing the renormalization group flow around the $c=3/2$ fixed point. The axes are labeled by the perturbations to the $c=3/2$ point (The operator $\cos2\phi$ is replaced by $\cos4\tilde\phi$ when considering $\mathcal{H}^{+}$).

One could add to the relevant perturbation $\epsilon_3$ some dangerously irrelevant term $\epsilon_1 \epsilon_2 s_3$ as well as other irrelevant parameters of the third Ising chain. Those perturbations set the VEV of $s_3$ in the symmetry breaking phase, and the result is that the flow ends up at some point in the gapless phase, not necessarily the point (c) in Fig. \ref{fig3/2}. If one allows the coefficient of the irrelevant terms to be large, it is possible that the flow ends up outside of the stable gapless region of Fig. \ref{figgenonphasediagram}, in one of the gapped phases.

\subsection{Multicritical Points with $c=7/5$ and the Genon Duality}\label{subsecgenonduality}

Finally, let us connect our description of the domain wall phase diagram to the discussion in Section \ref{secKSLtricrit}. Recall in the case of a single KSL layer, the cut open domain wall hosts a (non-chiral) $c = 1/2$ Ising CFT, stabilized by a $\bZ_2$ symmetry and KW duality. The most relevant symmetric perturbation is the $T \bar T$ operator with scaling dimension 4, where $T$ is the stress-energy tensor of the chiral Ising CFT. This perturbation is the four-Majorana interaction in Eq. \eqref{eqn4fermipert}. When this perturbation is sufficiently strong, it drives the system into a $c=7/10$ critical point, beyond which a gap open.

\begin{figure}
    \centering
    \includegraphics[width=0.8\columnwidth]{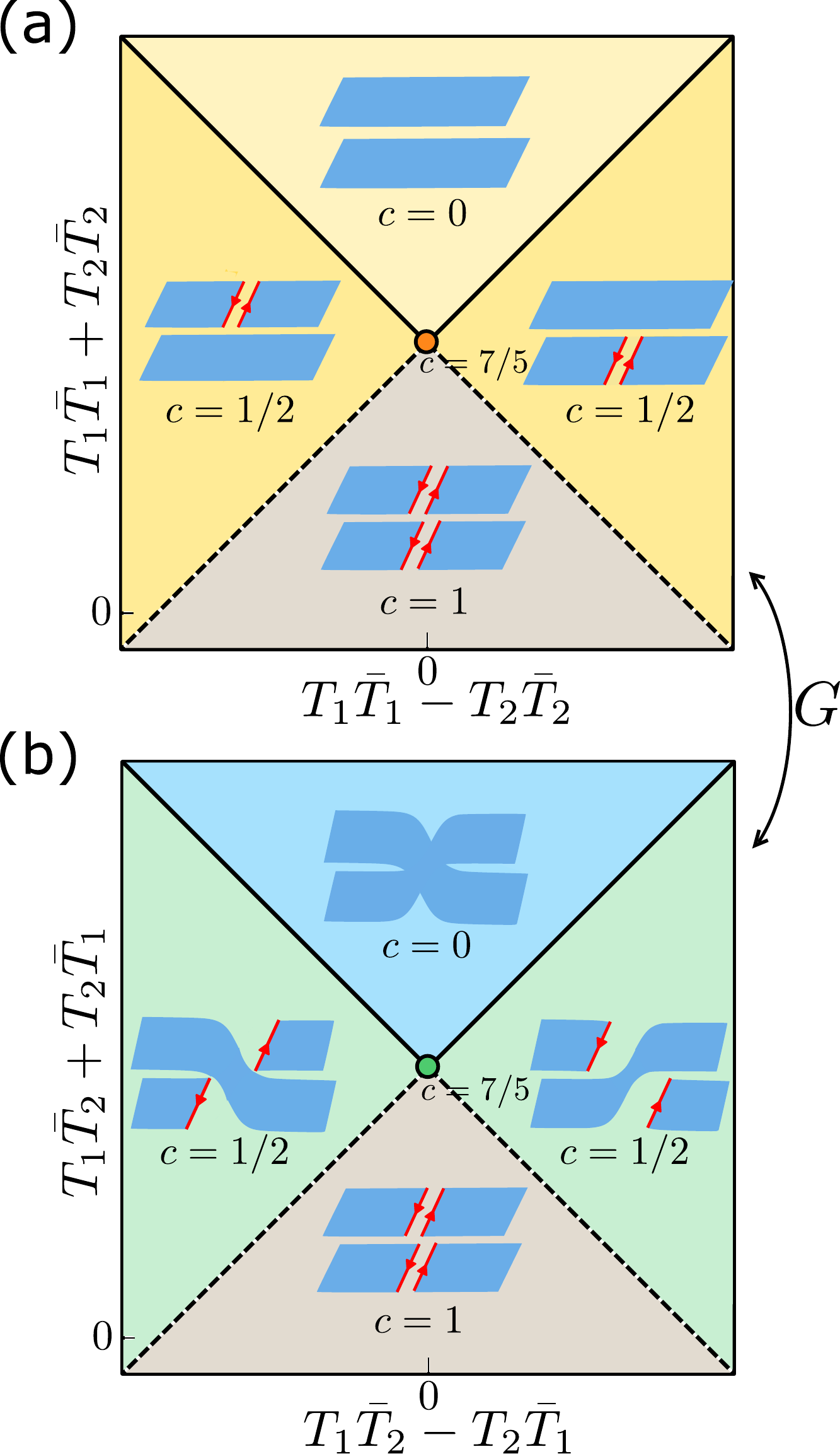}
    \caption{(a) The phase diagram in the vicinity of a $c=7/5$ point. The axes are labeled by the irrelevant deformations of the (Ising)$^2$ theory, which drive the gluing transitions of the layers. The two diagonal lines correspond to independent decoupling transitions at each layer, discussed in Sec. \ref{secKSLtricrit}. There are four stable phases in the vicinity, represented pictorially - Both layers may be either glued or cut. The transition lines between the $c=0$ and $c=1/2$ phases (black) have $c=7/10$, while the transition lines between the $c=1/2$ and $c=1$ phases (dotted black) have $c=6/5=1/2+7/10$. (b) A dual phase diagram related by application of a genon duality. The middle point consists of a dual $c=7/5$ theory. The nearby phases are obtained by fusing a genon along the domain wall, flipping the layers from one side. The gapless phases at the bottom of each are both the cut open phase, which is self-dual under the genon duality.}
    \label{figmulticrit}
\end{figure}

Likewise, in the cut open domain wall of a KSL bilayer, we have two decoupled Ising CFTs that can be coupled by $T_i \bar T_j$ for $i,j = 1,2$, where $T_i$ denotes the stress-energy tensor of the $i$th layer\footnote{We do not consider couplings such as $T_1 T_2$, as these cannot open a gap due to chirality. Meanwhile, the $T_j$ themselves only change the Fermi velocities.}. Such a perturbation is generated by a two-body spin interaction in the honeycomb model connecting the $i$th and $j$th layers (in analogy with the ZZ interaction for the horizontal cut in Fig. \ref{figtricrit}). It is clear from the analysis of a single layer that for each of the perturbations
\begin{equation}T_1 \bar T_1 + T_2 \bar T_2 \qquad {\rm and} \qquad T_1 \bar T_2 + T_2 \bar T_1\end{equation}
there is a critical strength where the system encounters a continuous transition and becomes a pair of two decoupled tricritical Ising models, with total central charge $c = 7/5$. Tuning above the critical value will result in the trivial domain wall for $T_1 \bar T_1 + T_2 \bar T_2$ and the genon domain wall for $T_1 \bar T_2 + T_2 \bar T_1$, see Fig. \ref{figmulticrit}.

The two $c = 7/5$ theories in Fig. \hyperref[figmulticrit]{\ref{figmulticrit}a,b} must correspond to two different critical points. This can be seen from the fact that there is only one symmetry allowed perturbation of the $c = 7/5$ theory that gaps the system completely, and this perturbation drives the system into different gapped phases in the the two $c=7/5$ theories. The gapping operator is $\epsilon_1' + \epsilon_2'$, where $\epsilon'_{1,2}$ is the self dual operator \cite{francesco_conformal_1997,chang_topological_2019} which appeared in the study of a single layer (cf. Fig. \ref{figtricrit}).

There is an interesting duality which we denote by $G$, which is realized by applying a genon defect to the domain wall, effectively interchanging the layers on one side of the domain wall\footnote{One could apply the defect to either side of the domain wall, resulting in two similar but distinct dualities, which differ by an overall swap of the two layers.}. This ``genon duality" relates the above tricritical points and their nearby phase diagrams. $G$ is a self-duality of the cut open phase, and acts on the antiholomorphic stress tensors by $\bar T_1 \leftrightarrow \bar T_2$, while the holomorphic stress tensors are invariant. The two $c = 7/5$ points are exchanged under this duality. See Fig. \ref{figmulticrit} for an illustration of the duality action on the critical points and their nearby phases.

Similar to the $e\leftrightarrow m$ duality of the toric code, we can write $G$ as an operator acting on the quasi-1d system, satisfying a fusion algebra related to the anyons of the bulk. This operator permutes the phase diagram so that the trivial and genon domain walls are interchanged, while the toric code gluing phase and the two $c=1$ gapless domain walls are self-dual. Applying the general expressions for the genon fusion rules in \cite{barkeshli_genons_2012,barkeshli_symmetry_2014}, $G$ satisfies
\begin{equation}
    G^{2} =1+\mathcal{L}_{\psi_{1}\psi_{2}}+\mathcal{L}_{\sigma_{1}\sigma_{2}}.
    \label{eqgenonfusion}
\end{equation}

While it is straightforward to understand how $G$ acts in the gapped phases\footnote{For instance, for states $\vert a_1,a_2\rangle$ in the trivial domain wall one has $G\vert a_1,a_2\rangle=\vert a_1\times a_2\rangle$ in the genon domain wall, where $a_1\times a_2$ is the result of fusing $a_1$ and $a_2$. For the other direction, one uses \eqref{eqgenonfusion}.}, determining the action on the gapless phases is much harder. We start in the cut open phase, identified with the orbifold CFT at $K = 1$ with $\bZ_2 \times \bZ_2$ symmetry action as in \eqref{eqnZ2symmetry2}. The stress tensors of the (Ising)$^2$ theory can be written in the orbifold variables as:
\begin{equation}\label{eqT1minusT2}
    T_1 - T_2 \sim \cos(2 \tilde\phi + \tilde\theta) \qquad \bar T_1 - \bar T_2 \sim \cos(2 \tilde\phi - \tilde\theta),
\end{equation}
which can be checked by matching the conformal spins (the combinations $T_1 + T_2$ and its conjugate are not vertex operators). 

By the action of $G$ on the stress energy tensors, the first of the operators in \eqref{eqT1minusT2} should be $G$-even while the second should be $G$-odd. We hence find that $G$ involves a transformation
\begin{equation}\label{eqngenondualityguess}
\begin{split}
\tilde\phi \mapsto \tilde\phi + \pi/4\\
\tilde\theta \mapsto \tilde\theta - \pi/2.
\end{split}
\end{equation}
This does not completely determine the transformation rules under $G$, but it is sufficient for determining the action of $G$ on the phase diagram of the genon chain. We leave the complete derivation of the action of $G$ to future work.

We can track this duality action through the $c = 3/2$ point and into the orbifold phase of the genon chain. In the latter, the symmetries $\mathcal{L}_{\sigma_j}$ are spontaneously broken and the two sectors are determined by the presence or absence of a $\sigma$ loop encircling the torus (cf. Section \ref{subsecsymmetrybreaking}). This property is clearly preserved under inserting a genon line, and thus $G$ preserves the sectors. Another way to see the above is to note that the order parameter that detects the sectors, $\mathcal{L}^Y_{\psi_1 \psi_2 \psi_{\bar{1}} \psi_{\bar{2}}}$, is $G$-invariant.

We conclude that $G$ has a well-defined action on the orbifold CFT in each sector. Furthermore, $G$ acts on the $c = 3/2$ point by acting trivially on the third Ising theory, which mediates the spontaneous duality breaking.

From these considerations, we can deduce the action of $G$ on the $c=1$ orbifold phase of the genon chain. To this end, we focus on the $\tilde{K}=1$ point in the $\mathcal{H}^{+}$ sector, which corresponds to two critical Ising theories. At this point, Eqs. \eqref{eqT1minusT2} and \eqref{eqngenondualityguess} hold. This implies that the $c=1$ critical line that separates the trivial and the genon domain walls is self dual. Furthermore, the operator $\cos4\tilde\phi$ that tunes from the critical line to nearby gapped phases is odd under $G$, and hence $G$ interchanges the trivial and genon domain walls, as expected. The operator $\cos8\tilde\phi$ that drives the system into the TC gluing phase is $G$ even, indicating that this phase is $G$ self-dual. In the $\mathcal{H}^{-}$ sector, related to $\mathcal{H}^{+}$ by T-duality, one arrives at similar conclusions by exchanging $\theta$ and $\phi$. 

Interestingly, this allows us to relate the relevant directions in the genon chain to microscopic operators acting on a cut open domain wall of a (KSL)$^2$. In the $\mathcal{H}^{+}$ sector, by Eq. \eqref{eqT1minusT2}, we have
\begin{eqnarray}
        (T_1 - T_2)(\bar T_1 - \bar T_2)&=&(T_1 \bar T_1 + T_2 \bar T_2)-(T_1 \bar T_2 + T_2 \bar T_1) \nonumber\\ &\sim& \cos 4\tilde \phi + \cos 2\tilde \theta.
\end{eqnarray}
In terms of the microscopic operators of the cut open domain wall, $T_1 \bar T_1$ corresponds to a spin-spin interaction across the domain wall in the first layer, $T_1 \bar T_2$ couples spins from one layer with spins from the other layer across the domain wall and so forth. From this, we see e.g. that the operator $\cos4\tilde\phi$ that tunes between the trivial and genon domain walls corresponds microscopically to the difference of the direct and crossed spin-spin interactions across the domain wall.

\section{Discussion}\label{secdiscuss}

In this work, we have analyzed gapped and gapless boundaries in topologically ordered systems by using symmetry principles. The relevant symmetries are not microscopic, but emergent, and encode the locality rules for boundary operators. For gapped boundaries the symmetry principle recovers results based on anyon-condensation. However, the emergent symmetries hold throughout the whole phase diagram of the boundary, and this fact allowed us to characterize gapless phases and phase transitions in several examples.

We focused on a cylinder geometry, where one boundary is fixed in some gapped phase and the other boundary is dynamical. The symmetry operators are bulk anyonic strings encircling the cylinder, while strings connecting the two boundaries serve as order parameters. Twist defects of the bulk topological order induce dualities of the edge phase diagram.

In the case of abelian topological states, the different edge phases are characterized by spontaneous symmetry broken states or SPT phases for the emergent symmetries. On the other hand, for non-Abelian states one must study the less familiar fusion category symmetries. Those may include non-invertible symmetries, as well as transformations that are usually treated as dualities in 1d systems.

There are a couple of questions left to address related to the Ising genon chain we studied. For instance, while we have shown that the stable gapless phase is connected to the cut-open domain wall through a $c = 3/2$ critical point, we have only done so from field theoretic considerations. It remains to realize this transition in a microscopic model, either extending the genon chain or starting from a honeycomb bilayer.

Another microscopic model enjoying the same fusion category symmetry is an (Ising)$^2$ anyon chain \cite{feiguin_interacting_2007,buican_anyonic_2017,gils_anyonic_2013}. We suspect that phase diagrams of such model have the same structure as the ones we obtained for the (KSL)$^2$ domain walls, but this remains to be seen. An interesting question in comparing the two is to understand the role of translation invariance, which appears to have a Lieb-Schultz-Mattis-type relationship with the fusion category symmetry \cite{Pfeifer_2012}.

It may be possible to adapt our methods to study 2d bulk transitions using coupled wire constructions \cite{kane_2002,Sagi_2015,mong_universal_2014,tupitsyn_topological_2010,bais_modulars-matrix_2012,schulz_topological_2013,Kane_2018,tam2019coupled}. Those system can be thought of as an array of domain walls (``wires") in some topologically ordered bulk. In these systems, one has a fusion category symmetry acting on each wire. This seems like a natural setting to try to extend our methods, and we hope to analyze it in the future.

Another interesting direction to explore is to study the zero dimensional edge modes between different gapped or gapless boundaries, \`a la \cite{Lindner2012,clarke2013exotic,Cheng2012,barkeshli_genons_2012,Scaffidi_2017,verresen_gapless_2019,kong_mathematical_2020}. Edge modes between SPT phases protected by fusion category symmetries were studied in \cite{thorngren2019fusion}. Interestingly, there is typically no fully symmetric ``trivial phase" in systems protected by such symmetries. For instance, the Tambara-Yamagami symmetry class we studied has a unique SPT, but no symmetric trivial phase. A closely related symmetry class ${\rm Rep}(D_8)$, related to domain walls in the all-fermion topological order $SO(8)_1$, has three SPT phases, and a triality acting on them \cite{thorngren2019fusion}. It would be fascinating to see what phase diagram appears for such domain walls.

Philosophically speaking, as the topology of the bulk phase is encoded in the identity of the local operators at the boundary, we may now appreciate a kind of global bulk-boundary correspondence. That is, while any given boundary state may not be able to characterize the topological bulk, the complete boundary phase diagram may do so. In particular, there may be certain features of the boundary phase diagram that are unique to a given bulk phase, and if they are observed, can diagnose the topological order of the bulk. A simple example is a $c = 1/2$ boundary transition which requires tuning only a single parameter, even in the absence of any global symmetry. This is a feature which is unique to a bulk topological order with an Abelian anyon of even order, e.g. if it is a $\bZ_2$ spin liquid~\cite{Barkeshli_2014}. Thus we expect the study of boundary critical points of low co-dimension to become central to the study of 2d topological order.


\section{Acknowledgements}
We thank Ari Turner for collaboration on the genon chain in early stages of the project. We also thank Hiromi Ebisu, Yuval Oreg, and Simon Trebst for useful discussions. TL thanks his friends, and especially Evyatar Tulipman for invaluable support. RT thanks Yifan Wang for related collaborations and many uplifting discussions, as well as Dominic Williamson and Dave Aasen for many insightful discussions about topological phases, and especially again Dave Aasen for explaining his unpublished work (following \cite{aasen_topological_2016}), which develops an equivalent symmetry principle for boundaries of topological phases, using different methods.
This research was supported by the Israel Science Foundation Quantum Science and Technology grant no. 2074/19. TL, EB, and AS acknowledge support from
CRC 183 of the Deutsche Forschungsgemeinschaft. 
AS was supported by the European Research Council (Project LEGOTOP) and the Israeli Science Foundation. RT was
supported by the Zuckerman STEM Leadership Program.
NL acknowledges financial support from the European Research Council (ERC) under the European Union Horizon 2020 Research and Innovation Programme (Grant Agreement No. 639172) and  the Israeli Center of Research Excellence (I-CORE) ``Circle of Light.'' 

\bibliography{References.bib}

\appendix
\onecolumngrid
\section{Dualities and Gauging of Discrete Symmetries}\label{appgauging}

\subsection{Kramers Wannier Duality}\label{appgaugingKW}

We illustrate the concepts of dualities and gauging of discrete symmetries in the transverse field Ising model, following Ref.~ \cite{Mcgreevyqft}. We consider a chain with $L$ sites and periodic
boundary conditions:
\begin{equation}
H=\sum_{j}J\sigma_{j}^{z}\sigma_{j+1}^{z}+h\sigma_{j}^{x}.
\end{equation}
The Hamiltonian has a global $\bZ_2$ symmetry $Q=\prod_{j=1}^{L}\sigma_{j}^{x}$.
To facilitate the duality transformation, we introduce an auxiliary spin $\tau^z_j$ on every link $(j,j+1)$, supplemented
by the constraints
\begin{equation}
G_{j}=\sigma_{j}^{z}\tau_{j}^{x}\sigma_{j+1}^{z}=1
\label{eq:constraint}
\end{equation}
for all $j$. Note that this means, in particular, that
\begin{equation}
Q_\tau \equiv \prod_{j}G_{j}=\prod_{j}\tau_{j}^{x}=1.
\end{equation}
We replace the Hamiltonian by
\begin{equation}
H'=\sum_{j}J\sigma_{j}^{z}\sigma_{j+1}^{z}+h\tau_{j}^{z}\sigma_{j+1}^{x}\tau_{j+1}^{z}.\label{eq:Hg}
\end{equation}
Note that $[H',G_{j}]=0$ and $[H',\tau^z_{j}]=0$ for all $j$, while $\{G_j, \tau^z_j\}=0$. Consequently, the spectrum of $H'$ is $2^L$-fold degenerate. This degeneracy is removed by projecting to the physical Hilbert space, defined by the constraints (\ref{eq:constraint}). We can choose a gauge where
$\tau_{j}^{z}=1$ for all $j$ and recover the original Hamiltonian
$H$, showing that $H$ and $H'$ have the same spectrum. 

The KW duality amounts to rewriting $H'$ in terms of the $\tau$ spins (i.e., ``integrating out'' the $\sigma$ spins). This is
done by choosing a different gauge where the state of the $\sigma$ spins
is fixed. The $\sigma$ spins in the first term in $H'$ can be eliminated using the constraint:
\begin{equation}
\sigma_{j}^{z}\sigma_{j+1}^{z}=\tau_{j}^{x}.
\label{eqmappingofoperators}
\end{equation}
The second term can be treated as follows. Using the $\sigma_{j}^{x}$
basis, consider a state in the gauge $\tau_{j}^{z}=1$: 
\begin{equation}
\vert\Psi\rangle=\vert\sigma_{1}^{x}=s_{1},\dots,\sigma_{L}^{x}=s_{L}\rangle\otimes\vert\tau_{1}^{z}=1,\dots\tau_{L}^{z}=1\rangle.
\label{eq:Psi}
\end{equation}
To perform the transformation, let us first assume that on this state, $Q=\prod_{j}\sigma_{j}^{x}=1$.
Then, we can apply a series of gauge transformations to $\vert\Psi\rangle$
and bring it to the form 
\begin{equation}
\vert\Psi\rangle\rightarrow\vert\tilde{\Psi}\rangle=\vert\sigma_{1}^{x}=1,\dots,\sigma_{L}^{x}=1\rangle\otimes\vert\tau_{1}^{z}=s_{1},\tau_{2}^{z}=s_{1}s_{2},\dots\tau_{L}^{z}=s_{1}s_{2}\dots s_{L}=+1\rangle,\label{eq:trans}
\end{equation}
in such a way that
\begin{equation}
\tau_{j}^{z}\tau_{j+1}^{z}=s_{j}=\sigma_{j}^{x}.
\label{eq:strelation}    
\end{equation}
Hence, in the new gauge, 
\begin{equation}
\tilde{H}=\sum_{j}J\tau_{j}^{x}+h\tau_{j}^{z}\tau_{j+1}^{z}.
\label{eq:Htilde}
\end{equation}

Eq.~(\ref{eq:Htilde}) is the KW dual of the original Hamiltonian, with $h$ and $J$ interchanged. 

Note that:
\begin{enumerate}

\item Keeping $\sigma_{j}^{x}$ in the Hamiltonian (\ref{eq:Hg}), we obtain
$\tilde{H}=\sum_{j}J\tau_{j}^{x}+h\tau_{j}^{z}\sigma_{j+1}^{x}\tau_{j+1}^{z}$,
where $\sigma_{j+1}^{x}$ plays the role of a gauge field coupled
to a matter field with charge $\tau_{j}^{x}$. $Q$ is the flux through
the ring, and is referred to in the gauge theory context as the ``magnetic symmetry''.  
The number of physical degrees of freedom in $H$ and $\tilde{H}$ is the same,
since $Q_\tau=1$, but the Hilbert space contains both states
with $Q=1$ and $-1$. The order parameter in the original
theory, $\sigma_{j}^{z}$, twists the boundary condition for the $\tau$ spins.

\item The transformation (\ref{eq:trans}) is done for states where
$Q=\prod_{j=1}^{L}s_{j}=1$, as can be seen by multiplying the relation (\ref{eq:strelation}) on all the bonds. Consistently, the $\tau$ spins also
satisfy a single $\bZ_{2}$ constraint: $Q_\tau=1$.
Had we started from a state with $Q=-1$, the application of the gauge transformation would inevitably leave an odd number of negative $\sigma^x_j$'s, since $[Q,G_j]=0$. This would leave the $h$ term in the Hamiltonian (\ref{eq:Hg}) with $\sigma^x_j=-1$ on a single site, i.e., the $\tau^z_j$'s satisfy twisted (anti-periodic) boundary conditions.  
Henceforth, we limit the transformation (\ref{eq:trans}) to states that satisfy $Q=1$.
States with $Q=-1$ are sent to zero.
\item
In the gauge theory, the disorder operator becomes a local operator $\tau^z_j$. The disordered phase ($h>J$) is the ferromagnetic phase of the $\tau$ spins. However, this phase has a unique ground state on a ring, due to the constraint $Q_\tau=1$. The ordered phase ($J>h$) is the paramagnetic phase of the $\tau$ spins, but it has two degenerate ground states with $Q=\pm 1$ (corresponding to periodic and anti-periodic boundary conditions for $\tau^z_j$). We can think of this phase as spontaneously breaking the \emph{magnetic symmetry} $Q$ of the gauge theory, whose $\bZ_2$ conserved charge measures the holonomy (or flux) of the gauge field around a circle and for which all twist operators $\sigma^z_j$ are charged \cite{dijkgraaf1989,bhardwaj_finite_2017}. 

\end{enumerate}

The KW duality transformation may also be regarded as an operator $D$ that maps between the Hilbert space of the $\sigma$'s and itself. Explicitly, 
\begin{equation}\label{eqDoperator}
D \Big\vert \{ \sigma^x_j = s_j\}\Big\rangle=(1+Q)\Big\vert \{\sigma^z_j=\prod_{i\le j}s_i \}\Big\rangle
\end{equation}
(cf. Eqs.~\ref{eq:Psi},\ref{eq:trans}). The transformation first maps the Hilbert space of the $\sigma$'s to that of the $\tau$'s that reside on the links of the lattice, then shifts the lattice by half a unit cell and finally renames the $\tau$'s back to $\sigma$'s. 

The gauging procedure is reminiscent of Kramers-Wannier duality in higher dimensions \cite{doi:10.1063/1.1665530,RevModPhys.51.659}, although one dimension is special because we can identify the gauge degrees of freedom (which live on the bonds) with the spin degrees of freedom (which live on the sites) by a ``half unit" translation. 

Along with the global symmetry $Q$, the operator $D$ satisfies an algebra dubbed the Ising fusion algebra \cite{frohlich_kramers-wannier_2004,aasen_topological_2016}:

\begin{equation}\label{eqnisingfusionapp}
\begin{gathered}
{Q}^2 = 1 \\
{Q} {D}= {D}{Q}={D} \\
{D}^2 = 1 + {Q}.
\end{gathered}
\end{equation}
For the interpretation of the duality and symmetry operators in the context of topological defects on the lattice, see \cite{aasen_topological_2016}. For a related discussion on $\bZ_2$ gauging see Ref.~ \cite{ji_categorical_2020}. 

\subsection{Gauging Symmetries in CFT}

Recall the operator--state correspondence in CFT's. Via a conformal mapping between the cylinder and the plane, one maps boundary states on a cylinder to local operators on the plane \cite{ginsparg_applied_1988,francesco_conformal_1997}. One could consider a twisted version of this correspondence, relating operators at the end of defect lines to states in the twisted sector associated with this defect. This is done by inserting a defect line at some point in space, extending in the time direction. By applying the same conformal mapping to the plane, a correspondence between states in the twisted sector and operators at the end of defect lines (twist operators) is established.

We will now gauge the $\bZ_2$ symmetry of the Ising CFT, complementing our description on the lattice in Appendix \ref{appgaugingKW}. This amounts to including states in the twisted sector of the symmetry, as well as symmetrizing all the states. By operator state correspondence, we the result is a CFT where all charged operators are projected out and twist operators are included. 

This procedure is implemented neatly at the level of partition functions: to include the twisted sector, one adds to the original partition function the sector with twisted boundary conditions in the space direction. To remain only with symmetric states, one adds two additional partition function, with the symmetry line inserted in the time direction (in red). 

\begin{equation}
    \includegraphics[width=8cm]{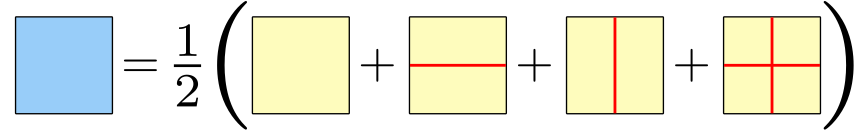}
\end{equation}
The first and second terms combine into:
\begin{equation}{\rm Tr}_\mathcal{H} \frac{1}{2}(1 + \mathcal{L}_\epsilon) e^{-\beta H}\end{equation}
which projects onto symmetric operators in the untwisted sector. The operator $\mathcal{L}_\epsilon$ is the Verlinde line for $\epsilon$, which implements the $\bZ_2$ global symmetry $Q$ in the CFT. The third and fourth terms combine into
\begin{equation}{\rm Tr}_{\mathcal{H}_{\rm twisted}} \frac{1}{2}(1 + \mathcal{L}_\epsilon) e^{-\beta H},\end{equation}
which projects onto symmetric operators in the twisted sector $\mathcal{H}_{\text{twisted}}$ of the Hilbert space. This matches the expected operator content we mentioned above.

\subsection{Ising CFT and its Duality Defect}\label{appKWduality}

The Ising CFT has three primary fields, $1,\sigma,\epsilon$. The spin field $\sigma$ is charged under the $\bZ_2$ symmetry, and represents the scaling limit of the lattice operator $Z_{i}$. The energy field $\epsilon$ is uncharged and relevant, and perturbs off criticality. The CFT also contains a non-local disorder field $\mu$, which is the scaling limit of the finite symmetry string:

\begin{eqnarray}\label{eqnlatticecontcorresp}
\sigma^z_{i}&\sim&\sigma(x)\nonumber\\
\sigma^z_{i}\sigma^z_{i+1}-\sigma^x_{i}&\sim&\epsilon(x)\nonumber\\
\prod_{j < i}\sigma^x_{i}&\sim&\mu(x)
\end{eqnarray}

The Ising CFT has topological defect lines (TDL) with the Ising fusion algebra \cite{frohlich_kramers-wannier_2004} (The Verlinde lines for the primary fields \cite{VERLINDE1988360}). When a TDL is dragged over local bulk fields, there are two possible scenarios. Either the TDL is associated with an invertible operator, meaning it maps local operators to local operators, or it is non-invertible, and some local operators are sent to non-local ones attached to the end-point of a string in the fusion of the TDL with its dual. In the Ising case, $\mathcal{L}_{\epsilon}$ is the invertible $\bZ_2$ symmetry line, while $\mathcal{L}_{\sigma}$ annihilates all $\bZ_2$-charged states. The operator $\mathcal{L}_{\sigma}$ should be thought of as the continuum counterpart of the operator $D$ in \eqref{eqDoperator} - it is the KW duality when considered as a transformation on the Hilbert space. When the $\mathcal{L}_{\sigma}$ line is dragged over a spin field, it changes it to the disorder operator, which lives at the end-point of a $\bZ_2$ string \cite{Fr_hlich_2007}. The energy operator is charged under this duality line.

The fusion rules of these TDLs
\begin{equation}\mathcal{L}_\sigma^2 = 1 + \mathcal{L}_\epsilon\end{equation}
along with their F-moves allow us to make contact with the picture of Kramers-Wannier duality as $\bZ_2$ gauging. Indeed, following Section 5.1 of \cite{aasen_topological_2016}, we imagine nucleating a small $\mathcal{L}_\sigma$ loop (in green) and then dragging it around the system. This does not affect the partition function because it is a symmetry, and one can derive the identity:
\begin{equation}
    \includegraphics[width=11cm]{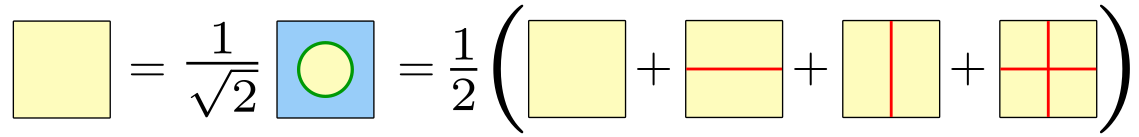}.
\end{equation}
On the LHS we have the usual partition function of the Ising CFT, while on the RHS we have the $\bZ_2$ gauged partition function. When we expand either side in terms of the Ising characters, we find this equality corresponds to the mapping of operators in Eqs. \eqref{eq:strelation},\eqref{eqmappingofoperators}. For instance, from the untwisted sector the $\bar \chi_\sigma \chi_\sigma$ piece corresponding to the order parameter is projected out (occurring with opposite sign in the first and third terms), while from the untwisted sector we regain a $\bar \chi_\sigma \chi_\sigma$ piece from the disorder operator, but the fermion towers $\bar \chi_I \chi_\psi$ and $\bar \chi_\psi \chi_I$ are projected out.

\subsection{Shift Symmetries of the Compact Boson}\label{appshiftsymmetry}

The compact boson has an anomaly free shift symmetry
\begin{equation}
    S_{\theta}=\begin{cases} \theta\mapsto\theta+\pi\\ \phi \mapsto \phi
\end{cases}
\end{equation}
the gauging of which leads to a compact boson of different radius, with fields
\begin{equation}\label{appeqnshiftsymmgauged}
    \begin{gathered}
    \tilde \theta = 2\theta \\
    \tilde \phi = \phi/2,
    \end{gathered}
\end{equation}
effectively rescaling the Luttinger liquid parameter by $\tilde K = 4K$. Indeed, gauging $S_\theta$ projects out all vertex operators $\sin n\theta$, $\cos n\theta$ with $n$ odd. Meanwhile, the operators in the $S_\theta$-twisted sector are $\phi$ vertex operators with half-integer coefficient. This is since the open symmetry string operator for the $\theta$ shift symmetry is:
\begin{equation}\label{appeqntwistop}\exp \left( i \frac{\alpha}{2\pi} \int_{0}^x \partial \phi\right ) = e^{-i\frac{\alpha}{2\pi} \phi(0)}e^{i\frac{\alpha}{2\pi} \phi(x)}.\end{equation}
Thus, the periodicity of $\theta$ is effectively doubled while that of $\phi$ is effectively halved, hence we find \eqref{appeqnshiftsymmgauged}. See \cite{ginsparg_applied_1988} for more details.

The magnetic symmetry we obtain by gauging by definition gives $-1$ to operators in the twisted sector and $+1$ to all other operators. We see in these variables that it acts as
\begin{equation}
    M=\begin{cases} \tilde\theta\mapsto\tilde\theta\\ \tilde\phi \mapsto \tilde\phi + \pi,
\end{cases}
\end{equation}
as this picks out those operators such as $\cos(\phi/2)$ which come from the twisted sector, and are non-local in the ungauged theory. We observe that while $S_\theta$ shifts $\theta$, $M$ shifts $\tilde \phi$, as we have observed in the two sectors in Section \ref{secKSLgenon}. This is related to $T$-duality in Section \ref{subsecisingcubed}.

\section{The Genon Chain}\label{app:genonchain}

\subsection{Different Bases and Gapped Phases}\label{appgappedphases}

\begin{figure}
    \centering
    \includegraphics[width=10cm]{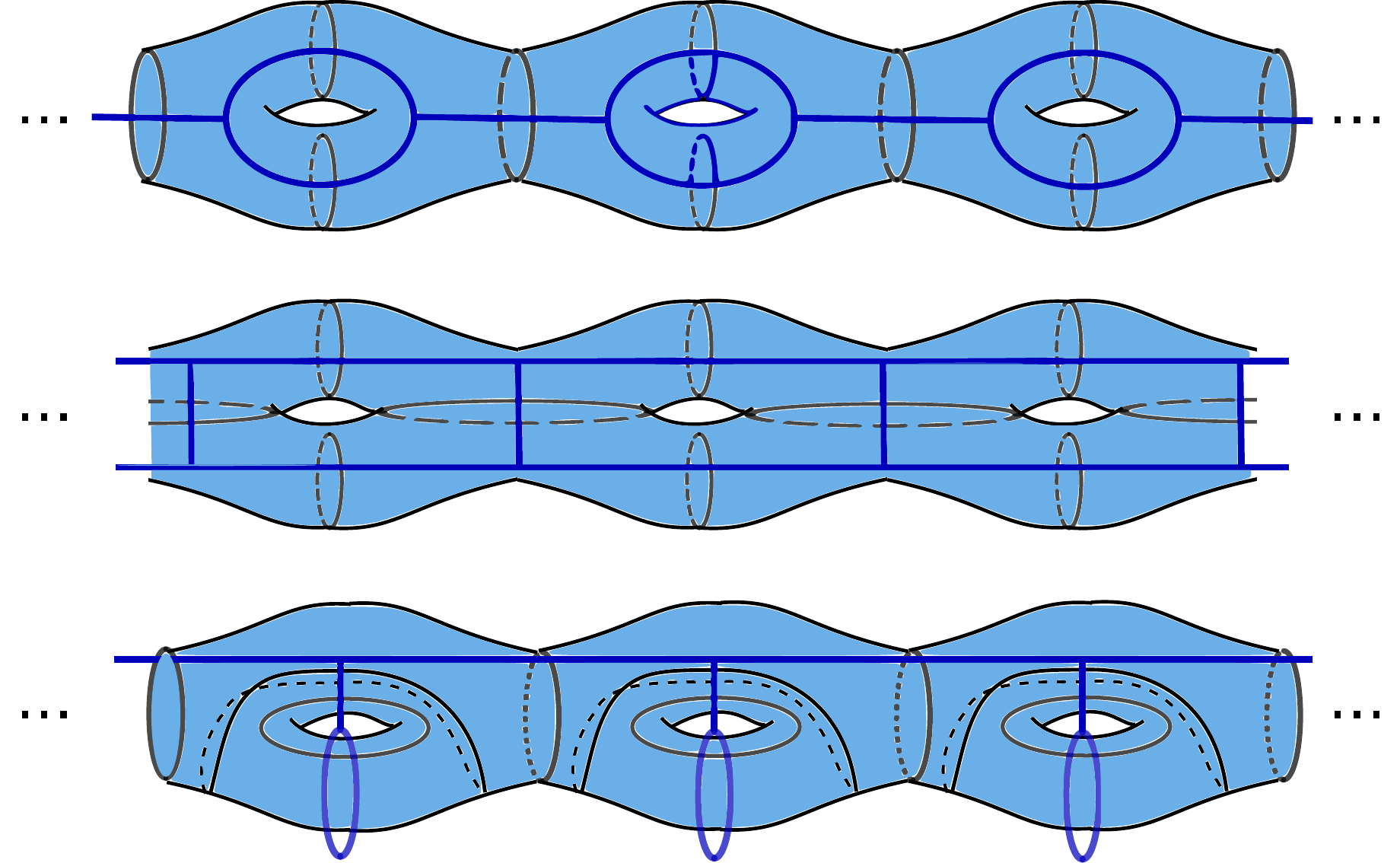}
    \caption{Three convenient pants decompositions of the infinite genon double cover. The surface is cut on the black circles, and split into pair of pants. Each line in the blue skeleton carries an anyon label, and fusion rules are enforced on each vertex.}
    \label{Pants}
\end{figure}

As discussed in \ref{subsec5genonchain}, different pants decompositions of high genus surfaces give rise to different bases for the Hilbert space of the genon chain. In Fig. \ref{Pants}  we present three different pants decompositions along with their respective fusion graphs. The first two decompositions are related by an F move. The first and the third are related by an F move, followed by an S move \cite{hatcher1999pants}.

The ground states in the three gapped phases of the genon chain can be written explicitly using those three bases. Conveniently, states which are difficult to write in one pants decomposition are easy to write in another. We list the ground state wave-functions, sticking to the notation of Sec. \ref{subsec5bilayerphases}. The 9 ground states in the trivial domain wall are:
\begin{equation}
   \includegraphics[width=5cm]{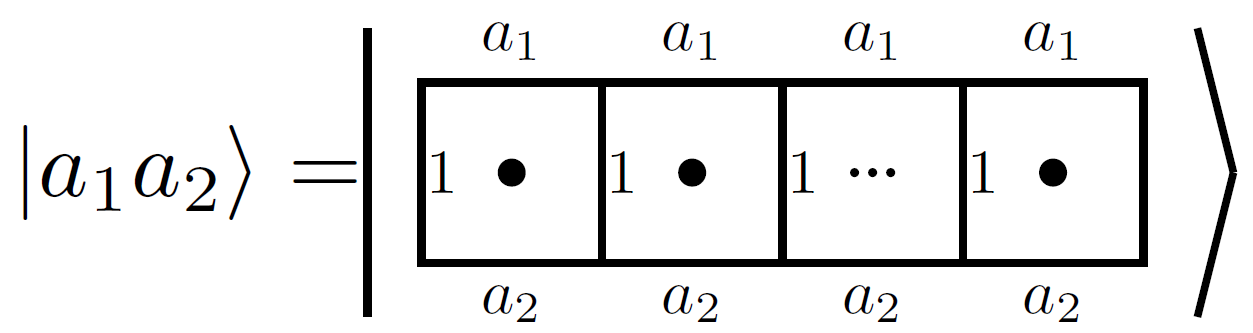}
   \label{eq:gstrivial}
\end{equation}
The 3 ground states of the genon domain wall are:
\begin{equation}
   \includegraphics[width=5cm]{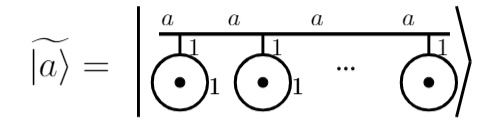}
\end{equation}
The 6 ground states in the toric code gluing domain wall are:
\begin{equation}
   \includegraphics[width=12cm]{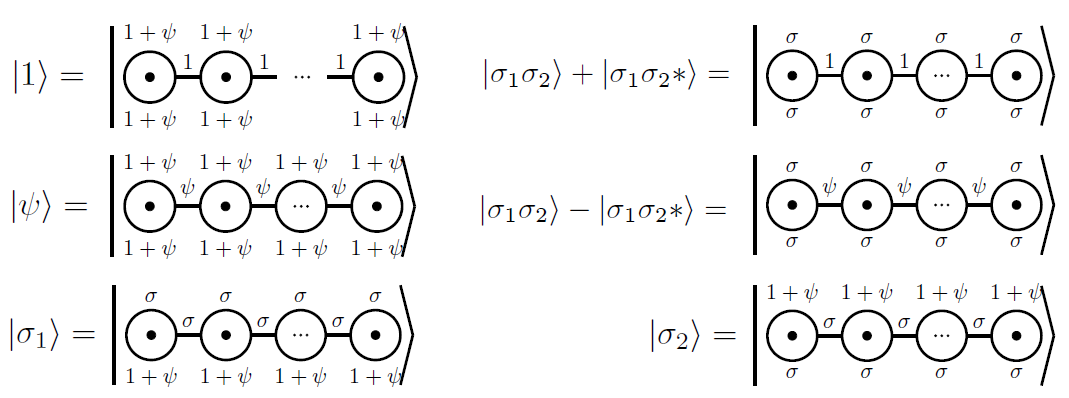}
\end{equation}

Each of these ground states belongs to one of the sectors $\mathcal{H}^{\pm}$. Together, the ground state subspace forms the phases listed in the first three rows of Table~\ref{tableTYPhases}. For example, consider the nine ground states of the trivial domain wall (Eq.~\ref{eq:gstrivial}). The four states $\vert \sigma, a\rangle$ and $\vert a, \sigma\rangle$ with $a=1,\psi$ are in the $\mathcal{H}^-$ sector; by applying $\mathcal{L}_{\psi_1}$ and $\mathcal{L}_{\psi_2}$ to these states, one can verify that these are the ground states of $I_x + I_y$. Similarly, the other five ground states (where either both $a_1$, $a_2$ are equal to $\sigma$, or neither is equal to $\sigma$) are in $\mathcal{H}^+$, and these states form the phases Brok$+$Symm. 

\begin{figure}
    \centering
    \includegraphics[width=16cm]{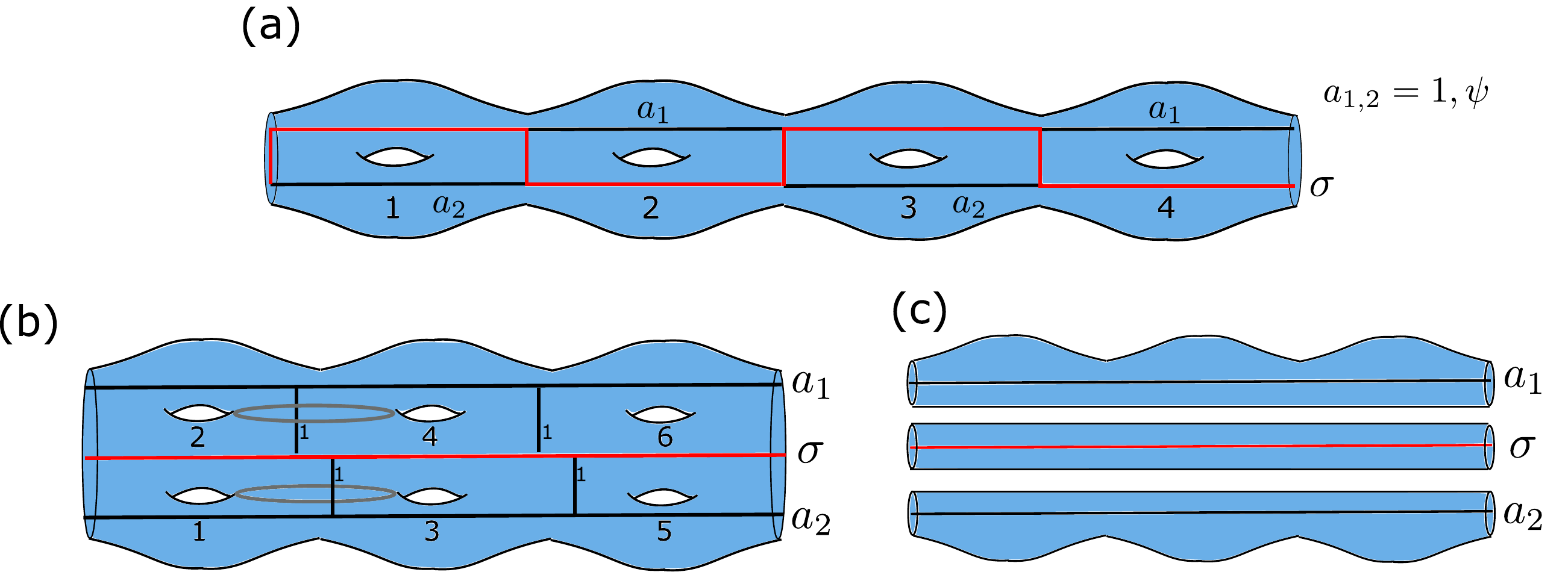}
    \caption{Microscopic construction of the phases in the last three rows in Table \ref{tableTYPhases}. 
    We illustrate the construction for the Brok phase in the $\mathcal{H}^-$ sector. The corresponding states are shown in (a). To obtain these states, we first deform our surface so that every even-numbered hole is moved on top of its left neighbor. This deformation, and a convenient ``three-leg ladder'' basis corresponding to it, are shown in (b). In this setup, global symmetries still act by fusion at the top and bottom legs of the ladder. The phase Brok in $\mathcal{H}^-$ is realized by passing a $\sigma$ string through the central leg, 1's on all rungs, and 1 or $\psi$ on the outer legs. This state corresponds to the disconnected surface shown in (c). By separating the part containing $\sigma$ this way, we have an effective two-leg ladder skeleton in the $\mathcal{H}^+$ sector, and we can apply the constructions above to find the other phases. deforming the holes back to their original position, we find that the $\sigma$ string zig-zags through the system, as shown in (a).}
    \label{figsectorswtiching}
\end{figure}

This naturally raises the question whether the phases in the first column of Table~\ref{tableTYPhases} [realized by the Hamiltonian~\eqref{eqH} in the $\mathcal{H}^-$ sector] can also be realized by a different Hamiltonian in the $\mathcal{H}^+$ sector, and vice versa. 
E.g., can the phase Brok$+$Symm be realized in the $\mathcal{H}^-$ sector by a different Hamiltonian? The field theory construction suggests that should be possible. E.g., as argued at the end of Sec.~\ref{subsubseccritline}, when the Luttinger parameter $K$ is small enough, the operator $\cos4\theta$ becomes relevant and drives the system into Brok$+$Symm if its coefficient is negative. It turns out that this phase can indeed be accessed microscopically, if one allows for longer-range interactions in the genon chain model. Such a construction is shown in Fig.~\ref{figsectorswtiching}.  

\subsection{Mapping to the Ashkin-Teller Model}\label{appATmapping}

In the $\mathcal{H}^-$ sector we have four states per two genons \cite{gils_ashkinteller_2009}, which we represent in the bubble basis with an angular variable $\alpha$:
\begin{equation}
    \includegraphics[width=8cm]{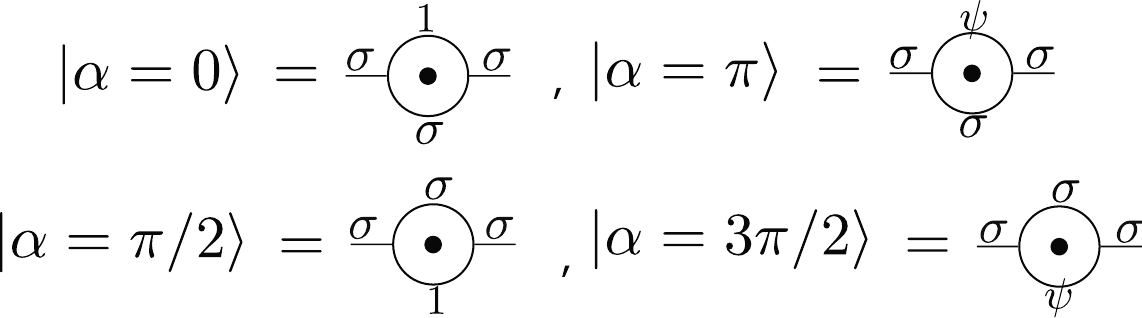}
\end{equation}
We also define a dual operator $e^{i\beta}$ which rotates $\alpha$ by $\pi/2$.

Recall that the genon chain Hamiltonian has 4 terms:
\begin{equation}
\begin{split}
H=\sum_{R}J_{\sigma}^{R}W_{\sigma}^{R}+J_{\psi}^{R}W_{\psi}^{R}\\
+\sum_{P}J_{\sigma}^{P}W_{\sigma}^{P}+J_{\psi}^{P}W_{\psi}^{P}
\end{split}
\end{equation}
Those operators are represented pictorially on the surface as:
\begin{equation}
    \includegraphics[width=8cm]{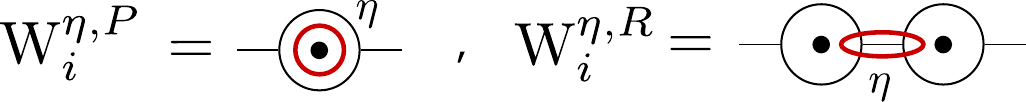}
\end{equation}
In addition, we define the operators:
\begin{equation}
    \includegraphics[width=8cm]{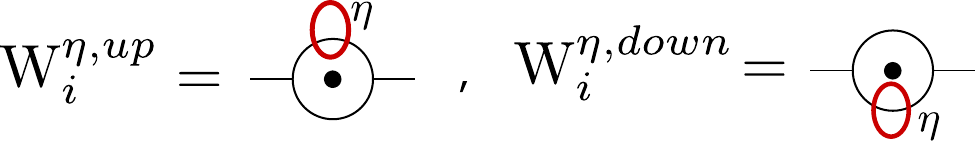}
\end{equation}
Gils shows \cite{gils_ashkinteller_2009} that in term of 4-states clock variables the operators are:

\begin{equation}
\begin{matrix}

W_{\psi}^{P_{i}}&=\cos(2\beta_{i}) & W_{\sigma}^{P_{i}}=\frac{1}{\sqrt{2}}\cos(\beta_{i})\\

W_{\psi}^{R_{i}}&=\cos(2(\alpha_{i}-\alpha_{i+1})) & W_{\sigma}^{R}=\frac{1}{\sqrt{2}}\cos(\alpha_{i}-\alpha_{i+1})\\

W_{\psi}^{up_{i}}&=\cos(2\alpha_{i}) & W_{\sigma}^{up_{i}}=\frac{1}{\sqrt{2}}\cos(\alpha_{i})\\

W_{\psi}^{down_{i}}&=-\cos(2\alpha_{i}) & W_{\sigma}^{down_{i}}=\frac{1}{\sqrt{2}i}\sin(\alpha_{i})\\

\end{matrix}
\end{equation}
To get the Ashkin-Teller spin chain as defined in Ref. \cite{kohmoto_hamiltonian_1981}, one sets $J_{\sigma}^{P,R}=\lambda J_{\psi}^{P,R}$, $J_{i}^{P}=\beta J_{i}^{R}$ and identifies $\beta,\lambda$, as the two parameters in the Ashkin Teller model and  and set the overall coefficient.

One can show using the pictorial calculus that in this basis the $\bZ_2 \times \bZ_2$ symmetries act as: 

\begin{equation}
\begin{split}
 \mathcal{L}_{\psi_{1}}&: |0\rangle\leftrightarrow|\pi\rangle \\  
 \mathcal{L}_{\psi_{2}}&: |\pi/2\rangle\leftrightarrow|3\pi/2\rangle  
\end{split}
\end{equation}

\begin{figure}
    \centering
    \includegraphics[width=9cm]{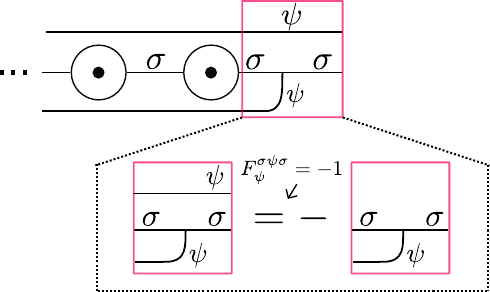}
    \caption{By using pictorial calculus, we find that each $\psi$ string is charged under the other, so that the presence of the $\sigma$ line induces an SPT twist (aka discrete torsion for the $\bZ_2 \times \bZ_2$ symmetry). This is a consequence of $F_{\psi}^{\sigma\psi\sigma}=-1$.}
    \label{Fmoveapp}
\end{figure}

However, we point out an important subtlety: If we consider finite symmetry strings, we see that the string end of one symmetry is charged under the other symmetry and vice versa. In other words, the $\bZ_2 \times \bZ_2$ symmetry has discrete torsion (see Fig. \ref{Fmoveapp}).

We now focus on the $J_{\psi}=0$ point. Perform a change of variables $|\alpha_{i}=0,\pi/2,\pi,3\pi/2\rangle\to|\uparrow\uparrow\rangle,|\uparrow\downarrow\rangle,|\downarrow\downarrow\rangle,|\downarrow\uparrow\rangle$. We can write the spin operators in terms of the angular variables.
\begin{equation}
\begin{split}
Z^{1}_{i}=\cos({\alpha_{i}})+\sin({\alpha_{i}})\\
Z^{2}_{i}=\cos({\alpha_{i}})-\sin({\alpha_{i}})
\end{split}
\end{equation}

In terms of those variables, the Hamiltonian takes the form of two decoupled Ising models:

\begin{equation}
\begin{split}
H=\sum_{i}J_{\sigma}^{P}Z^{1}_{i}Z^{1}_{i+1}+J_{\sigma}^{R}X^{1}_{i}+J_{\sigma}^{P}Z^{2}_{i}Z^{2}_{i+1}+J_{\sigma}^{R}X^{2}_{i}
\end{split}
\end{equation}
For $J_{\sigma}^{P}=J_{\sigma}^{R}$ both Ising models are critical, and we obtain the $K=1$ point along the orbifold line.
We identify the $\bZ_2 \times \bZ_2$ symmetry action from equations (B3) and (B4) as:
\begin{equation}
\begin{split}
        \mathcal{L}_{\psi_{1}}&:(Z_1,Z_2)\mapsto (Z_2,Z_1),\\ \mathcal{L}_{\psi_{2}}&:(Z_1,Z_2)\mapsto (-Z_2,-Z_1).
\end{split}
\end{equation}

The above equations are used to read off the action on the (Ising)$^2$ CFT scaling fields (Eq. \eqref{eqsymmetryIsing2} in the main text).

\subsection{KW Duality Action on Local Operators}\label{appdualmapping}

Let us focus on the $\mathcal{H}^{-}$ sector. Under $\mathcal{L}_{\psi_1}$,$\mathcal{L}_{\psi_2}$, the operators of the orbifold transform as in Eqs. \eqref{eqnZ2symmetry}. By considering the symmetry action on minima of potentials such as $\cos n \phi$, $\cos n \theta$, one can figure out to which $\bZ_{2}\times \bZ_{2}$ phase they belong. For example, the state $\phi=\pi/2$ is invariant under both $\bZ_{2}$ generators, but as argued above, the ends of the two $\bZ_2$ strings anti-commute and the system is in the SPT phase. At the fixed points $\phi = 0, \pi$ or $\theta = 0,\pi$, we must take into account the twisted sectors---there are two states associated with each of these values of $\phi$ or $\theta$, leading to a spontaneously breaking of the magnetic symmetry $M$ (cf. \eqref{eqnZ2symmetry}). We find that the six $\bZ_2 \times \bZ_2$ gapped phases are represented as: 
\begin{equation}
    \begin{gathered}
    |\phi=\pi/2 \rangle \in {\rm SPT} \\ 
    |\theta = \pi/2 \rangle \in {\rm Sym} \\
    |\phi=0 \rangle \in I_x \\
    |\phi=\pi \rangle \in I_y \\
    |\theta=\pi/4 \rangle, |\theta = 3\pi/4\rangle \in I_z \\
    |\theta=0\rangle, |\theta = \pi\rangle \in {\rm Brok}.
    \end{gathered}
\end{equation}

To find the charges of the operators $\cos n \phi$, $\cos n\theta$ under $\mathcal{L}_{\sigma_{1}\sigma_{2}}$, we can check to which gapped phases the operator perturbs the orbifold (for either sign), and then consider the action of $\mathcal{L}_{\sigma_{1,2}}$ on this phase. If this results in the same phase, the operator is neutral, otherwise it is charged. For example, $\pm \cos\phi$ perturbs into $I_x$ or $I_y$, depending on the sign of its coefficient. Under $\mathcal{L}_{\sigma_{1}\sigma_{2}}$, the phases $I_x$ and $I_y$ are exchanged (cf. Table II), therefore $\mathcal{L}_{\sigma_{1}\sigma_{2}}$ maps $\cos\phi$ to $-\cos\phi$. 
In contrast, consider $\pm \cos 2\phi$ tunes between the SPT and $I_x + I_y$. Both phases are invariant under $\mathcal{L}_{\sigma_{1}\sigma_{2}}$, so $\pm \cos 2\phi$ is uncharged. We summarize the result of this analysis in Table V. 

For each operator $\hat{O}$ in the table, we present the eigenvalue of the symmetry when acting on its corresponding state $\hat{O}|0\rangle$. Note that for non-unitary symmetries, symmetric operators are characterized by this eigenvalue being equal to the quantum dimension, or VEV of the symmetry operator (which is 1 for unitary symmetries) \cite{chang_topological_2019} (this is a generalization of the Perron-Frobenius theorem on the dominant eigenvalue, and in fact provides an equivalent definition for the quantum dimension \cite{etingof2015tensor}).

\begin{table}
\caption{Orbifold charge assignments of vertex operators (i.e. the action of the symmetries on the corresponding states). For the duality symmetry $\mathcal{L}_{\sigma_{1}\sigma_{2}}$, eigenstates with eigenvalues 2 correspond to symmetric operators. Twist field are transformed as in Eqs. \eqref{eqnZ2symmetry},\eqref{eqndualityactionminussector}. The operators $\mathcal{L}_{\sigma_{1}},\mathcal{L}_{\sigma_{2}}$ map the first column to the second and vice versa. Symmetry-allowed perturbations are highlighted in boldface.}
\begin{tabular}{ll} 
\begin{tabular}{ |c|c|c|c|c| } 
 \hline
$\mathcal{H}^{-}$ Operator & $\mathcal{H}^{+}$ Operator &$\mathcal{L}_{\psi_{1}}$ &$\mathcal{L}_{\psi_{2}}$  & $\mathcal{L}_{\sigma_{1}\sigma_{2}}$ \\ \hline
  &$\cos\tilde{\phi}$&$-$1&$-$1&0\\
$\cos\phi$  & $\cos2\tilde{\phi}$& 1&1&$-$2 \\
&$\cos3\tilde{\phi}$&$-$1&$-$1&0\\
$\bm{\cos 2\phi}$  & $\bm{\cos 4\tilde{\phi}}$ &1&1&2 \\
&$\cos5\tilde{\phi}$&$-$1&$-$1&0\\
$\cos3\phi$  & $\cos 6\tilde{\phi}$& 1&1&$-$2 \\
&$\cos7\tilde{\phi}$&$-$1&$-$1&0\\
$\bm{\cos 4\phi}$  & $\bm{\cos 8\tilde{\phi}}$&1&1&2 \\
$\cos\theta$  &  &$-$1&$-$1&0\\
$\cos2\theta$  &$\cos\tilde{\theta}$&  1&1&$-$2\\
 \hline
\end{tabular} &

\end{tabular}
\label{tablecharges}
\end{table}

To obtain a complete list of charge assignments for all operators, one must apply modular bootstrap techniques to compute the torus partition function twisted by $\mathcal{L}_{\sigma_1\sigma_2}$. This was solved for the $\mathcal{H}^+$ sector in \cite{thorngren2019fusion} at the (Ising)$^2$ CFT point $\tilde K = 1$. The general formula will appear in \cite{thorngren2019fusion2}:
\begin{equation}{\rm Tr}_{\mathcal{H}^+} \mathcal{L}_{\sigma_1 \sigma_2} q^{L_0-1/24}\bar q^{\bar L_0 - 1/24} = {1\over  |\eta|^2} \left(
\sum_{m,n\in \bZ }  
(-1)^{m+n}q^{{1\over 2}\left({2n\over R}+{mR \over 2}\right)^2}\bar q^{{1\over 2}\left({2n\over R}-{mR\over 2 }\right)^2}
+
\sum_{m,n\in \bZ}  (-1)^{m+n}q^{ m^2}\bar q^{ n^2}
\right),\end{equation}
where $R^2 = 4 \tilde K$. For comparison, the untwisted partition function is \cite{francesco_conformal_1997}
\begin{equation}{\rm Tr}_{\mathcal{H}^+} q^{L_0-1/24}\bar q^{\bar L_0 - 1/24}\end{equation}\begin{equation} = \frac{1}{|\eta|^2}\left(\frac{1}{2}\sum_{m,n \in \bZ} q^{\frac{1}{2}\left(\frac{n}{R} + \frac{mR}{2}\right)^2}\bar q^{\frac{1}{2}\left(\frac{n}{R} - \frac{mR}{2}\right)^2} + \sum_{j=0}^{j=3} \sum_{m,n \in \bZ}q^{(8n+2j+1)^2/16} \bar q^{(8m+2j+1)^2/16} + \frac{1}{2} \sum_{m,n \in \bZ} (-1)^{m+n} q^{m^2} \bar q^{n^2}\right).\end{equation}
The first part of this partition function counts the $C$-invariant vertex operators $\cos n \tilde\phi$, $\cos m \tilde\theta$ and their $C$-invariant products $\cos \tilde\phi \cos \tilde\theta$, $\sin \tilde\phi \sin \tilde\theta$, etc. The second term counts the twist operators $\sigma_i$ (corresponding to $j = 0,2$) and $\tau_i$ (corresponding to $j = 1,3$). The third term is there for the proper counting of $C$-invariant operators made only from the currents $\partial \theta$, $\partial \phi$, and no vertex operator.

We see by comparing with the above that the operators with $n$ odd, corresponding to $\cos n \tilde\phi$, have been projected out. The charge of $\cos 2n \tilde\phi$ is $(-1)^n$ and the charge of $\cos m \tilde\theta$ is $(-1)^m$. An operator such as $\sin 2\tilde\phi \sin \tilde\theta$ which we could not easily analyze by the method above is seen to be even. Meanwhile the lowest surviving non-vertex operator is the exactly marginal $(1,1)$ operator, as we expected. In general, we have
\begin{equation}
    V^+_{n,m} \mapsto \begin{cases} 0 & n {\rm\ is\ odd} \\ 2(-1)^{n/2+m}V^+_{n,m} & {\rm otherwise}
    \end{cases},
\end{equation}
where $V^+_{n,m}$ is any $C$-invariant vertex operator with $\tilde\phi$ index $n$ and $\tilde \theta$ index $m$. All twist operators are sent to zero, while all $C$-invariant Schur polynomials in the compact boson currents have eigenvalue 2. More succinctly,
\begin{equation}
\mathcal{L}_{\sigma_1 \sigma_2}^X = \frac{1}{2}(1 + (-1)^n)(1 + M)i^{n+2m} \qquad \mathcal{H}^+ {\rm\ sector}.
\end{equation}
Using $\mathcal{L}_{\psi_1}^X = (-1)^n$, $\mathcal{L}_{\psi_2}^X = (-1)^n M$, we find
\begin{equation}
    (\mathcal{L}_{\sigma_1 \sigma_2}^X)^2 = (1 + (-1)^n)(1 + M) = 1 + \mathcal{L}_{\psi_1}^X + \mathcal{L}_{\psi_2}^X + \mathcal{L}_{\psi_1 \psi_2}^X,
\end{equation}
as required by \eqref{eqnTYfusionalg}. The action in the $\mathcal{H}^-$ sector is obtained by applying $T$-duality:
\begin{equation}
\mathcal{L}_{\sigma_1 \sigma_2}^X = \frac{1}{2}(1 + (-1)^m)(1 + M)i^{m+2n} \qquad \mathcal{H}^- {\rm\ sector}.
\end{equation}

\end{document}